\documentclass[12pt]{article}
\usepackage{epsfig}
\usepackage{enumerate}
\usepackage{amsmath}
\usepackage{amssymb}
\usepackage{url}
\usepackage{hyphenat}
\usepackage{verbatim}
\usepackage{accents}
\usepackage{graphicx}
\usepackage{subcaption}
\usepackage{listings}
\usepackage{epstopdf}
\usepackage{svg}
\usepackage[utf8]{inputenc}
\usepackage{todonotes}
\usepackage{booktabs}
\usepackage{authblk}
\usepackage{float}

\usepackage[
backend=biber,
style=nature,
]{biblatex}
\usepackage{algorithmicx}
\usepackage{algorithm}
\algnewcommand\algorithmicinput{\textbf{Input: }}
\algnewcommand\algorithmicoutput{\textbf{Output: }}
\usepackage{algpseudocode}
	\algnewcommand\algorithmicforeach{\textbf{for each}}
	\algdef{S}[FOR]{ForEach}[1]{\algorithmicforeach\ #1\ \algorithmicdo}
	\algnewcommand\algorithmictimes{\textbf{times}}
	\algdef{E}[REPEAT]{Times}[1]{#1\ \algorithmictimes}
	\algrenewcommand\textproc{\textsf}
	\algnewcommand{\IfThen}[2]{\State \algorithmicif\ #1\ \algorithmicthen\ #2}
 
\usepackage{hyperref}
\hypersetup{colorlinks=true}

\usepackage[capitalise, noabbrev, nameinlink]{cleveref}

\everymath{\displaystyle}

\newcommand\black[1]{\textcolor{black}{#1}}

\title{GrainPaint: A multi-scale diffusion-based generative model for microstructure reconstruction of large-scale objects}

\author[a]{Nathan Hoffman$^{*,\dag,}$}
\author[b]{Cashen Diniz$^{*,}$}
\author[c]{Dehao Liu}
\author[d]{Theron Rodgers}
\author[d]{Anh Tran$^{\dag,}$}
\author[a,b]{Mark Fuge}
\affil[a]{Department of Mechanical Engineering, University of Maryland, College Park, Maryland 20742, United States}
\affil[b]{Department of Mechanical and Process Engineering, ETH Zürich, Rämistrasse 101, 8092 Zürich, Switzerland}
\affil[c]{Department of Mechanical Engineering, State University of New York at Binghamton, Binghamton, NY 13902, United States}
\affil[d]{Sandia National Laboratories, Albuquerque, NM 87123, United States}

\begin{document}

\maketitle
\def\thefootnote{*}\footnotetext{These authors contributed equally to this work.}
\def\thefootnote{\dag}\footnotetext{Corresponding authors: nhoffma1@umd.edu, anhtran@sandia.gov.}

\begin{abstract}
\begin{figure}[!htbp]
\includegraphics[width=\textwidth]{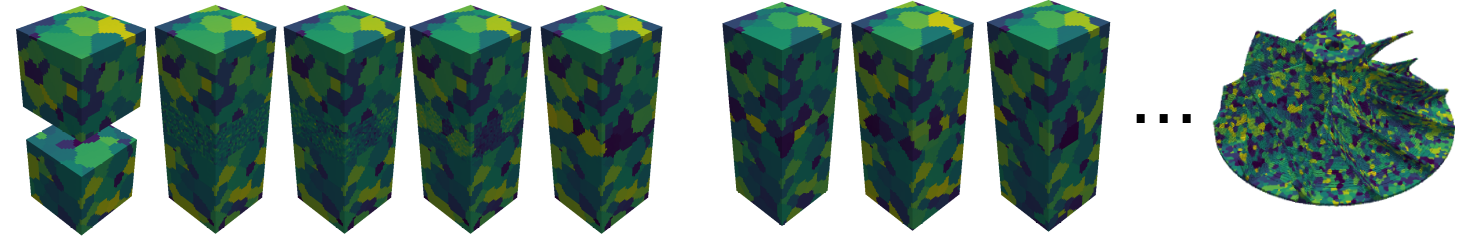}
\caption{GrainPaint -- inpainting microstructure for large-scale CAD objects with diffusion-based generative model.}
\end{figure}
Simulation-based approaches to microstructure generation can suffer from a variety of limitations, such as high memory usage, long computational times, and difficulties in generating complex geometries. Generative machine learning models present a way around these issues, but they have previously been limited by the fixed size of their generation area. We present a new microstructure generation methodology leveraging advances in inpainting using denoising diffusion models to overcome this generation area limitation. We show that microstructures generated with the presented methodology are statistically similar to grain structures generated with a kinetic Monte Carlo simulator, SPPARKS. 


\end{abstract}



\clearpage
\tableofcontents
\clearpage







\section{Introduction} \label{sec:intro}



The primary goal of computational materials science is to construct insightful process-structure-property (PSP) relationships to better understand materials behavior and facilitate inverse materials design~\cite{national2011materials,holdren2014materials,lander2021materials}. In the PSP relationships, modeling the process-structure linkage is an important research subject, as it is naturally linked with manufacturing. 
For example, varying temperature and time in annealing will result in a completely different microstructure that may perform completely differently. 
To that end, many integrated computational materials engineering (ICME)~\cite{horstemeyer2012integrated, allison2013implementing} models dedicated to the process-structure linkage have been developed and implemented over the last two decades, including phase-field simulations, kinetic Monte Carlo (kMC), and cellular automata. Despite much effort in parallelizing computation across nodes and cores from a computational perspective, these ICME models are often computationally expensive, even with large high-performance computing clusters. This has led to attempts to mimic the process-structure linkage through a computationally cheaper model, specifically through machine learning (ML) approaches, where the ICME model leads the ML model in a teacher-student paradigm~\cite{de2019new}. 

In the era of high-throughput computational materials science, the integration of microstructure characterization and reconstruction with ML approaches, alongside materials modeling and simulation, plays a crucial role in unveiling the PSP linkages. 
In this context, the microstructure reconstruction problem aims to generate statistically equivalent representative volume elements (SERVEs), given some target statistical microstructure characterization. 
Based on characterization methods, microstructure characterization and reconstruction methods can be divided into statistical functions, physical descriptors, spectral density functions, multi-point statistics, and machine learning~\cite{bostanabad2018computational}. Among these methods, ML has attracted much attention in the field of inverse materials design because of its flexibility, simplicity, and efficiency. 

Generative models have the following advantages over physics-based simulation models. 
Firstly, simulation software packages, such as SPPARKS, are physics-based, so the physics of the problem must be known. Generative models are data-driven, so they require no knowledge of physics, only data to train on. Secondly, physics-based simulations such as phase field, cellular automata, and kinetic Monte Carlo may be computationally expensive, and the computational cost depends on the complexity of the physical process or what physical process is being simulated. The computational cost of a generative model depends on the complexity of the features in the data. For this reason, generative models have a lower computational cost in some scenarios. Lastly, the geometry requirements inherent to physics-based simulations make the generation of some complex geometries not feasible. In contrast, generative models can handle more flexible classes of geometries.

Generative models are a class of ML models that generate samples similar to those drawn from a dataset. In the case of microstructure reconstruction with a generative model, the task is to generate microstructures statistically equivalent to those in a training set, in the sense that their microstructure characterization statistics match up to a tolerance. 
Recently published works have used various types of generative models including variational autoencoders (VAEs)~\cite{kingma2013auto, vahdat2020nvae}, generative adversarial networks (GANs)~\cite{gulrajani2017improved, mirza2014conditional}, and denoising diffusion probabilistic models (DDPMs)~\cite{buehler2023computational,fernandez4698278denoising, fernandez-zelaia_digital_2024, dureth_conditional_2023, vlassis_denoising_2023, lee_data-driven_2024, azqadan_predictive_2023, buzzy_statistically_2024}.

VAEs learn to represent input data in a lower-dimensional latent space as a probabilistic distribution and sample from this distribution to generate new samples. VAEs have been applied to the design of bioinspired composite structures~\cite{chiu2023designing}, anechoic coating~\cite{sun2022variational}, nanostructured materials~\cite{attari2023towards}, dual-phase steel~\cite{kim2021exploration}, and multi-material 3D-printed composite solids~\cite{xue2020machine}. In GANs, a generator model learns to generate samples while a discriminator model decides if they are realistic. GANs have demonstrated outstanding abilities in producing diverse and realistic structures for metamaterials~\cite{xue2020machine}, composite materials~\cite{mao2020designing, qian2022design, liu2023reconstruction}, and microstructures~\cite{lambard2023generation, nguyen2022synthesizing}, fostering exploration within the design space. However, the latent spaces of VAEs and GANs may be unstable, \textit{i.e.}, small changes in the latent space produce large changes in the output. This instability can cause problems for optimization problems solved in the latent space~\cite{woldseth2022use}. Moreover, GANs are difficult to train due to issues such as mode collapse, instability, and sensitivity to hyperparameters~\cite{arjovsky2017towards, salimans2016improved}. GANs also require a trade-off that sacrifices diversity for fidelity and hence might not have good coverage of the entire data distribution. 
These shortcomings of GANs have provided diffusion models the opportunity to surpass GANs as the new state-of-the-art algorithm for image synthesis on several metrics and data sets~\cite{dhariwal2021diffusion}. Consequently, there has been a surge of denoising diffusion probabilistic models (DDPMs)~\cite{ho2020denoising, rombach2022high, ramesh2022hierarchical}, that are replacing many of these state-of-the-art models. Recently, Vlassis and Sun~\cite{vlassis2023denoising} trained a diffusion model by embedding the 1D target stress-strain curve as the feature vector to guide the generation of 2D microstructures. Buehler~\cite{buehler2023computational} used a VAE to obtain the latent features of 2D hierarchical microstructures and built a DDPM to design metamaterials. ML models have also been applied to a variety of optimization problems, including topology optimization~\cite{rastegarzadeh2022multi, wang2022ih}, airfoil shape optimization~\cite{diniz2024optimizing}, genetic algorithms~\cite{chang2022machine}, and Bayesian optimization~\cite{xue2020machine} to guide the design process. This integration signifies a broader and more holistic approach to inverse materials design.

GANs and VAEs must be trained to inpaint in a region of a specific shape in a specific position. Recent works with diffusion models have overcome both of these limitations, allowing inpainting over arbitrary regions with realistic results~\cite{repaint}. Such capability presents the possibility of using a diffusion model to progressively generate a large microstructure out of small pieces. However, current literature lacks exploration into this microstructure generation approach and  its application in reconstruction of large-scale computer-aided design (CAD) objects with arbitrary shape.  Furthermore, all of these types of generative models have a common limitation\textemdash the size of the output is fixed. This limits the use of ML based microstructure reconstruction to tasks that only require small microstructure samples. Inpainting, which is a procedure for filling in part of an image with contextually appropriate generated content, presents a way around this limitation. GANs and VAEs have been applied to inpainting tasks, but they are limited in both quality and flexibility. Addressing these gaps could significantly propel the field of microstructure design forward, especially in domains necessitating stochastic three-dimensional microstructures. Such domains include, but are not limited to, the development of scaffolds for tissue engineering~\cite{kanwar_novel_2022}, the enhancement of additive manufacturing processes~\cite{wang_data-driven_2019}, and the optimization of components for batteries~\cite{karaki_optimizing_2023}. This advancement could be pivotal in overcoming the present limitations and fostering innovation in these critical areas of research.
\textcolor{black}{
Recent work has also shown DDPMs are capable of generating statistically accurate microstructures in both 2D and 3D. For example, D\"{u}reth et al.~\cite{dureth_conditional_2023} found that DDPMs are effective in the generation of high-quality 2D microstructures for a diverse variety of materials. Other work has also demonstrated how diffusion models can be leveraged for generating 3D microstructures. For example, DDPM generated 3D microstructures have been shown to match experimental data of fuel cell microstructures \cite{bentamou20253d}. Diffusion models have also been used in different implementations to generate 3D microstructures from 2D images, with superior performance compared to previously used methods such as GANs \cite{lee2024multi, phan2024generating}. 
}


To address the above challenges, we propose a diffusion model called GrainPaint to generate arbitrarily sized 3D grain structures. Specifically, this paper contributes the following:
\begin{enumerate}
\item A 3D diffusion model trained on microstructures generated by SPPARKS. \textit{To the best of our knowledge, this is the first 3D diffusion model trained by SPPARKS-generated microstructures.}
\item A parallelization scheme to generate arbitrarily sized grain structures using diffusion models via an inpainting procedure. \black{\textit{To the best of our knowledge, this is the first application of a 3D diffusion model to generate microstructures of arbitrary shape and size.}}
\item A comparison of microstructure statistics between microstructures generated by the diffusion model and SPPARKS.
\item A methodology for generating microstructures with the diffusion model for any arbitrary, generalized 3D geometries.
\end{enumerate}

\section{Methodology}\label{sec:Methodology}

\subsection{Microstructure generation with SPPARKS}

SPPARKS~\cite{plimpton2009crossing,mitchell2023parallel} \textemdash an open-source parallel simulation code developed at Sandia National Laboratories \textemdash is used to generate a 3D microstructure dataset. 
Beside normal grain growth, SPPARKS can also be used to model metal additive manufacturing~\cite{rodgers2017simulation}, grain evolution during welding~\cite{rodgers2017monte}, electron beam welding~\cite{rodgers2016predicting}, thermal sprays~\cite{rodgers2021fast}, among many other processes.
The physics underpinning the grain growth model~\cite{garcia2008three} is summarized as follows. 

In on-lattice kMC~\cite{mitchell2023parallel}, each lattice site has an integer spin value $S_i$ from 1 to a user-defined value $Q$. Setting $Q=2$, we re-obtain the canonical Ising model. 
the Hamiltonians of the Potts model for the energy of a site $i$ with $M$ neighbors can be written as
\begin{equation}
H_i = \sum_{j=1}^M \delta(S_i, S_j),
\end{equation}
where the energy of the entire system is simply $H_i$ summed over $N$ sites, and 
\begin{equation}
\delta (S_i, S_j) = 
\begin{cases}
0 \quad \text{ if } S_i = S_j,\\
1 \quad \text{ if } S_i \neq S_j.
\end{cases}
\end{equation}

In the grain growth simulation, the Potts model~\cite{anderson1989computer,garcia2008three} is used to simulate curvature-driven grain growth. 
\textcolor{black}{Three stochastic numerical solvers for kMC are implemented in SPPARKS~\cite{slepoy2008constant}, which scale as $\mathcal{O}(N)$, $\mathcal{O}(\log N)$, and $\mathcal{O}(1)$~\cite{mitchell2023parallel}, respectively, where $N$ is the number of possible next sites. For kMC applications, uniform sampling remains the most commonly used tool to generate exponentially and uniformly distributed. }
Grain microstructures are represented by an integer value, called grain identifier (grain ID), stored at each voxel. 
\textcolor{black}{In materials science, grain ID refers to the unique identification assigned to each individual grain in a polycrystalline microstructure during materials characterization. It allows researchers to track and study the properties, orientations, and behaviors of specific grains within the material to better understand its overall performance.}

\textcolor{black}{The SPPARKS simulations are performed on a high-performance computing cluster, utilizing a single node. Each node is equipped with 192 GB of memory and dual sockets, each housing 18 Intel Broadwell E5-2695 cores clocked at 2.1 GHz. The nodes are interconnected via Omni-Path for high-speed communication. A training dataset is constructed from 1,000 SPPARKS stochastic simulations, each initialized with a unique integer seed for the pseudo-random number generators to capture microstructure-induced aleatory uncertainty.}

\subsection{Diffusion models}
Diffusion models are part of a greater family of models, all of which are based on the idea of maximizing the likelihood, $p(\mathbf{x})$, of all known data, $\mathbf{x}$. In practical problems, the ground-truth function describing $p(\mathbf{x})$ is often complex, and $\mathbf{x}$ can also be quite high-dimensional. As such, learning $p(\mathbf{x})$ exactly can be computationally infeasible. Therefore, likelihood-maximizing models instead introduce a random latent variable, $\mathbf{z}$, of lower-dimensional, and or lower complexity, which can be used to describe the joint distribution with $\mathbf{x}$,
\begin{equation}
p(\mathbf{x}) = \int p(\mathbf{x}, \mathbf{z}) d\mathbf{z}. 
\end{equation}

To realize the benefits of introducing the lower complexity $\mathbf{z}$,  a tractable Evidence Lower Bound (ELBO) can be defined to approximate the joint distribution integral
\begin{center}
\begin{eqnarray}
\log \; p(\mathbf{x}) &=& \log \int p(\mathbf{x}, \mathbf{z}) d\mathbf{z} \\
&=& \log \int \frac{p(\mathbf{x}, \mathbf{z})q_\mathbf{\theta}(\mathbf{z} | \mathbf{x})}{q_\mathbf{\theta}(\mathbf{z} | \mathbf{x})} \\
&=& \log \; \mathbb{E}_{q_\mathbf{\theta}(\mathbf{z} | \mathbf{x})} \left[\frac{p(\mathbf{x}, \mathbf{z})}{q_\mathbf{\theta}(\mathbf{z} | \mathbf{x})}\right] d\mathbf{z}\\
&\geq& \mathbb{E}_{q_\mathbf{\theta}(\mathbf{z} | \mathbf{x})} \left[\frac{p(\mathbf{x}, \mathbf{z})}{q_\mathbf{\theta}(\mathbf{z} | \mathbf{x})}\right], \qquad \qquad \text{(by Jensen's Inequality)}
\end{eqnarray}
\end{center}
where $q_\mathbf{\theta}(\mathbf{z} | \mathbf{x})$ is the variational distribution with learnable model parameters $\mathbf{\theta}$~\cite{luo2022understanding}.

Diffusion models differ from other related likelihood maximizing approaches in that $z$, has the same cardinality as the data $x$, but is noised according to a variance schedule parameterized by a hyperparameter, $\beta_{t}$. The index $t$ describes the data-to-noise ratio and ranges from 0 to $T$, with 0 representing the original, un-noised data, and $T$ representing maximally noised data. In the limit as $T$ increases, the data approaches an isotropic Gaussian distribution
\begin{equation}
    q(\mathbf{x}_{t} | \mathbf{x}_{t-1}) = \mathcal{N} (\mathbf{x}_{t}; \sqrt{\alpha_{t}} \mathbf{x}_{t-1}, 1-\alpha_{t}),
\end{equation}
where $\alpha_{t} = 1 -\beta_{t}$. The act of injecting Gaussian noise into the original data is known as the \emph{forward} process, whereas the \emph{reverse} process for denoising can be computed using the model predictions as
\begin{equation}
    p_{\theta}(\mathbf{x}_{0:T}) = p(\mathbf{x}_{t}) \prod_{t=1}^{T} p_{\theta}(\mathbf{x}_{t-1} | \mathbf{x}_{t}),
\end{equation}
where
\begin{equation}
    p_{\theta}(\mathbf{x}_{t-1} | \mathbf{x}_{t}) = \mathcal{N} (\mathbf{x}_{t-1}; \mu_{\theta}(x_t,t), \sigma_{\theta}(x_t,t)).
\end{equation}
Commonly, $\sigma_{\theta}$ is set equal to $\beta_t$. The final ELBO loss function can be written as 
\begin{equation}
\mathcal{L}_{ELBO}(\theta) =\mathbb{E}_{q(\mathbf{x}_{0:T} | \mathbf{x}_{0})} \left[\frac{p(\mathbf{x}_{0:T})}{q(\mathbf{x}_{1:T} | \mathbf{x}_{0})}\right].
\end{equation}
In many cases, and as is done in this paper, the MSE between the predicted and actual noise added to the data can be used as a much simpler estimation of the ELBO~\cite{ho2020denoising}. 

\subsection{Outpainting}


In the context of generative models, inpainting is the process of generating new data in masked regions of existing data. Typically, inpainting can be implemented as a supervised approach. 
\textcolor{black}{In supervised inpainting, parts of the ground-truth data are masked (hidden), and the model is then trained to reconstruct these masked regions. The masked portions of the data can be random, or strategically chosen in order to better suit specific tasks (e.g, masking only the upper or lower half ground-truth images). }
In general, these supervised approaches to inpainting can be computationally expensive and may generalize poorly in diverse masking scenarios if trained inadequately. Unlike VAE and GANs, diffusion models have the capability to perform inpainting completely unsupervised, without the need for any additional training. This is because after a diffusion model learns a distribution in training, the model can be conditioned on known pieces to data to perform inpainting. \black{In contrast to inpainting, outpainting describes the process of extending data generation beyond a models original context window. The outpainting process begins by using prior model-generated data on the edge of a new context window. The remaining portion of this context window represents the region outside of the original boundaries of the generation, and is masked. Using the same method as in inpainting, the model can then generate smooth continuations into this masked area~\cite{zhang2023towards}.}

In our work, we follow the RePaint approach proposed by Lugmayr et al.~\cite{repaint}. Consider a context window, $\mathbf{x}$, composed of known data, $\mathbf{x}^{\text{known}}$, and unknown data, $\mathbf{x}^{\text{unknown}}$, masked by $m$ such that,
\begin{equation}
\mathbf{x} = m \odot \mathbf{x}^{\text{known}} + (1-m) \odot \mathbf{x}^{\text{unknown}}.
\label{eq:mask}
\end{equation}
In this scenario, we would like to generate data in the unknown, masked region, of the context window. To do this, Lugmayr \textit{et al.}~\cite{repaint} suggests that starting from pure random noise, $\mathbf{x}_{T} \sim \mathcal{N}(0, I)$, the next step, $\mathbf{x}_{t-1}$,  can be computed by running the forward process on the known data,
\begin{equation}
    \mathbf{x}^{\text{known}}_{t-1} \sim \mathcal{N} (\sqrt{\bar{\alpha}_{t}} \mathbf{x}_{0}, 1-\bar{\alpha}_{t}),
    \label{eq:forward_mask}
\end{equation}
and then the reverse process on the unknown data,
\begin{equation}
    \mathbf{x}^{\text{unknown}}_{t-1} \sim \mathcal{N} (\mu_{\theta}(\mathbf{x}_t,t), \sigma_{\theta}(x_t,t)),
\end{equation}
where $\bar{\alpha}_{t} = \prod_{t=1}^{t} \alpha_{t}$. 
Using \cref{eq:mask} we can write $x_{t-1}$ as
\begin{equation}
\mathbf{x}_{t-1} = m \odot \mathbf{x}^{\text{known}}_{t-1} + (1-m) \odot \mathbf{x}^{\text{unknown}}_{t-1}.
\end{equation}
It is clear from these equations that although $\mathbf{x}^{\text{unknown}}_{t-1}$ is dependent on $\mathbf{x}^{\text{known}}_{t}$ and $\mathbf{x}^{\text{unknown}}_{t}$, $\mathbf{x}^{\text{known}}_{t-1}$ is solely dependent on $\mathbf{x}_{0}$, which itself is only dependent on $\mathbf{x}^{\text{known}}_{0}$. 
\textcolor{black}{As a result, any conditioning can be diminished by $\mathbf{x}^{\text{known}}_{t-1}$ generating forward process. }
To ameliorate this issue, the forward process can be applied to the combined $\mathbf{x}_{t-1}$ such that 
\begin{equation}
\mathbf{x}_{t} \sim \mathcal{N} (\sqrt{\alpha_{t}} \mathbf{x}_{t-1}, 1-\alpha_{t}).
\end{equation}
To better ensure conditioning, additional repeating or resampling this process $n$ times are suggested. Whereas generating a sample using a diffusion model without resampling involves sampling each step in the schedule once, resampling involves running the reverse process and then the forward process $n$ times at each step.

\begin{figure}[H]
    \centering
    \includegraphics[width=\linewidth]{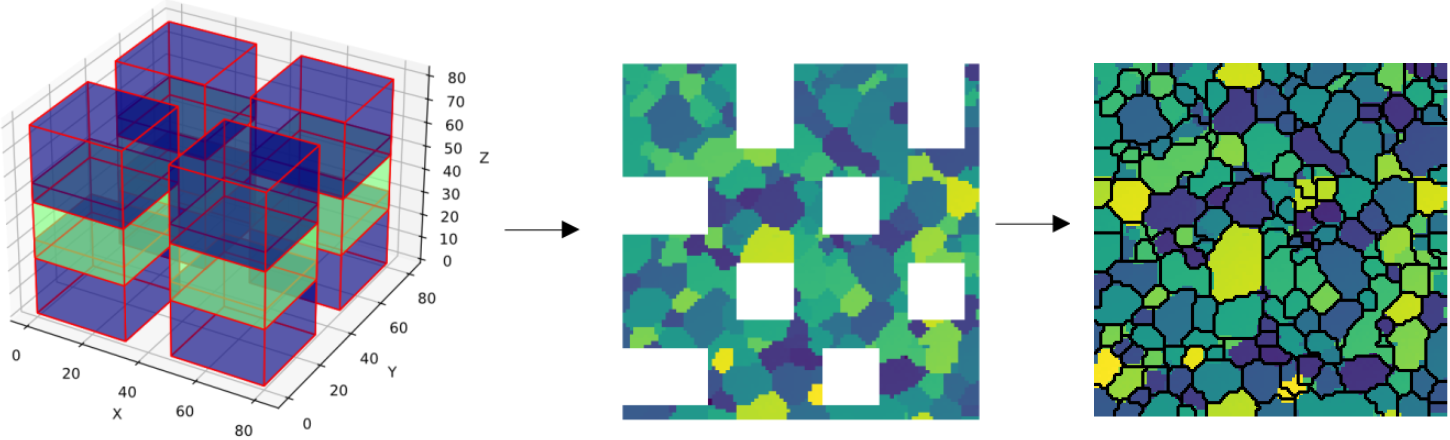}
    \caption{A representation of the three steps of our microstructure generation process: Planning, Inpainting, and Segmentation}
    \label{fig:flowchart}
\end{figure}

\textcolor{black}{\cref{fig:flowchart} shows how outpainting is applied in the context of  microstructure generation. The process begins by planning the cubic context windows in which to generate new microstructures (\cref{fig:flowchart}, left). This involves determining which regions to fill in first, and how much overlap each region will have with subsequent generations. Any overlapping portions of prior generated microstructures form the unmasked, ``known'', parts in the outpainting procedure. Next, the ``unknown'', masked, parts of each region (\cref{fig:flowchart}, middle, masked in white) are filled in by applying the RePaint algorithm to match the unmasked parts. Finally, the resulting microstructure is segmented into distinct grains with unique IDs (\cref{fig:flowchart}, right).}

\subsection{Model Design}
\begin{figure}[H]
\centering
\includegraphics[width=\textwidth]{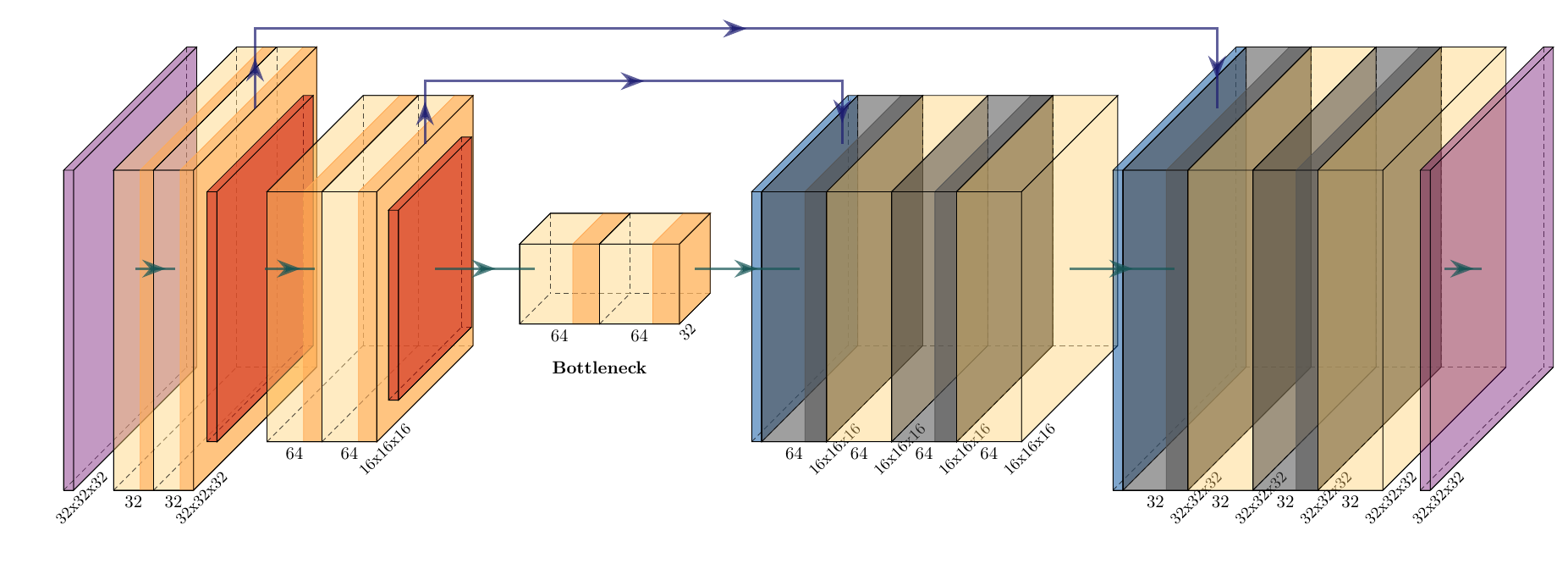}
\caption{3D U-net deep learning architecture used in this work.}
\label{fig:unet}
\end{figure}

The data used to train the GrainPaint model consists of 949 100$\times$100$\times$100 geometries generated from SPPARKS. The GrainPaint model used in this work is based on a 3D U-Net~\cite{ronneberger_u-net_2015} which operates on 32$\times$32$\times$32 blocks. \cref{fig:unet} shows the architecture of a 3D U-Net that is employed in this work. 

The geometries from SPPARKS are each split into 27 32$\times$32$\times$32 non-overlapping blocks plus 56 additional blocks centered on the boundaries between the first 27 blocks. \black{We use 27 as this is the number of non overlapping 32$\times$32$\times$32 blocks that fit within a 100$\times$100$\times$100 block. 56 is more arbitrary, it is a set of blocks within the same 100$\times$100$\times$100 region, overleaping the first 27 blocks.} This gives a total of 78767 training samples. Our DDPM was trained on a 250 step schedule for 10 epochs. 
\textcolor{black}{We used a linear variance schedule, as is suggested in one of the original DDPM implementations by Ho et al~\cite{ho2020denoising}. }
Training took around 59 hours on a single RTX 3090. 


Our microstructure DDPM model leverages the RePaint approach with resampling to generate new voxels in a masked region of the $32^3$ context window given, known, previously generated voxels. In this way, full CAD geometries can be generated which have seamless boundaries between context windows. We found that a good level of quality is achieved on a 250 step schedule with a jump size of 1 and 10 resamplings. The last 25 steps in the schedule were performed with no resampling. The number of resamplings was selected by qualitatively comparing the quality of different numbers of resamplings. 

\begin{figure}[!htbp]
    \centering
    \begin{subfigure}[b]{0.32\textwidth}
        \centering
        \includegraphics[height=135px,keepaspectratio]{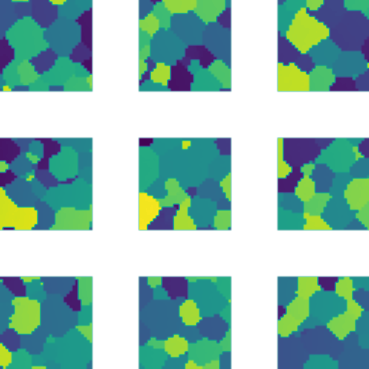}
        \caption{Step 1 -- no resamples}
    \end{subfigure}
    \begin{subfigure}[b]{0.32\textwidth}
        \centering
        \includegraphics[height=135px,keepaspectratio]{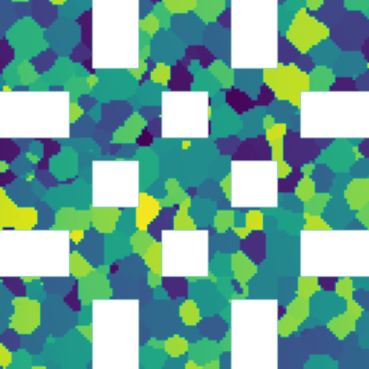}
        \caption{Step 4 -- no resamples}
    \end{subfigure}
    \begin{subfigure}[b]{0.32\textwidth}
        \centering
        \includegraphics[height=135px,keepaspectratio]{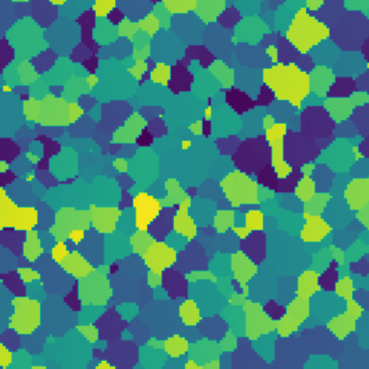}
        \caption{Steps 7 \& 8 -- no resamples}
    \end{subfigure}
    
    \begin{subfigure}[b]{0.32\textwidth}
        \centering
        \includegraphics[height=135px,keepaspectratio]{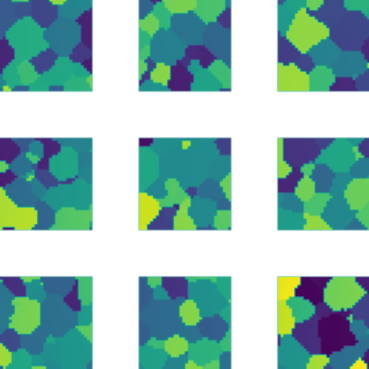}
        \caption{Step 1 -- 5 resamples}
    \end{subfigure}
    \begin{subfigure}[b]{0.32\textwidth}
        \centering
        \includegraphics[height=135px,keepaspectratio]{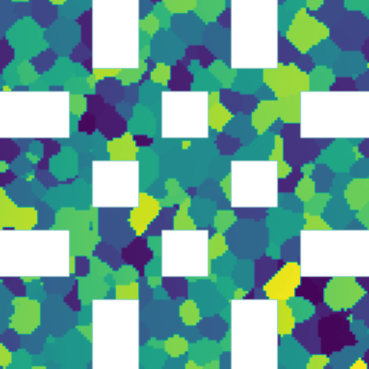}
        \caption{Step 4 - 5 resamples}
    \end{subfigure}
    \begin{subfigure}[b]{0.32\textwidth}
        \centering
        \includegraphics[height=135px,keepaspectratio]{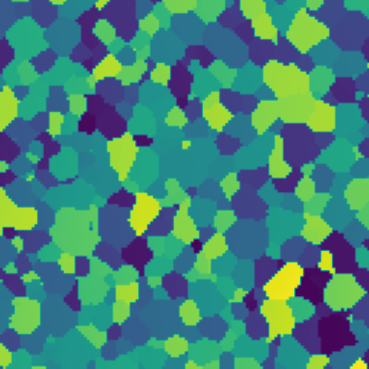}
        \caption{Steps 7 \& 8 -- 5 resamples}
    \end{subfigure}
    \begin{subfigure}[b]{0.32\textwidth}
        \centering
        \includegraphics[height=135px,keepaspectratio]{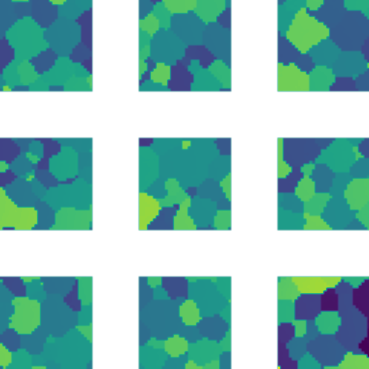}
        \caption{Step 1 -- 10 resamples}
    \end{subfigure}
    \begin{subfigure}[b]{0.32\textwidth}
        \centering
        \includegraphics[height=135px,keepaspectratio]{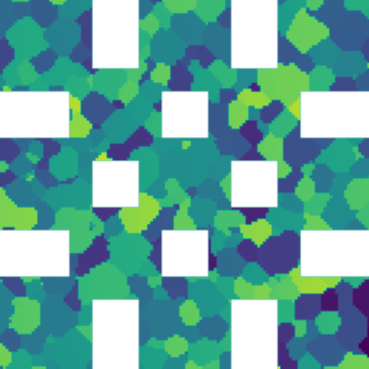}
        \caption{Step 4 -- 10 resamples}
    \end{subfigure}
    \begin{subfigure}[b]{0.32\textwidth}
        \centering
        \includegraphics[height=135px,keepaspectratio]{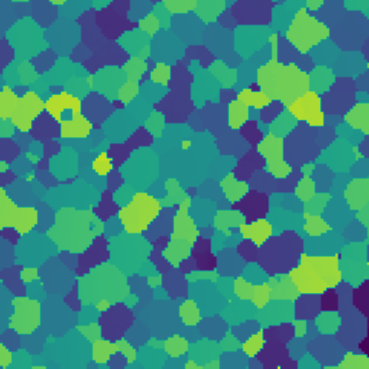}
        \caption{Steps 7 \& 8 -- 10 resamples}
    \end{subfigure}
    \begin{subfigure}[b]{0.32\textwidth}
        \centering
        \includegraphics[height=135px,keepaspectratio]{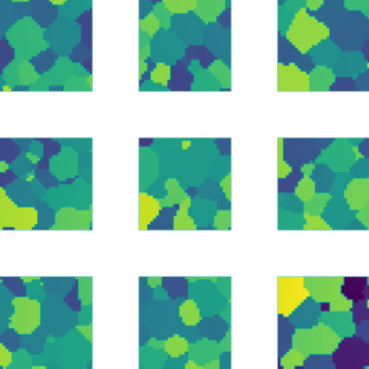}
        \caption{Step 1 -- 20 resamples}
    \end{subfigure}
    \begin{subfigure}[b]{0.32\textwidth}
        \centering
        \includegraphics[height=135px,keepaspectratio]{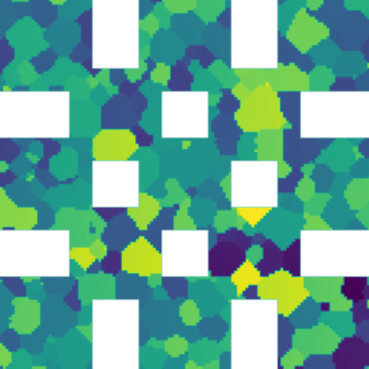}
        \caption{Step 4 -- 20 resamples}
    \end{subfigure}
    \begin{subfigure}[b]{0.32\textwidth}
        \centering
        \includegraphics[height=135px,keepaspectratio]{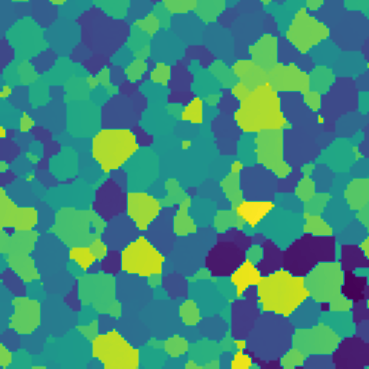}
        \caption{Steps 7 \& 8 -- 20 resamples}
    \end{subfigure}
\caption{Comparison of 128$\times$128 slices of a microstructure cube generated with different numbers of resamplings.}
\label{fig:resample_grid}
\end{figure}

Varied numbers of resamplings present a trade off between quality and computational cost, where twice as many resamplings takes about twice as long to run. A comparison between 0, 5, 10, and 20 resamplings is shown in a slice of 128$\times$128$\times$128 geometries in \cref{fig:resample_grid}. For the purpose of comparison, we use the same unconditioned samples (Step 1), except for the \black{block} in the lower right of the slice. We observe that at lower numbers of resamples, the grain boundaries tend to be lined up on the boundaries between the areas where the model generates, creating a series of lines in the grain boundaries. Lugmayr et al. observed a similar phenomenon, where low numbers of resamplings would lead the RePaint algorithm to match the texture of the patch but not the context ~\cite{repaint}. We did not observe the line features with 10 or more resamplings, which led us to choose 10 resamplings. 

\black{In addition to a qualitative evaluation of different numbers of resamplings, we also perform a quantitative analysis by measuring the grain volume, aspect ratio, and nearest neighboring centroid distance distributions. This comparison for 1 and 10 resamplings is shown in \cref{fig:grain_size_dist_main}, \cref{fig:aspect_ratio}, and \cref{fig:neigh_centroid_dist_main}. More detail on the microstructure statistics is provided in \cref{sec:iso_eval}. \cref{fig:grain_size_dist_main} shows that the grain volume distribution produced by GrainPaint with 10 resamplings is more similar to the distribution produced by SPPARKS. \cref{fig:aspect_ratio} shows that the magnitude of difference in distributions is larger for 1 resampling (f) than for 10 resamplings (c). \cref{fig:neigh_centroid_dist_main} shows that 1 resampling produces a distribution of nearest neighboring centroid distances shifted towards smaller values compared to 10 resamplings. In addition, we calculate the Kullback-Leibler divergence between these distributions. The results are shown in \cref{tab:KLD1} and match our qualitative observations. These three grain statistics further inform our choice of 10 resamplings.}

\begin{table}[!htbp]
\centering
    \caption{Kullback-Leibler divergence for grain statistics between structures simulated by SPPARKS and generated by GrainPaint}
    \label{tab:KLD1}
    \begin{tabular}{cccc}
        \toprule
        Statistic & Number of Resamplings & KL Divergence\\
        \midrule
        Grain Volume & 1 & 0.0827 \\
        Grain Volume & 10 & 0.056 \\
        Aspect Ratio & 1 & 1.421 \\
        Aspect Ratio & 10 & 0.0384 \\
        Centroid Distance & 1 & 0.127 \\
        Centroid Distance & 10 & 0.033 \\
        \bottomrule
    \end{tabular}
\end{table}

\clearpage

\subsection{Parallelization Approach}

The microstructure generation process used in this work uses an eight-stage process in order to take advantage of single GPU (batch) and multi-GPU parallelism. This eight-stage process is necessary to both enable parallelism and to ensure that overlapping regions are not generated at the same time. While the parallelization approach described in this work results in cuboid regions for inpainting, the RePaint algorithm works with regions of arbitrary shape, as well as non-contiguous regions. Also note that while ``seeds'' generated by SPPARKS could be used in step 1, this would present a few issues: First, the geometry would not be fully generated by a diffusion model in that case. Second, nearby samples from a SPPARKS geometry could not be used as these samples would be correlated. 

\begin{figure}[H]
    \centering
    \begin{subfigure}[b]{0.32\textwidth}
        \centering
        \includegraphics[width=1\linewidth]{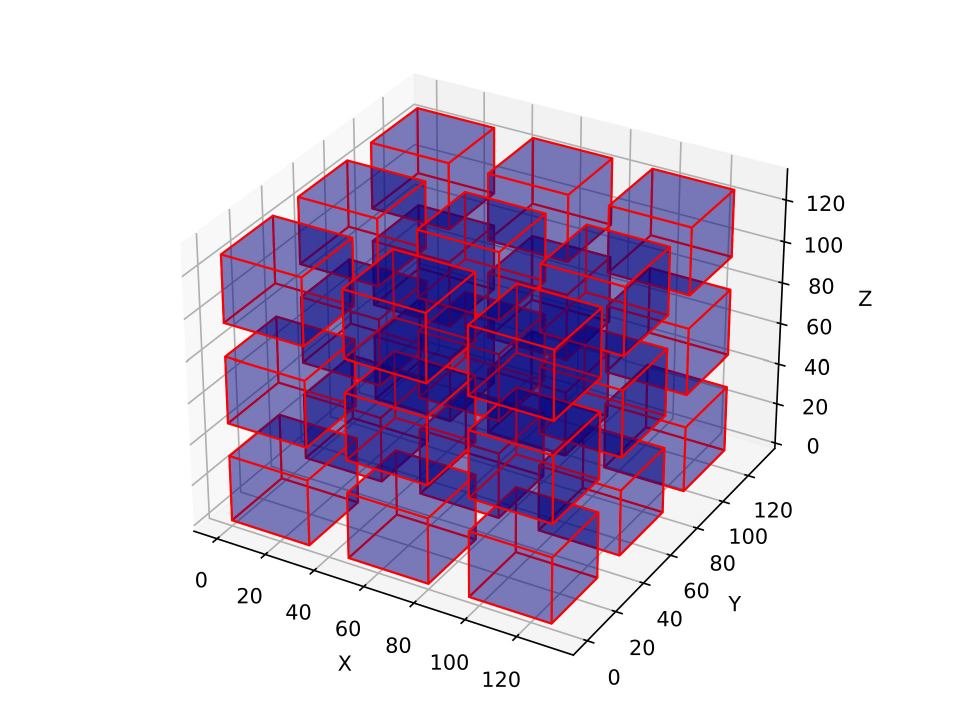}
        \caption{Step 1}
    \end{subfigure}
    \begin{subfigure}[b]{0.32\textwidth}
        \centering
        \includegraphics[width=1\linewidth]{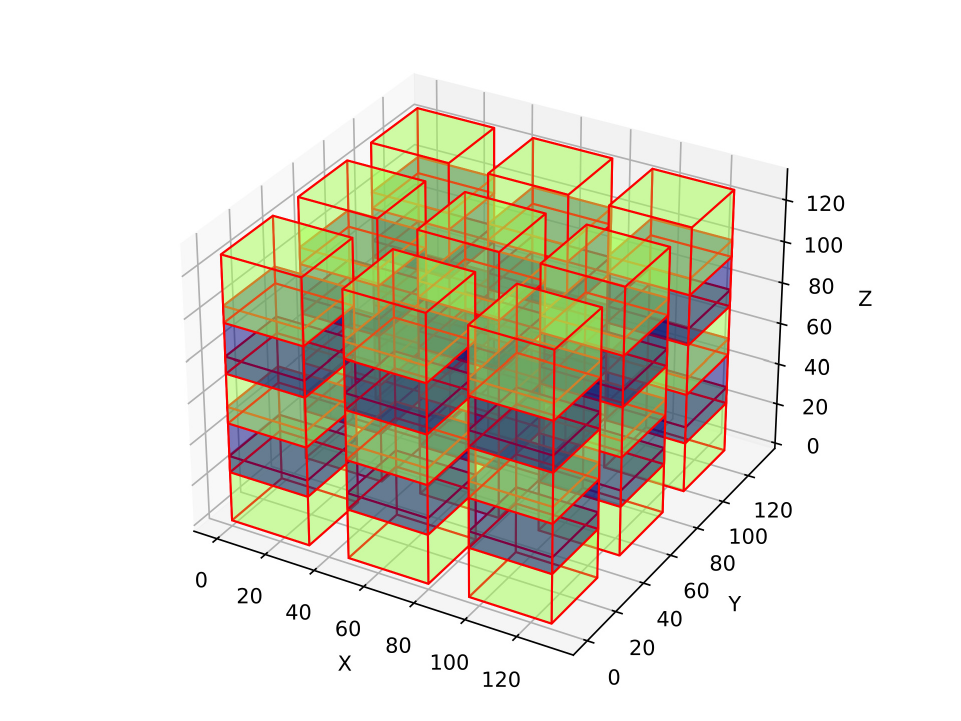}
        \caption{Step 2}
    \end{subfigure}
    \begin{subfigure}[b]{0.32\textwidth}
        \centering
        \includegraphics[width=1\linewidth]{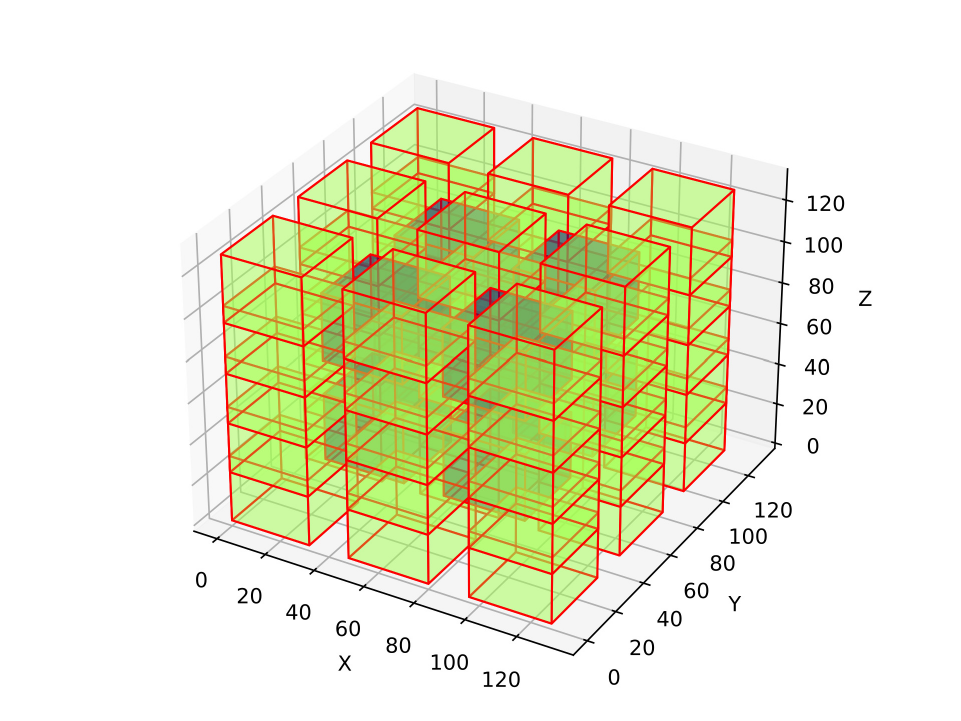}
        \caption{Step 3}
    \end{subfigure}
    
    \begin{subfigure}[b]{0.32\textwidth}
        \centering
        \includegraphics[width=1\linewidth]{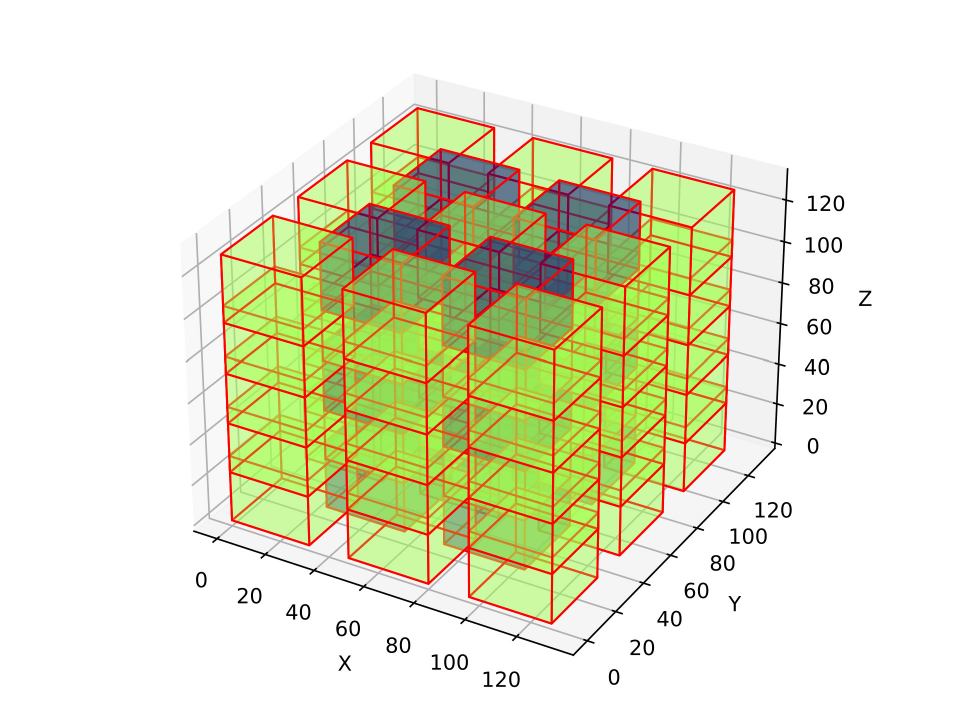}
        \caption{Step 4}
    \end{subfigure}
    \begin{subfigure}[b]{0.32\textwidth}
        \centering
        \includegraphics[width=1\linewidth]{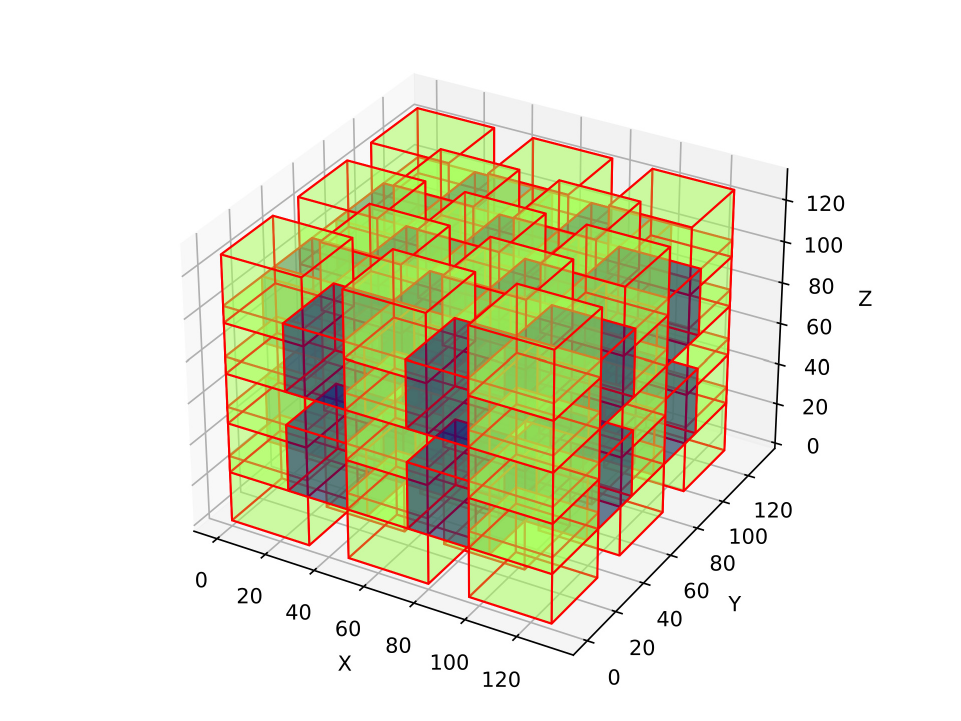}
        \caption{Steps 5 \& 6}
    \end{subfigure}
    \begin{subfigure}[b]{0.32\textwidth}
        \centering
        \includegraphics[width=1\linewidth]{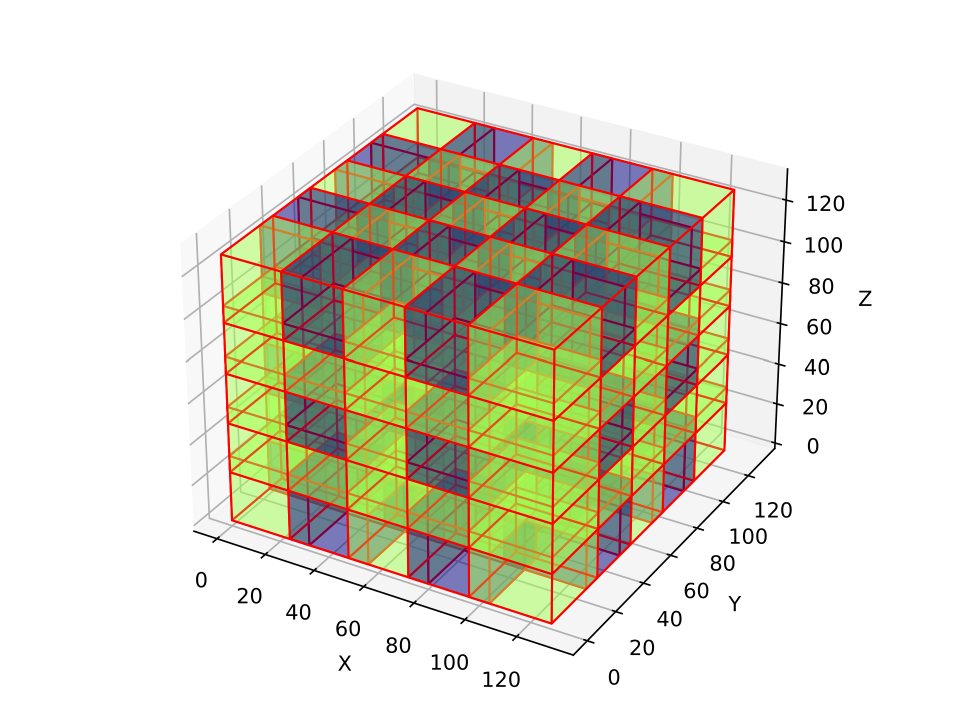}
        \caption{Steps 7 \& 8}
    \end{subfigure}
\caption{Microstructure generation plan. \black{\black{Blocks} added in the current step are shown in blue and \black{blocks} added in previous steps are shown in yellow.}}
\label{gen_steps}
\end{figure}

This process begins with the creation of a plan for generating the microstructure. Each stage uses the point generation algorithm (\cref{alg:genpoints}). Each of these points is the corner of a $32^3$ \black{block shaped} region where the GrainPaint model will be run. Each stage has a different offset and limit, shown in \cref{alg:all_stage}. All distances in the algorithm are expressed in multiples of the GrainPaint model generation region, in this case a distance of 1 in the algorithm corresponds to 32 voxels. Each point generated by the algorithm has a list of dependencies associated with it. These dependencies ensure that overlapping \black{blocks} are not generated at the same time. \cref{alg:genpoints} is run for each stage in \cref{alg:all_stage}. An example of the process is shown in \cref{gen_steps}. The gaps between the blocks in Step 1 are 16 voxels in all directions. This work does not evaluate different gap sizes, but the following considerations are likely significant in selecting a gap size: 
\begin{itemize}
\item A larger gap size will be more computationally efficient because fewer total blocks will need to be generated. However, too large of a gap size may not provide enough information for the inpainting to produce a realistic output.
\item A smaller gap will provide more information to the inpainting process which might improve quality. However, too small of a gap size will constrain the inpainting process too much and not allow it to place realistic output in the gap.
\end{itemize}
After all \black{blocks} and dependencies have been generated, the generation planning algorithm generates a series of batches of a specified size or smaller that respect the dependencies of each \black{block}. Once the plan is produced, the diffusion model generates in batches according to the plan, distributed across one or more GPUs.  Under the configuration used in this work, the GPU memory usage of a batch of size 1 is about 1.5GB. \black{To provide a hardware-equivalent comparison between SPPARKS and GrainPaint, we have run GrainPaint using only a CPU. On a 64-core AMD EPYC 7713p, GrainPaint takes 32 minutes, or 34 core-hours to generate a 32$\times$32$\times$32 block. In comparison, GrainPaint took 83 seconds to generate the same region on a single Nvidia A100.} The throughput of the generation algorithm increases for larger batch sizes up to the size where the batches of the generation plan are always the maximum size. For example, on 2 NVIDIA A100s, a 100$\times$100$\times$100 geometry can be generated in about 3.5 hours and a 200$\times$200$\times$200 geometry can be generated in about 7.4 hours. More examples are shown in \cref{tab:gen_time}. The observed reduction in throughput for larger geometries is likely due to increased overhead from saving checkpoints, rather than a decrease in performance of the diffusion model. Note that the diffusion model can only achieve this throughput when generating in batches (\textit{i.e.}, multiple \black{blocks} at the same time) as the generation of a single \black{block} will not fully load the GPU. SPPARKS has been observed to exhibit strong scaling in similar problems, on dozens of CPUs across dozens of nodes~\cite{mitchell_parallel_2023}. We expect that our GrainPaint model would also exhibit strong scaling as the most computationally demanding parts of the generation process are independent. 

\begin{table}
\centering
    \caption{Generation Time on 2$\times$ NVIDIA A100}
    \label{tab:gen_time}
    \begin{tabular}{cccc}
        \toprule
        RVE size & Time & Throughput (voxels/min)\\
        \midrule
        224$\times$224$\times$224 & 7h 23m & 25.3k \\
        368$\times$368$\times$224 & 20h 51m & 24.2k \\
        416$\times$416$\times$224 & 26h 41m & 24.2k \\
        464$\times$464$\times$224 & 34h 30m & 23.2k \\
        \bottomrule
    \end{tabular}
\end{table}

\begin{table}
\centering
    \caption{Offsets and Limit reductions for \cref{alg:all_stage}.}
    \label{tab:off_lim}
    \begin{tabular}{cccc}
        \toprule
        Stage & Offset & limitReduction\\
        \midrule
        1 & (0.00, 0.00, 0.00) & (0, 0, 0) \\
        2 & (0.00, 0.00, 0.75) & (0, 0, 1) \\
        3 & (0.75, 0.75, 0.75) & (1, 1, 1) \\
        4 & (0.75, 0.75, 0.00) & (1, 1, 0) \\
        5 & (0.75, 0.00, 0.75) & (1, 0, 1) \\
        6 & (0.00, 0.75, 0.75) & (0, 1, 1) \\
        7 & (0.75, 0.00, 0.00) & (1, 0, 0) \\
        8 & (0.00, 0.75, 0.00) & (0, 1, 0) \\
        \bottomrule
    \end{tabular}
\end{table}

\begin{algorithm}
\caption{Generate a grid of points}
\label{alg:genpoints}
\begin{algorithmic}
\Function{GeneratePoints}{$\mathbf{limit}$, $\Delta$, $\mathbf{offset}$, $prevPoints$}
    \State $(x_{\text{off}}, y_{\text{off}}, z_{\text{off}}) \gets \mathbf{offset} $
    \State $
    G = \left\{ (i \cdot \Delta + x_{\text{off}}, j \cdot \Delta + y_{\text{off}}, k \cdot \Delta + z_{\text{off}}) 
    \text{ for } 0 \leq i, j, k \leq \textbf{limit}[n]\} \right\}
    $
    \Comment{A grid of points corresponding to the corner with lowest coordinate of each cube}
    \State $\textbf{dependencies}(p) = \{ q \in prevPoints \mid \max(|q_x - p_x|, |q_y - p_y|, |q_z - p_z|) \leq 1 \}$
    \Comment{List of dependencies for each point}
    \State $points = \{ (p, \textbf{dependencies}(p)) \mid p \in G \}$
    \State \Return $points$
\EndFunction
\end{algorithmic}

\end{algorithm}

\begin{algorithm}
\caption{Generate a list of \black{block} for all stages}


\label{alg:all_stage}
\begin{algorithmic}
\State $\Delta \gets 1.5$
\Comment{Spacing $\Delta$ between each corner point, 1.5 gives 0.5 distance between each cube.}
\State $\mathbf{initialLimit} \gets (x_{\text{max}}, y_{\text{max}}, z_{\text{max}}) $ \Comment{Geometry size}
\State $prevPoints \gets \text{empty list}$
\Comment{Initialize}
\For{$stage \gets 1$ \textbf{to} 8}
    \State$(\mathbf{offset}, \mathbf{limitReduction})  \gets \text{set according to \cref{tab:off_lim}.}$

    \State $\mathbf{limit} \gets \mathbf{initialLimit} - \mathbf{limitReduction}$
    \State \textbf{append} $\Call{GeneratePoints}{\mathbf{limit}, \Delta, \mathbf{offset}, prevPoints}$ \textbf{to} $pointsWithDeps$
    \State $prevPoints \gets \text{extract points from } pointsWithDeps$
\EndFor
\end{algorithmic}
\end{algorithm}

\subsection{Segmentation}

\begin{figure}[!htbp]
    \centering
    \begin{subfigure}{0.49\textwidth}
        \centering
        \includegraphics[height=250px,keepaspectratio]{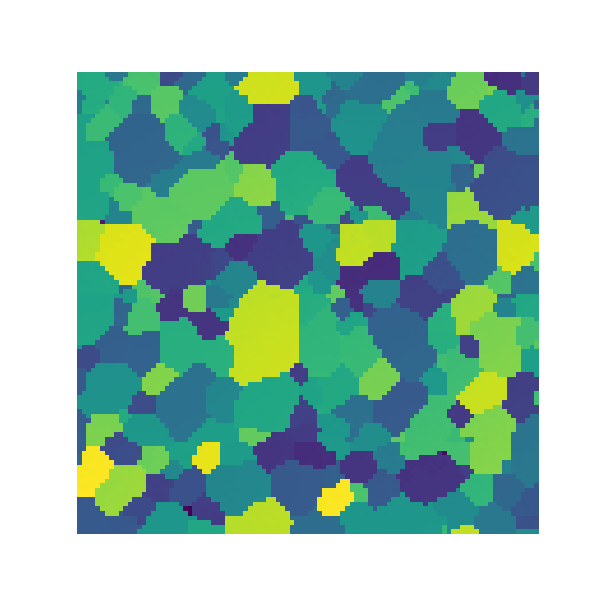}
        \caption{Before Segmentation}
    \end{subfigure}
    \hfill
    \begin{subfigure}{0.49\textwidth}
        \centering
        \includegraphics[height=250px,keepaspectratio]{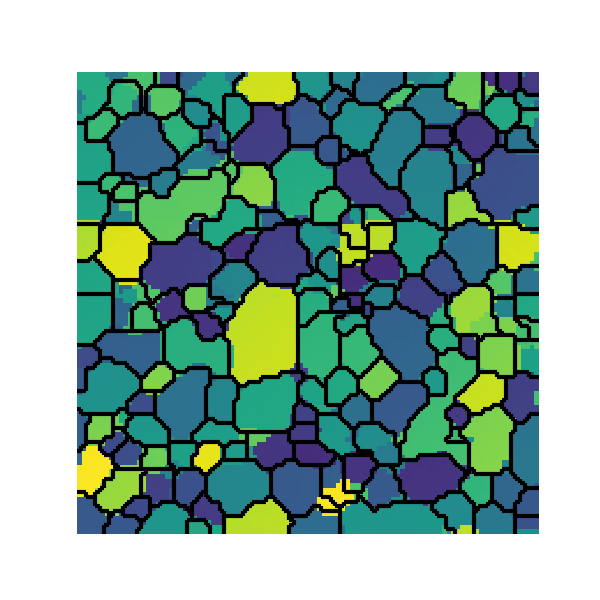}
        \caption{After Segmentation}
    \end{subfigure}

\caption{Example of the Results of the Segmentation Process.}
\label{fig:seg_results}
\end{figure}

The output of the GrainPaint model is an array of floats, so the elements of the output array must be clustered into grains before the grains can be analyzed. We perform clustering with the DBSCAN algorithm. DBSCAN does not require the number of clusters to be known before running, which provides an advantage over supervised clustering algorithms such as k-means clustering. DBSCAN also classifies some data points as noise which is helpful in dealing with noise in the diffusion model output. Clustering is performed using the DBSCAN algorithm, with each voxel converted to a four-dimensional point ($x$, $y$, $z$, value)~\cite{ester1996density}. DBSCAN uses a minimum cluster size parameter and an epsilon parameter controlling the maximum distance between two points for them to be placed in the same cluster. These parameters were manually tuned to epsilon=1.9 and min samples=15 which produces an output visually similar to the input and performed well on grain quality benchmarks. \black{An example showing the output of the segmentation algorithm is shown in \cref{fig:seg_results}. }The runtime of DBSCAN scales super-linearly as a function of the number of voxels, so we developed a hierarchical algorithm that clusters with DBSCAN on overlapping sub-sections of the array and then combines these into a clustering of the entire array. The DBSCAN epsilon parameter normally needs to be tuned to different geometry sizes, however, our hierarchical algorithm allows for the same epsilon to be used for many geometry sizes as the parts of the geometry run through DBSCAN are always the same size. 

\section{Results} \label{sec:results}

\subsection{CAD-based Microstructure Comparison}

\begin{figure}[!htbp]
    \centering
    \begin{subfigure}{0.32\textwidth}
        \centering        \includegraphics[height=135px,keepaspectratio]{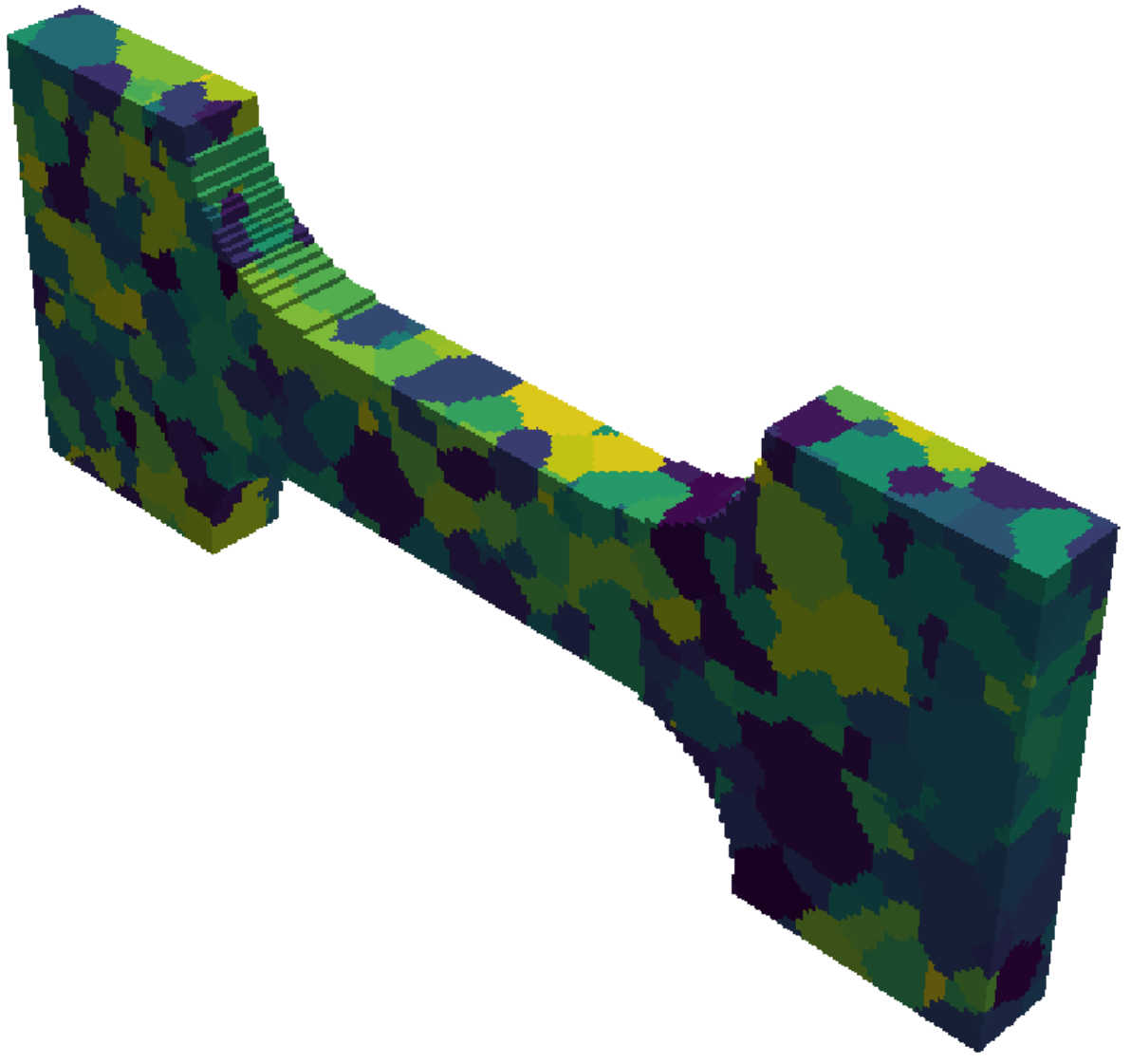}
        \caption{Dog Bone}
    \end{subfigure}
    \hfill
    \begin{subfigure}{0.32\textwidth}
        \centering
        \includegraphics[height=135px,keepaspectratio]{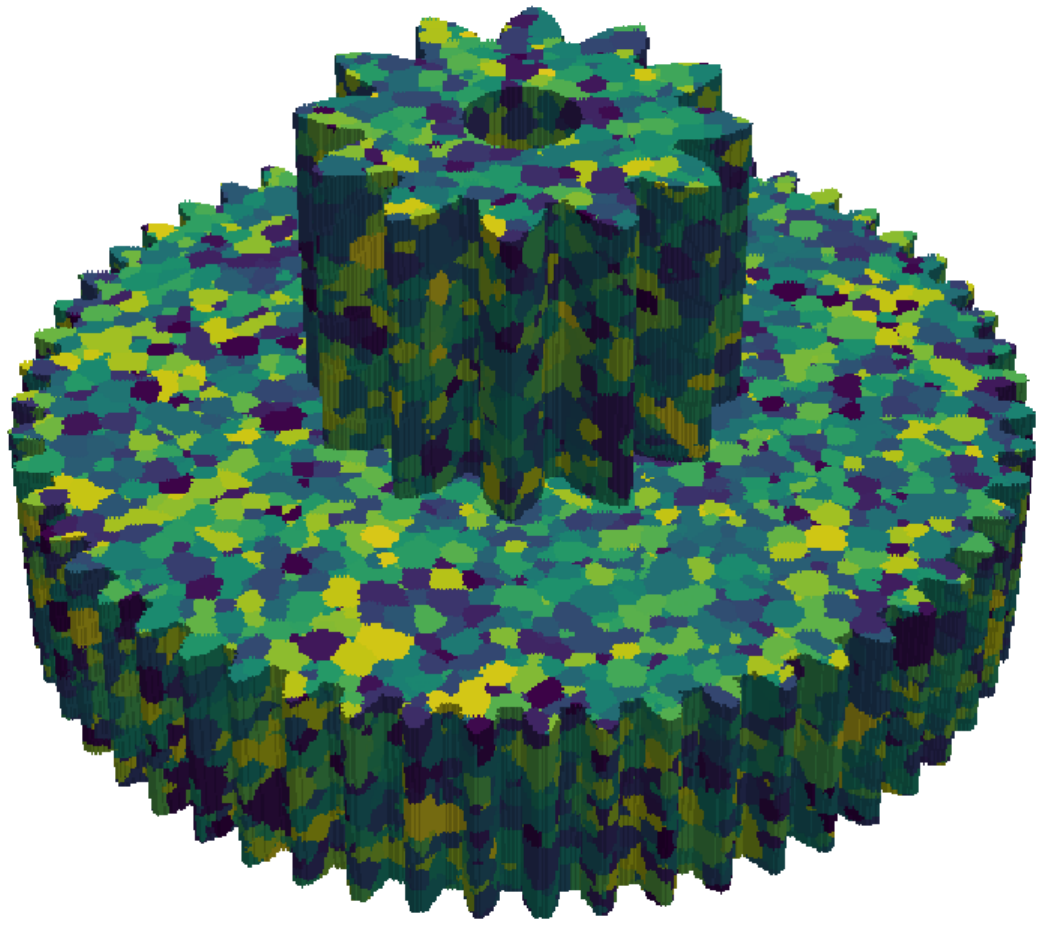}
        \caption{Gear}
    \end{subfigure}
    \hfill
    \begin{subfigure}{0.32\textwidth}
        \centering
        \includegraphics[height=135px,keepaspectratio]{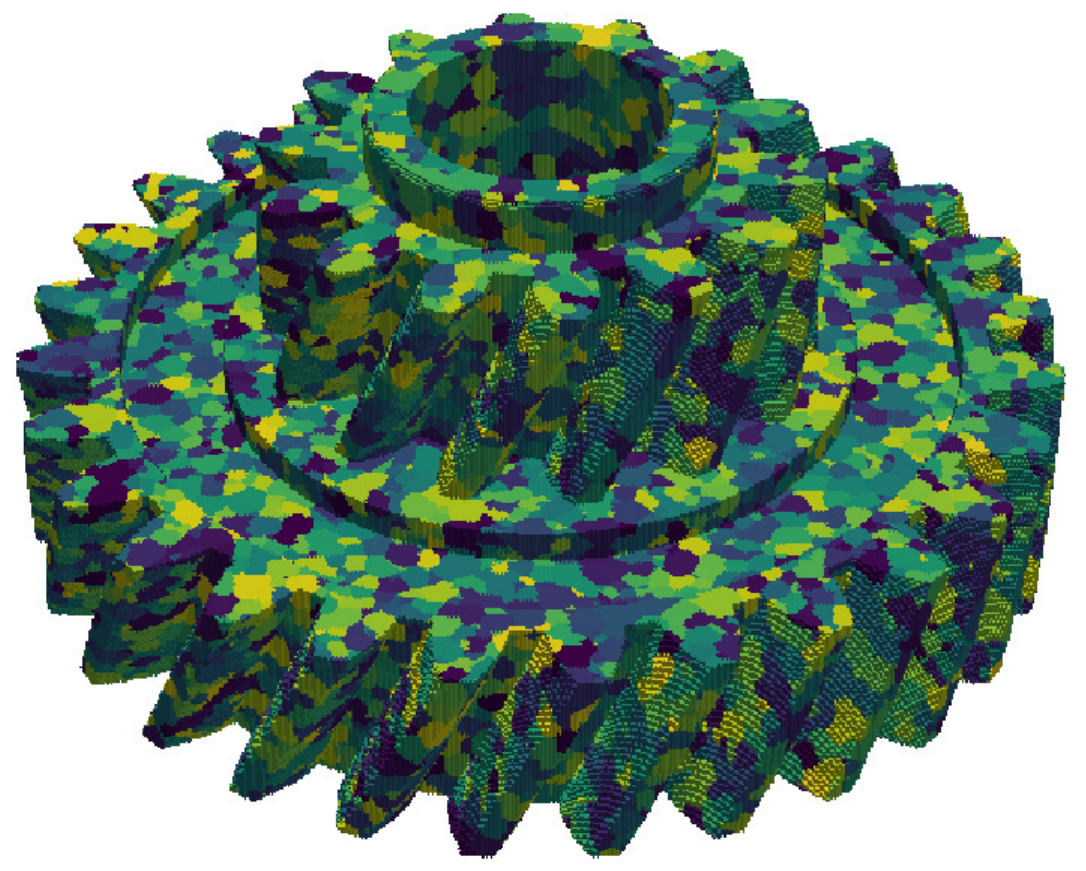}
        \caption{Helical Gear}
    \end{subfigure}

    \begin{subfigure}{0.32\textwidth}
        \centering
        \includegraphics[height=135px,keepaspectratio]{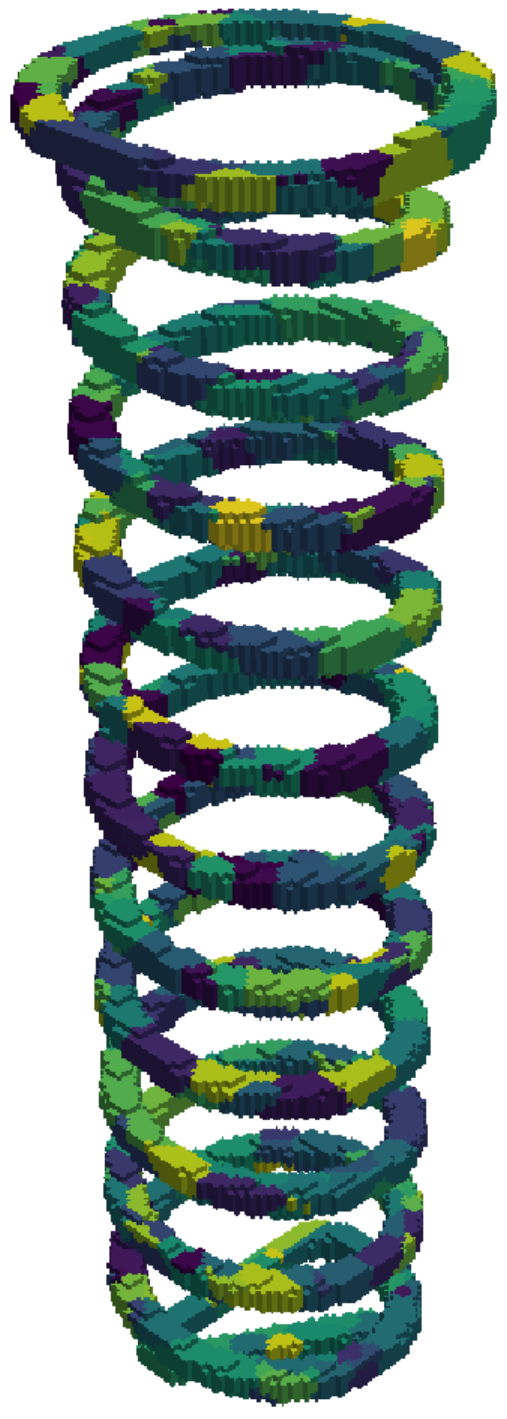}
        \caption{Spring}
    \end{subfigure}
    \hfill
    \begin{subfigure}{0.2\textwidth}
        \includegraphics[height=135px,keepaspectratio]{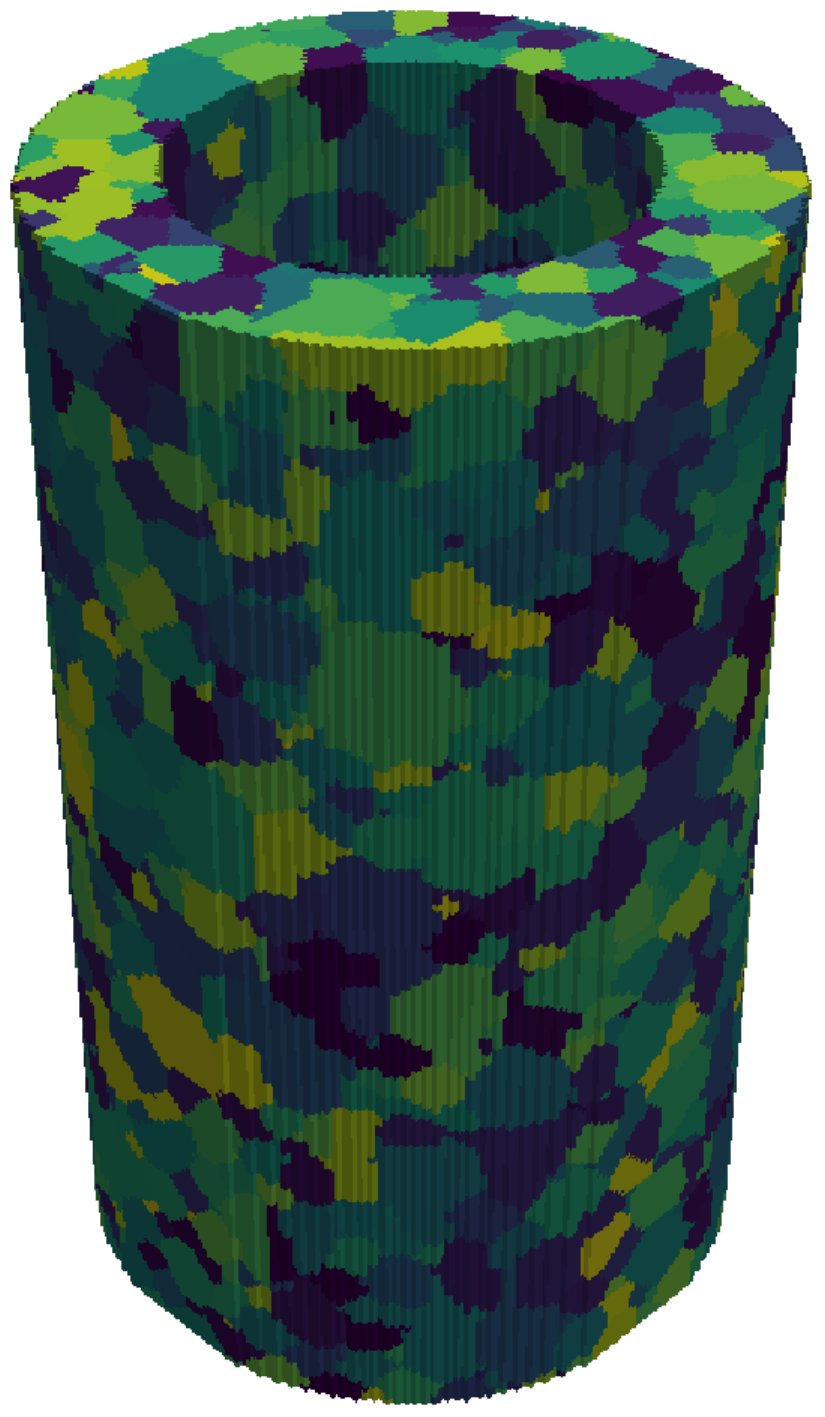}
        \caption{Tube}
    \end{subfigure}
    \hfill
    \begin{subfigure}{0.32\textwidth}
        \centering
        \includegraphics[height=135px,keepaspectratio]{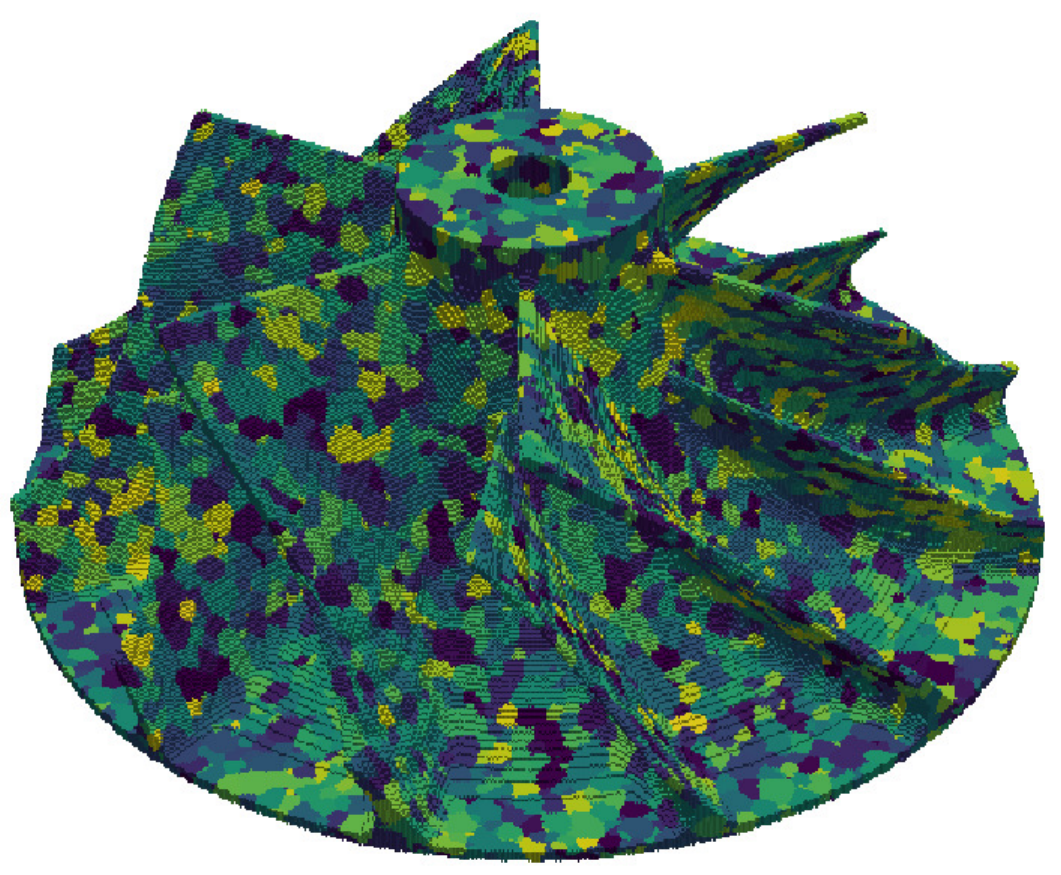}
        \caption{Turbo Blade}
    \end{subfigure}
\caption{Six CAD objects endowed with microstructures from the proposed diffusion-based generative model}
\label{fig:diffusionCAD}
\end{figure}

\begin{figure}[!htbp]
    \centering
    \begin{subfigure}{0.32\textwidth}
        \centering
        \includegraphics[height=135px,keepaspectratio]{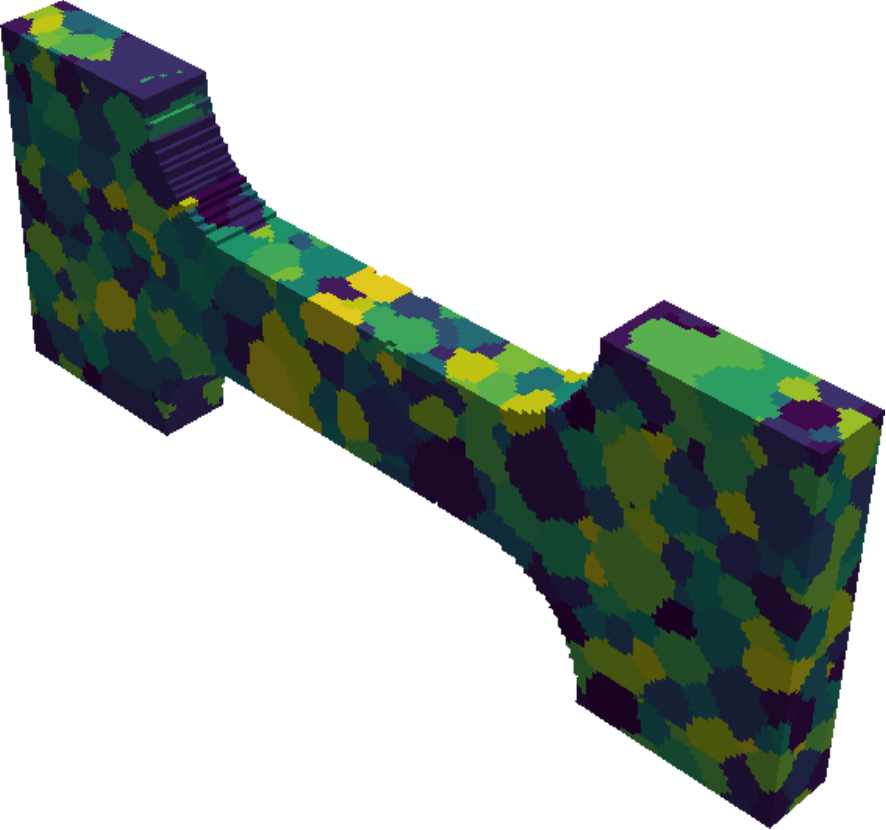}
        \caption{Dog Bone}
    \end{subfigure}
    \hfill
    \begin{subfigure}{0.32\textwidth}
        \centering
        \includegraphics[height=135px,keepaspectratio]{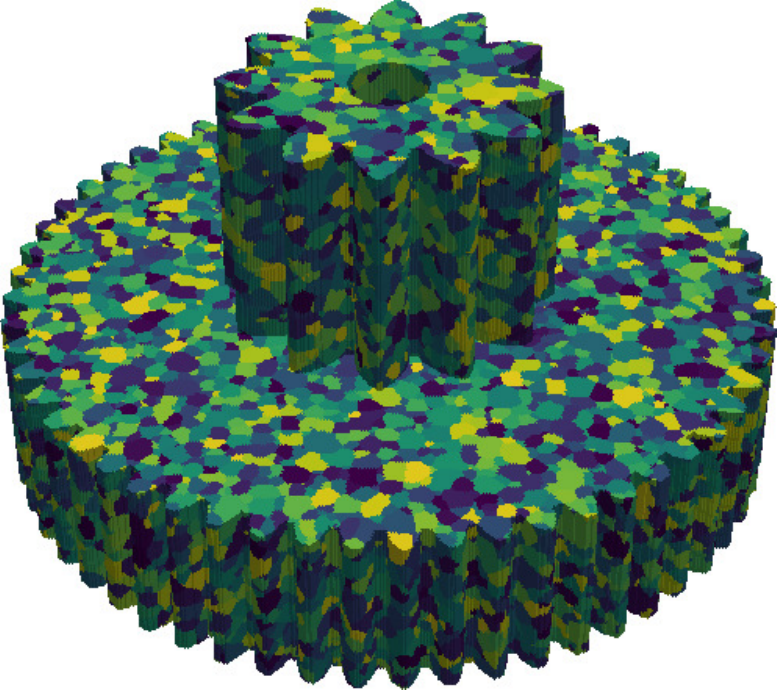}
        \caption{Gear}
    \end{subfigure}
    \hfill
    \begin{subfigure}{0.32\textwidth}
        \centering
        \includegraphics[height=135px,keepaspectratio]{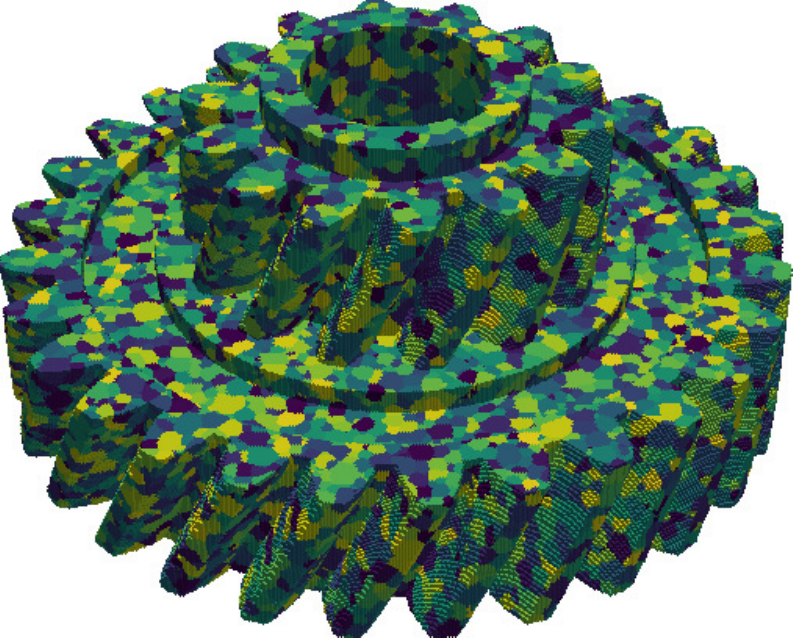}
        \caption{Helical Gear}
    \end{subfigure}

    \begin{subfigure}{0.32\textwidth}
        \centering
        \includegraphics[height=135px,keepaspectratio]{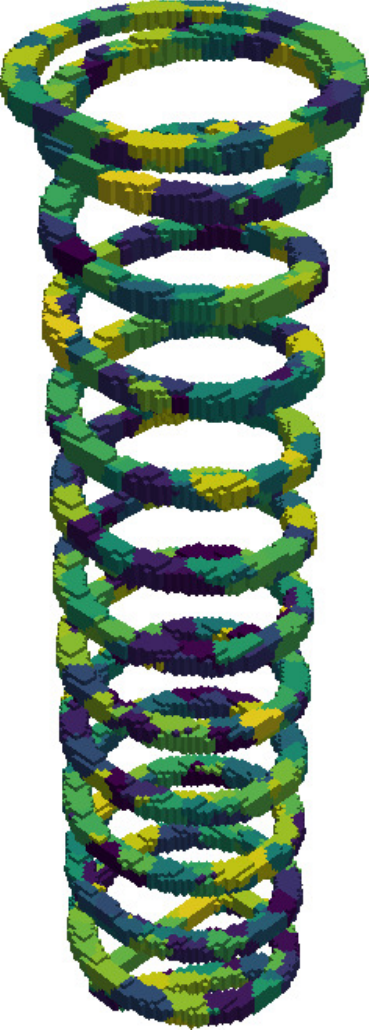}
        \caption{Spring}
    \end{subfigure}
    \hfill
    \begin{subfigure}{0.2\textwidth}
        \includegraphics[height=135px,keepaspectratio]{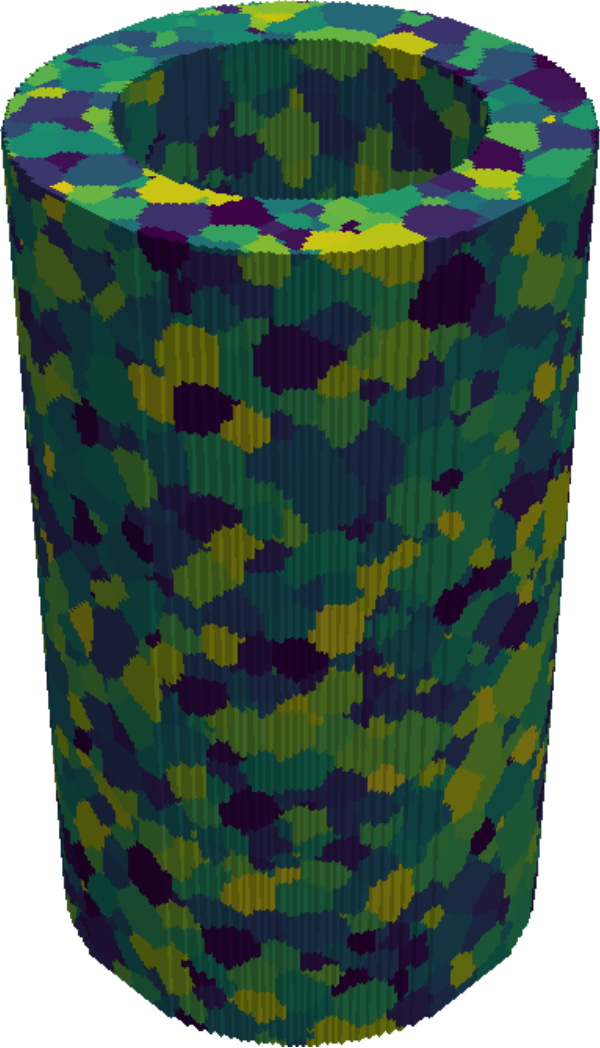}
        \caption{Tube}
    \end{subfigure}
    \hfill
    \begin{subfigure}{0.32\textwidth}
        \centering
        \includegraphics[height=135px,keepaspectratio]{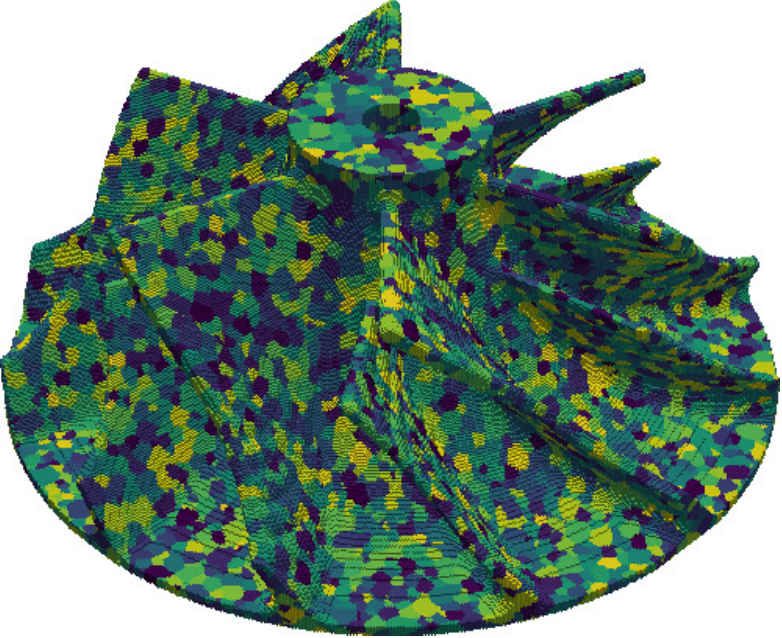}
        \caption{Turbo Blade}
    \end{subfigure}
\caption{Six CAD objects endowed with microstructures from SPPARKS.}
\label{fig:spparksCAD}
\end{figure}

The proposed diffusion model is used to generate 10 microstructures for each of 6 CAD objects. Before a microstructure can be applied to an STL file, the mesh of the STL file must first be voxelized. This procedure generates an empty voxel image of a resolution matching the generated microstructure and calculates if each voxel is outside the mesh. All of the voxels outside the mesh are used to create a mask that is applied to the generated geometry to produce a microstructure of the desired shape. \cref{fig:diffusionCAD} shows examples of microstructures generated by our diffusion model applied to CAD objects. \cref{fig:spparksCAD} shows examples of microstructures generated by SPPARKS applied to CAD objects. 
\textcolor{black}{We observe that the microsturctures generated by GrainPaint appear qualitatively similar to microstructures generated by SPPARKS, demonstrating that GrainPaint can be used to generate large microstructures in complex shapes.}

\subsection{Isotropic Microstructure Generation and Evaluation} 
\label{sec:iso_eval}

\textcolor{black}{To evaluate the performance of the proposed diffusion model on isotropic microstructures, we utilize SPPARKS to generate normal grain growth microstructures. }
During a Monte Carlo time-step, voxels in the computational domain are visited and their grain IDs are sampled probabilistically, with the probability $P$ of successful change in grain IDs as
\begin{equation}
P =
\begin{cases}
\exp\left( \frac{-\Delta E}{k_B T_s} \right) & \text{ if } \Delta E > 0, \\
1 &  \text{ if } \Delta E \leq 0,
\end{cases}
\label{eq:Metropolis}
\end{equation}
where $E$ is the total grain boundary energy calculated by summing all the neighbor interaction energies, $\Delta E$ can be regarded as the activation energy, $k_B$ is the Boltzmann constant, and $T_s$ is the simulation temperature. 
In the basic Potts model, the interaction energy between two \textcolor{black}{voxels} belonging to the same grain is zero, and $E$ is incremented by one for each dissimilar neighbor. From \cref{eq:Metropolis}, changes that decrease system energy are preferred, and the total system energy is monotonically decreased through grain coarsening.
It is worthy to note that the $T_s$ simulation temperature is not the real system temperature: $k_B T_s$ is an energy that defines the thermal fluctuation, \textit{i.e.}, noise, presented in the kMC simulation~\cite{garcia2008three}. 
\textcolor{black}{The higher the simulation temperature $T_s$ is, the higher the chance that voxels are flipping their grain membership in \cref{eq:Metropolis}. 
The effect of temperature $T_s$ on grain growth have been well-studied in~\cite{holm1991effects,holm2001computer,garcia2008three}. Specifically, increasing $T_s$ is linked to higher thermal fluctuations that causes rougher grain boundaries~\cite{holm2001computer} and monotonically decreasing kurtosis of the grain area distribution~\cite{holm1991effects}, which essentially results in rougher grain boundaries.}

We evaluate the similarity of the microstructures generated with the GrainPaint model with microstructures generated by SPPARKS using several microstructure statistics. \textcolor{black}{We selected aspect ratio, grain volume, and nearest neighboring centroid distance descriptors as they are perhaps the most commonly used in literature~\cite{bostanabad2018computational}.} We compare the microstructure descriptor probability density functions between these two sets of microstructures. 

\begin{figure}[!htbp]
\begin{subfigure}[b]{1.0\textwidth}
    \centering
    \includegraphics[width=\linewidth]{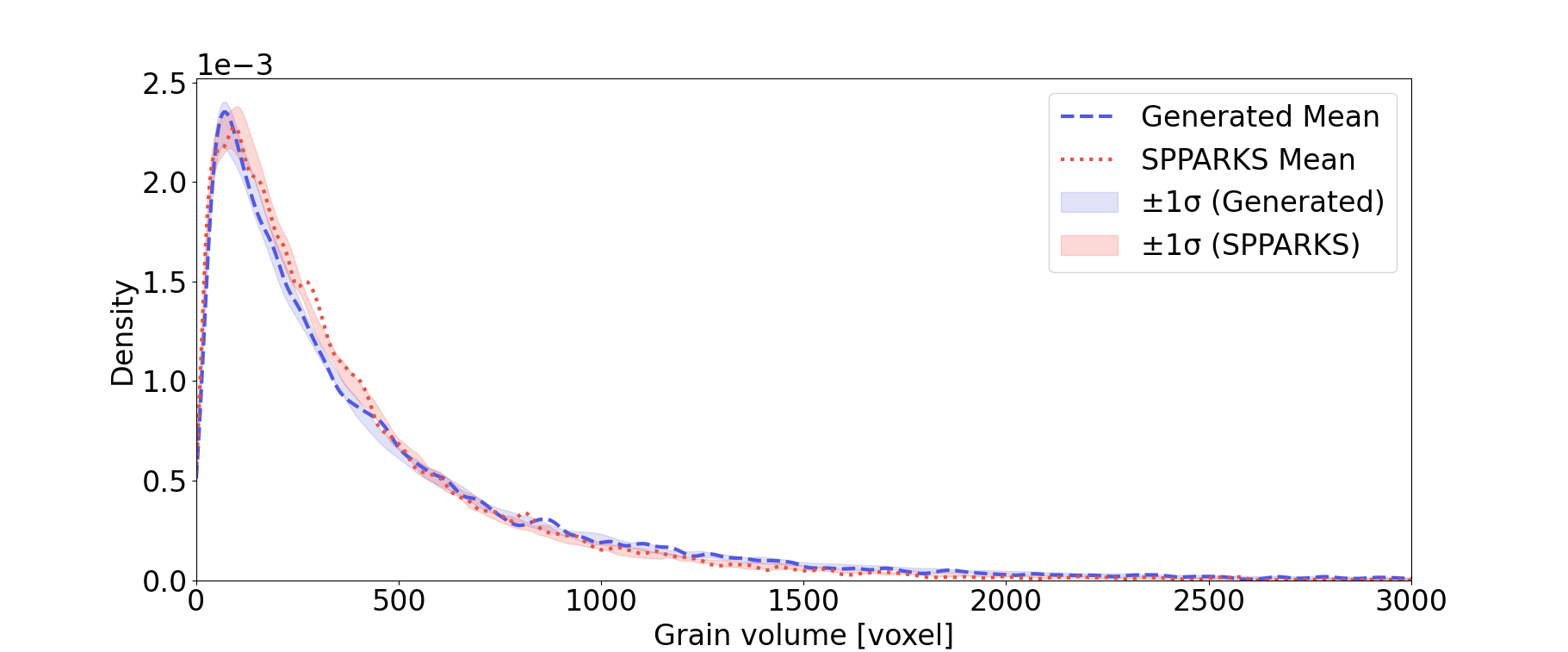}
    \caption{A comparison of the grain-size distribution shows an excellent agreement between SPPARKS and the proposed diffusion model for microstructure with 10 resampling steps.}
    \label{fig:grain_size_dist}
\end{subfigure}

\begin{subfigure}[b]{1.0\textwidth}
    \centering
    \includegraphics[width=1\linewidth]{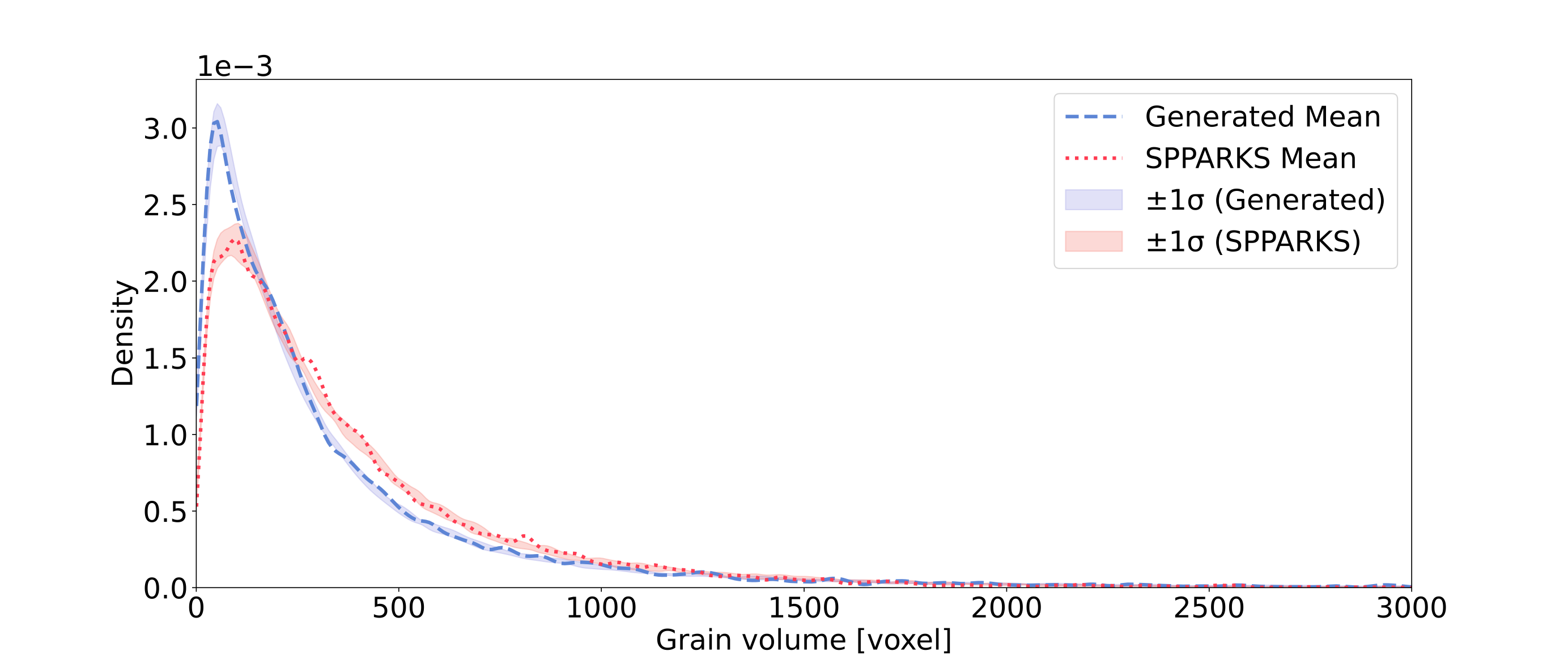}
    \caption{A comparison of the grain-size distribution between SPPARKS and GrainPaint run with 1 resampling (this figure) shows less agreement than SPPARKS and GrainPaint run with 10 resampling steps (\cref{fig:grain_size_dist}).}
    \label{fig:grain_size_dist1-10}
\end{subfigure}
\caption{Comparison of grain-size distributions between SPPARKS and GrainPaint for isotropic microstructures.}
\label{fig:grain_size_dist_main}
\end{figure}

The first descriptor we examine is grain volume. As all the grains are already assigned unique labels either by SPPARKS or by our segmentation algorithm, this evaluation can be performed by simply adding up the number of voxels with each index. This benchmark was calculated on two sets of geometries: 9 100$\times$100$\times$100 SPPARKS geometries and 16 100$\times$100$\times$100 GrainPaint model geometries. The average shown in \cref{fig:grain_size_dist} is the distribution for all the grains in all the geometries in each set and the standard deviation is calculated across all of the geometries in each set. The GrainPaint model and segmentation algorithm yield similar grain volume distributions, with SPPARKS having a slightly greater share of grains below about 500 voxels and the GrainPaint model having slightly more above about 1000 voxels. 

\begin{figure}[!htbp]
    \centering
    \begin{subfigure}{0.32\textwidth}
        \centering
        \includegraphics[height=135px,keepaspectratio]{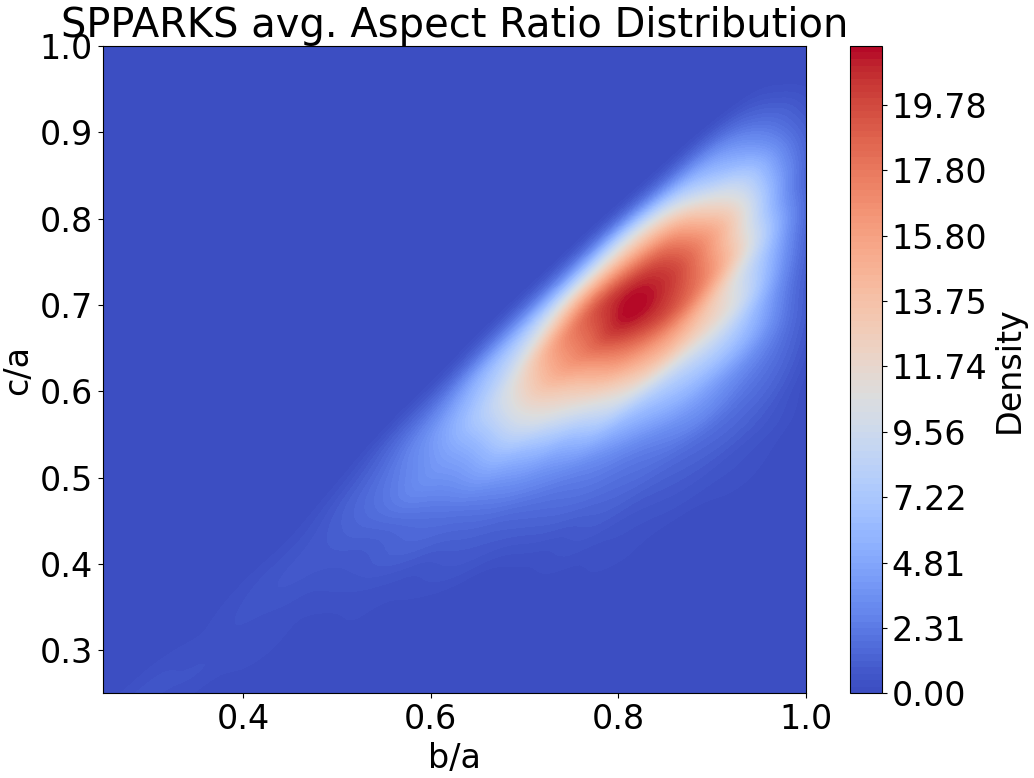}
        \caption{SPPARKS aspect ratio}
    \end{subfigure}
    \hfill
    \begin{subfigure}{0.32\textwidth}
        \centering
        \includegraphics[height=135px,keepaspectratio]{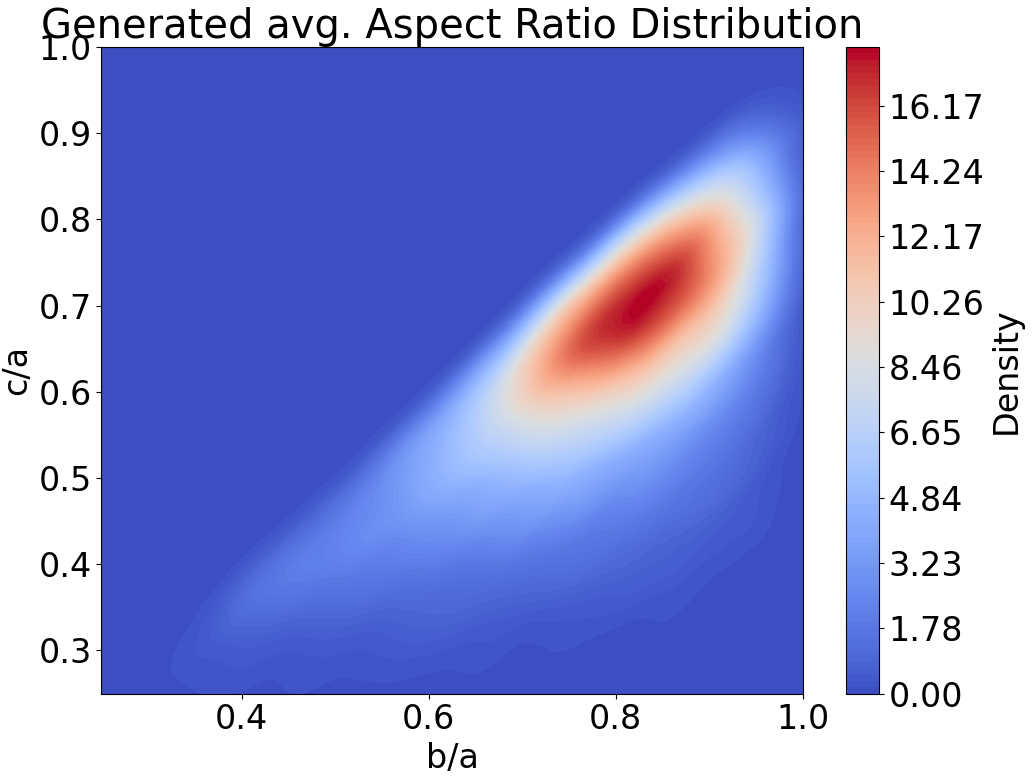}
        \caption{Diffusion aspect ratio}
    \end{subfigure}
    \hfill
    \begin{subfigure}{0.32\textwidth}
        \centering
        \includegraphics[height=135px,keepaspectratio]{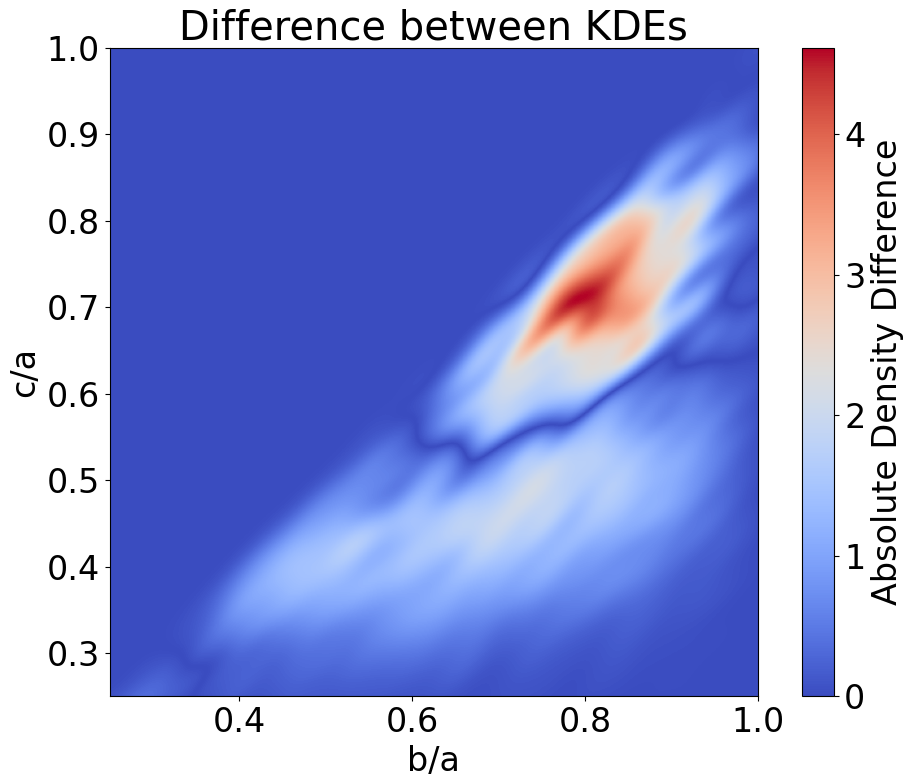}
        \caption{Absolute difference $L_1$.}
    \end{subfigure}

\centering
    \begin{subfigure}{0.32\textwidth}
        \centering
        \includegraphics[width=1\linewidth]{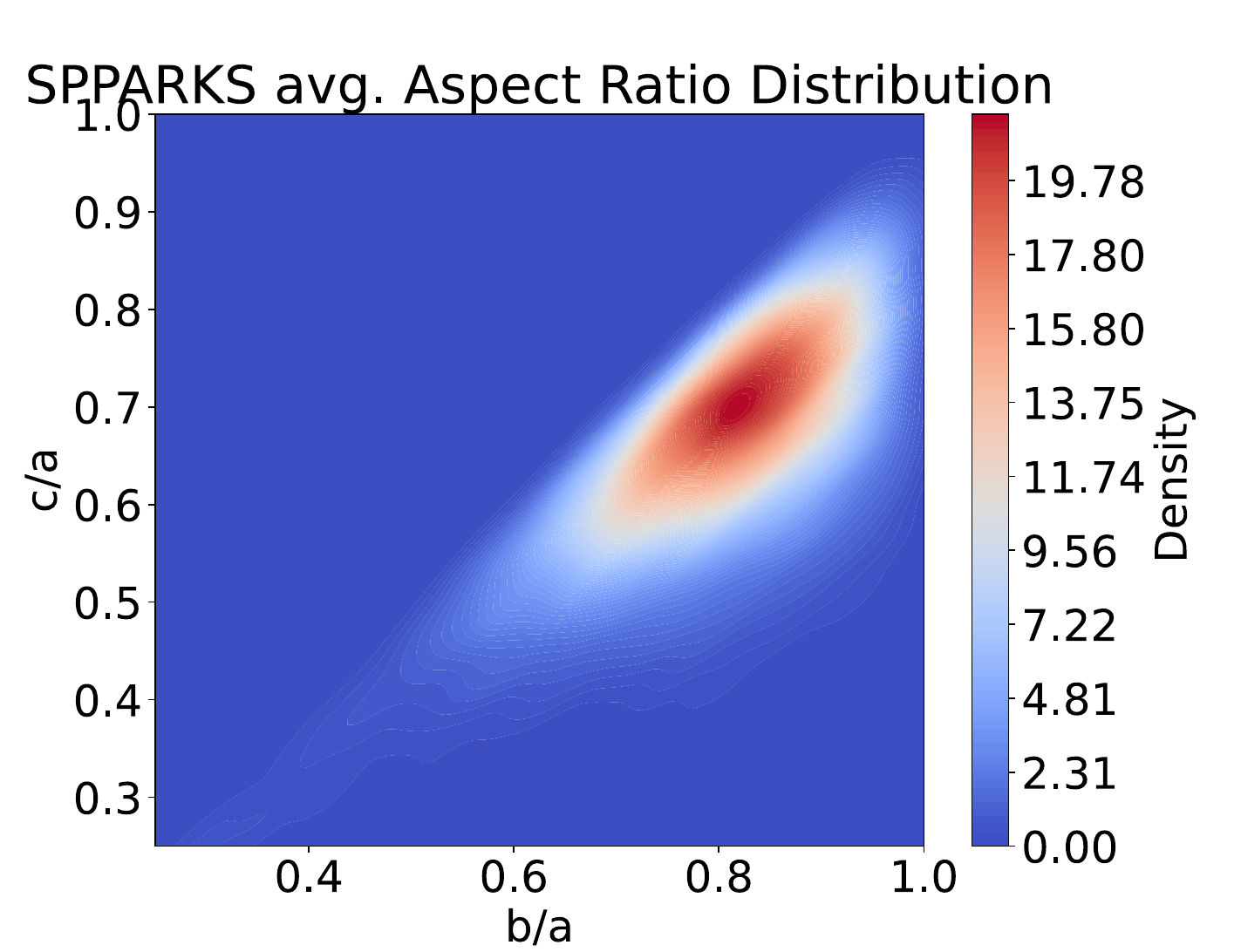}
        \caption{SPPARKS aspect ratio}
    \end{subfigure}
    \hfill
    \begin{subfigure}{0.32\textwidth}
        \centering
        \includegraphics[width=1\linewidth]{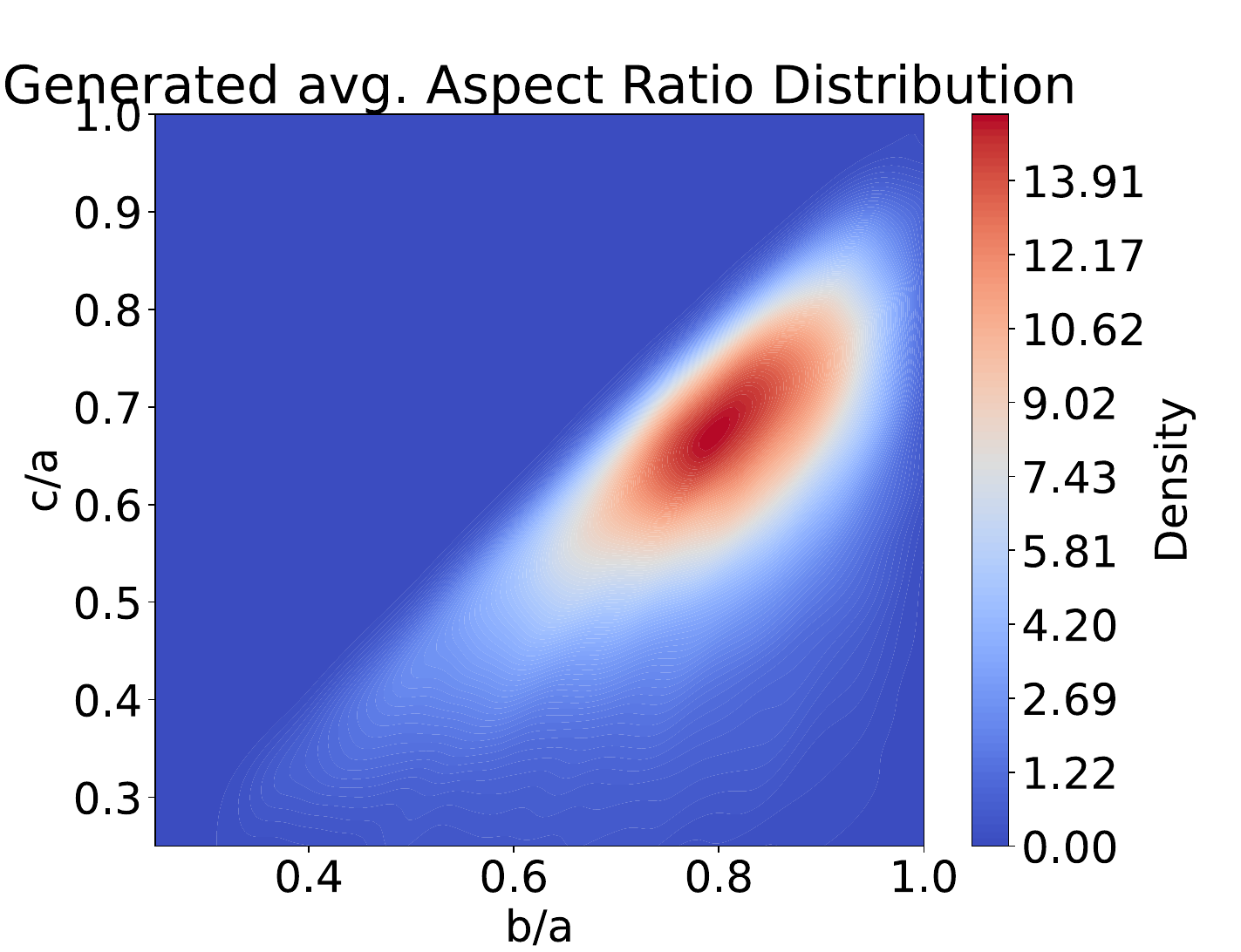}
        \caption{Diffusion aspect ratio}
    \end{subfigure}
    \hfill
    \begin{subfigure}{0.32\textwidth}
        \centering
        \includegraphics[width=1\linewidth]{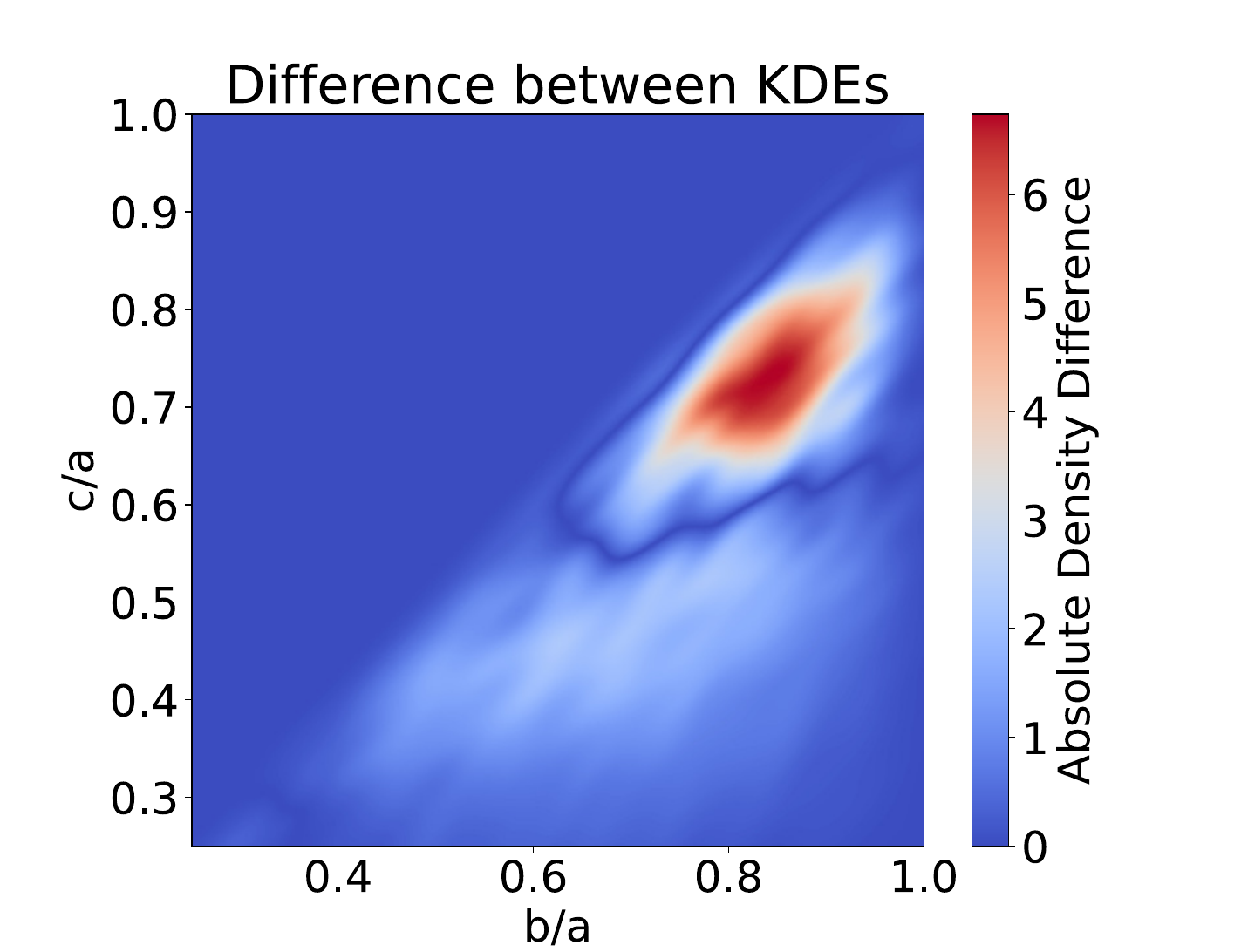}
        \caption{Absolute difference $L_1$.}
    \end{subfigure}
\caption{Grain aspect ratio comparison for isotropic microstructure dataset. (a), (b), and (c) show results for 10 resampling, (d), (e), and (f) show results for 1 resampling.}
\label{fig:aspect_ratio}

\end{figure}

The second descriptor we examine is the grain aspect ratio, \textcolor{black}{shown in \cref{fig:aspect_ratio}}. The grain aspect ratio is calculated using singular value decomposition (SVD), where the first dimension of coordinates transformed with SVD corresponds to the longest axis of the grain, the second dimension the second longest, and the third dimension the shortest. These lengths are denoted as $a$, $b$, and $c$, respectively, where $a \geq b \geq c$ are ordered dimensions of the major axes.

\begin{figure}[!htbp]
    \begin{subfigure}[b]{1.0\textwidth}
    \centering
    \includegraphics[width=1\linewidth]{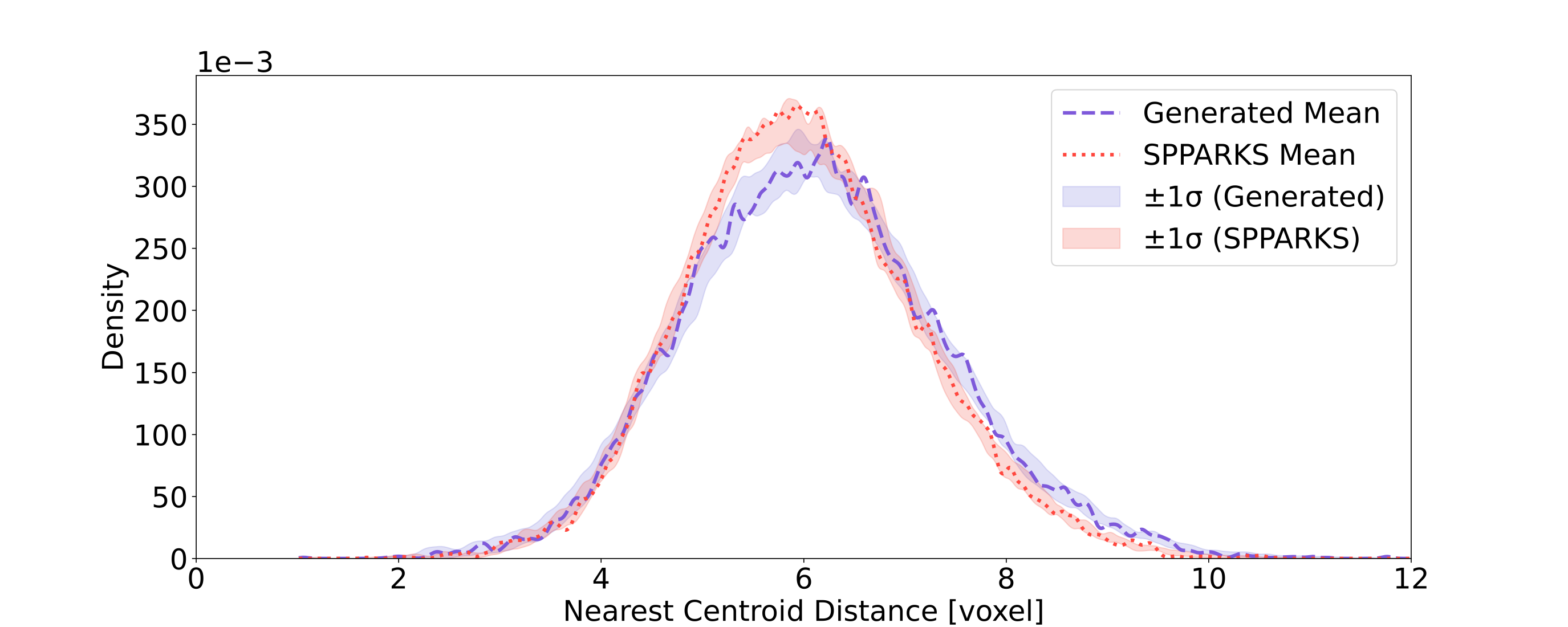}
    \caption{\black{A comparison of the nearest neighboring centroid distance distribution between microstructures simulated with SPPARKS and generated with our diffusion model, GrainPaint, for 10 resamplings.}}
    \label{fig:neigh_centroid_dist}
    \end{subfigure}

    \begin{subfigure}[b]{1.0\textwidth}
    \centering
    \includegraphics[width=1\linewidth]{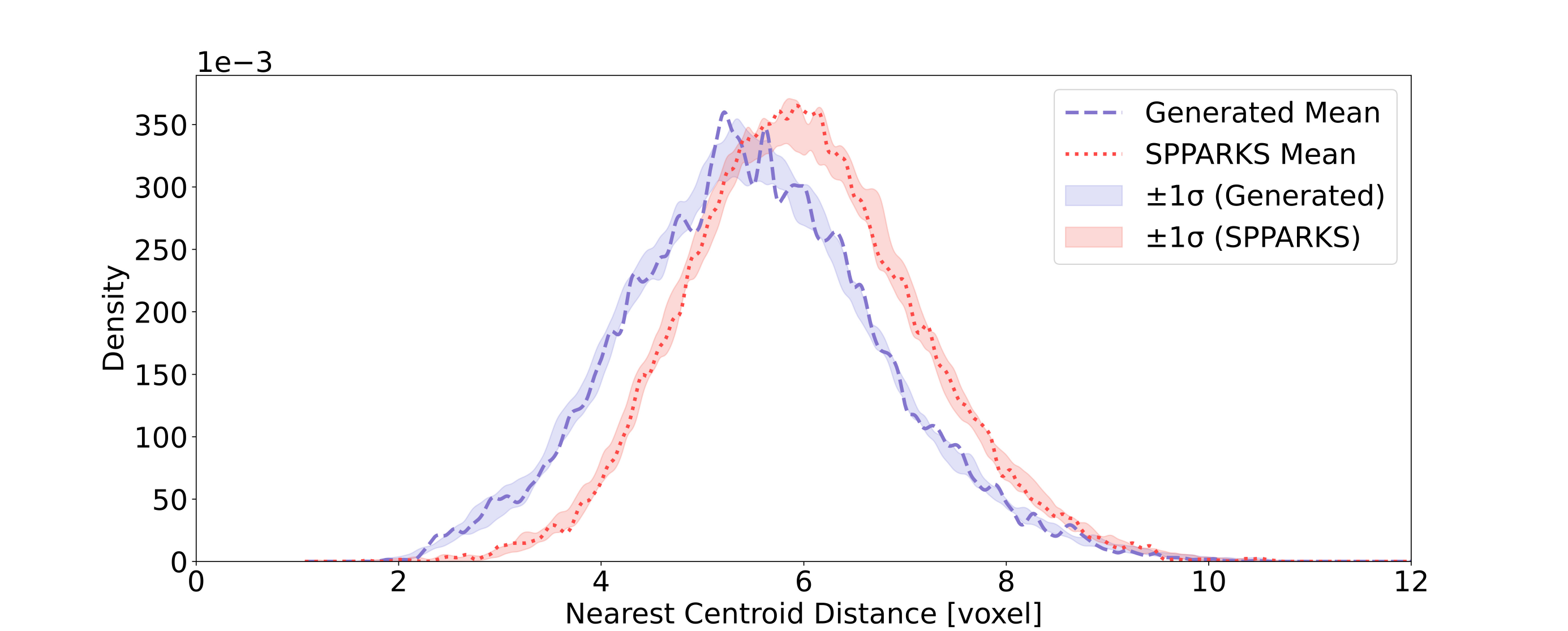}
    \caption{A comparison of the nearest neighbor distribution between SPPARKS and GrainPaint run with 1 resampling (this figure) shows less agreement than SPPARKS and GrainPaint run with 10 resampling steps (\cref{fig:neigh_centroid_dist}).}
    \label{fig:neigh_centroid_dist_1-10}
    \end{subfigure}
    \caption{\black{Nearest neighboring centroid distance distribution for microstructures simulated with SPPARKS and generated with our diffusion model, GrainPaint, for isotropic microstructures.}}
    \label{fig:neigh_centroid_dist_main}
\end{figure}

\black{The third descriptor we examine is distance to nearest neighboring centroid, shown in \cref{fig:neigh_centroid_dist}. The centroid is calculated as the mean coordinate of all voxels in the grain. The distributions are similar, with larger distances being slightly more likely in the microstructures generated by GrainPaint.}

\begin{figure}[!p]
    \centering
    \begin{subfigure}{0.49\textwidth}
        \centering        
        \includegraphics[width=1\linewidth]{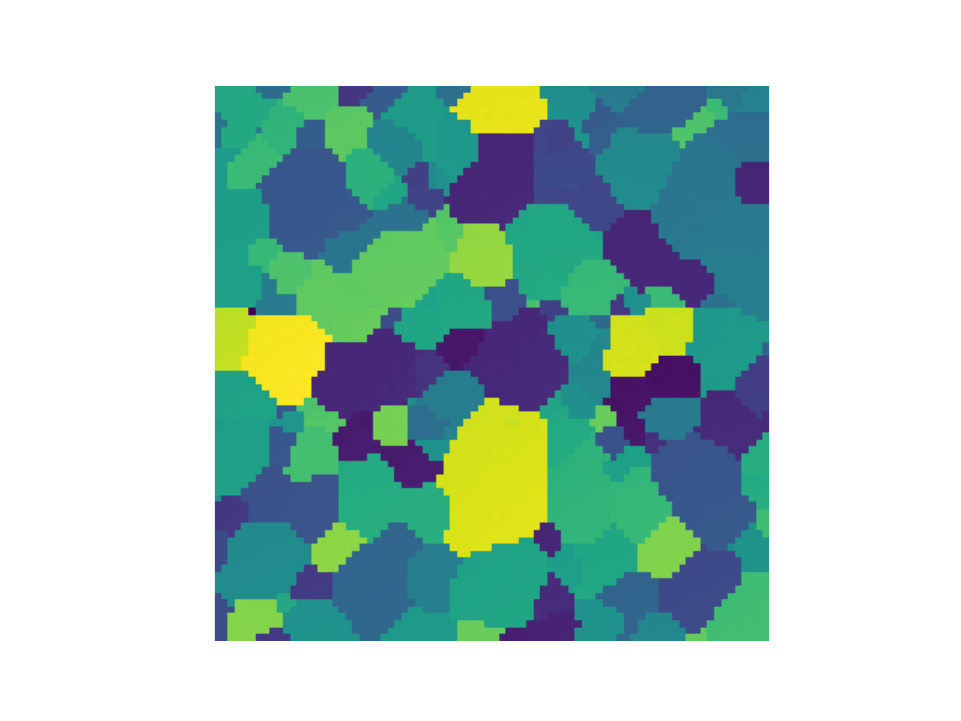}
        \caption{GrainPaint generated $x$ slice}
    \end{subfigure}
    \hfill
    \begin{subfigure}{0.49\textwidth}
        \centering        
        \includegraphics[width=1\linewidth]{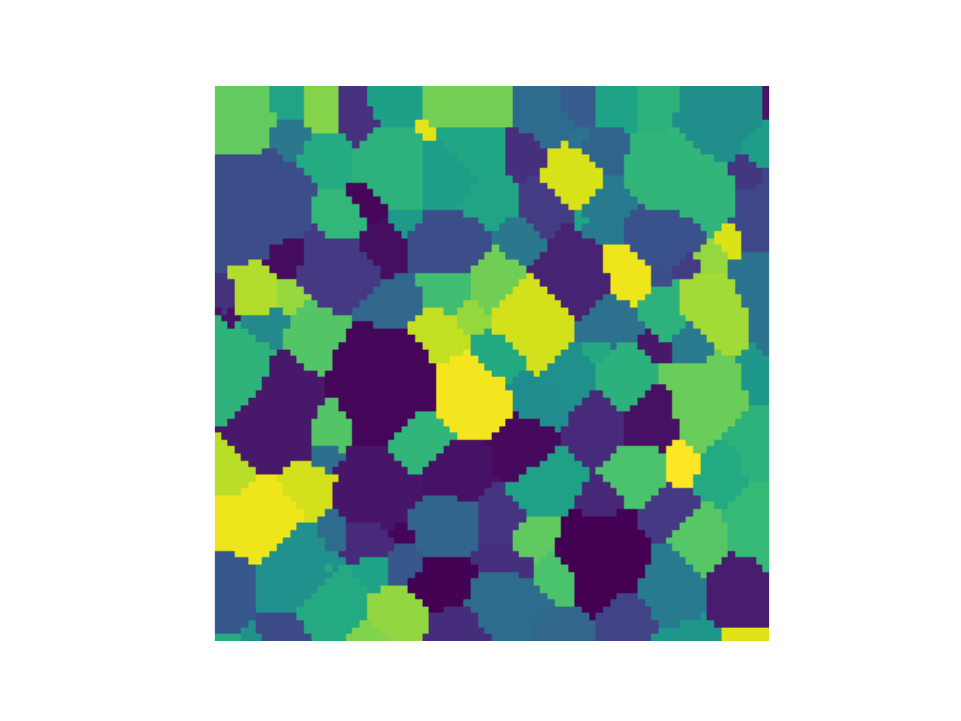}
        \caption{SPPARKS simulated $x$ slice}
    \end{subfigure}
    
    \begin{subfigure}{0.49\textwidth}
        \centering        
        \includegraphics[width=1\linewidth]{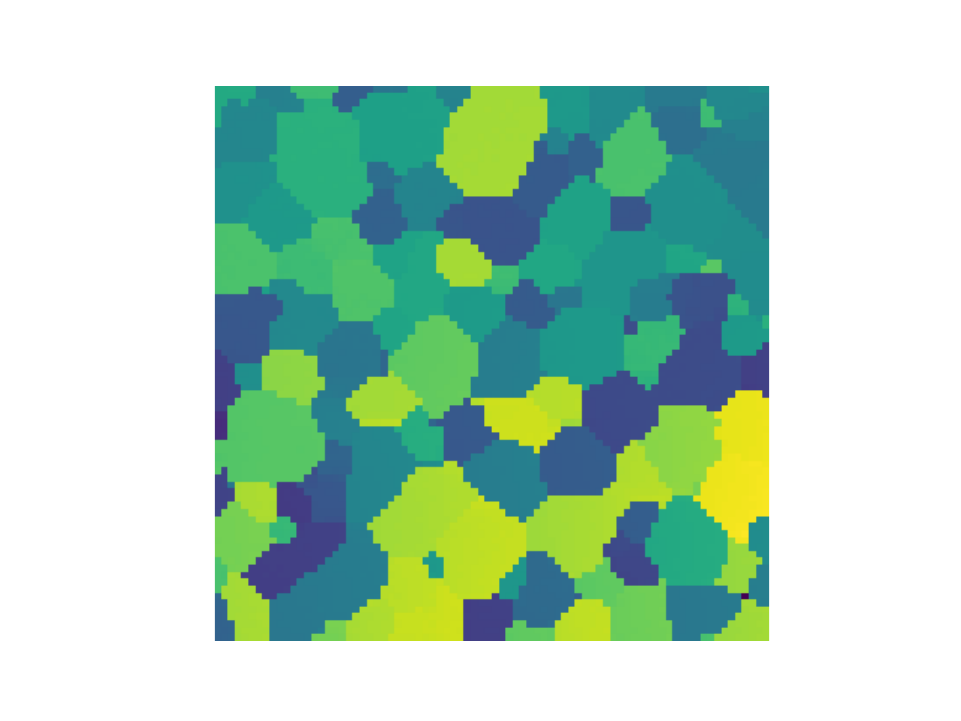}
        \caption{GrainPaint generated $y$ slice}
    \end{subfigure}
    \hfill
    \begin{subfigure}{0.49\textwidth}
        \centering        
        \includegraphics[width=1\linewidth]{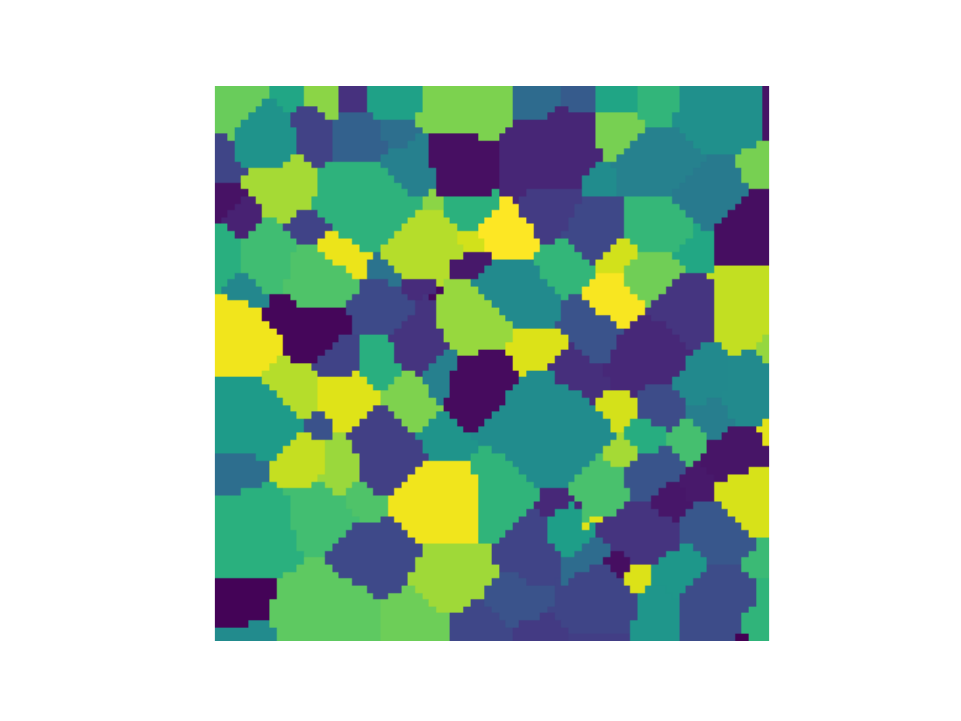}
        \caption{SPPARKS simulated $y$ slice}
    \end{subfigure}

    \begin{subfigure}{0.49\textwidth}
        \centering        
        \includegraphics[width=1\linewidth]{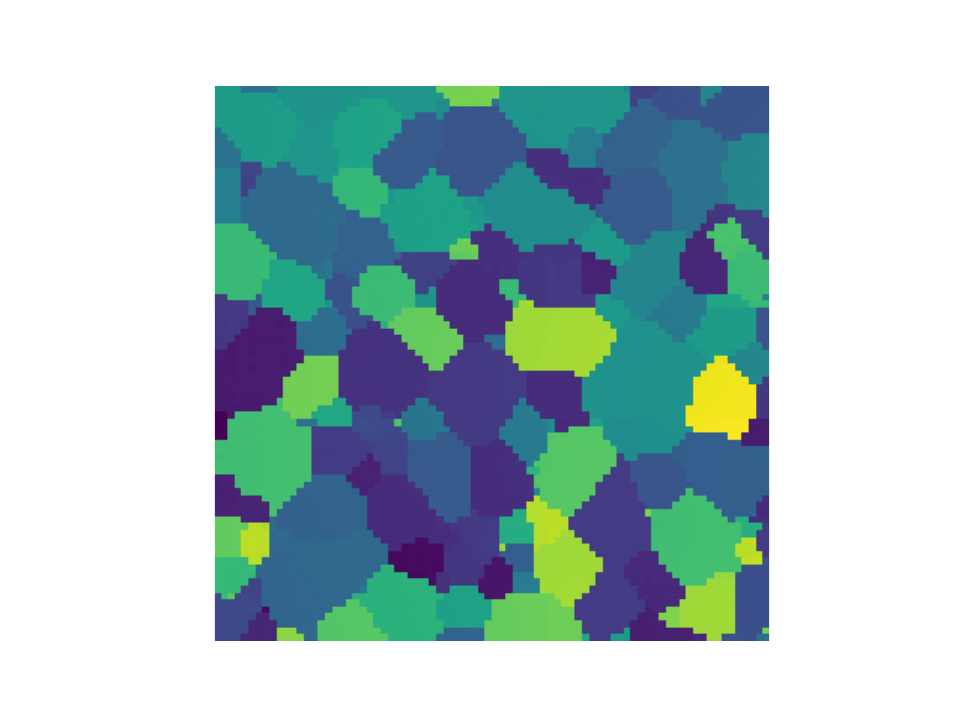}
        \caption{GrainPaint generated $z$ slice}
    \end{subfigure}
    \hfill
    \begin{subfigure}{0.49\textwidth}
        \centering        
        \includegraphics[width=1\linewidth]{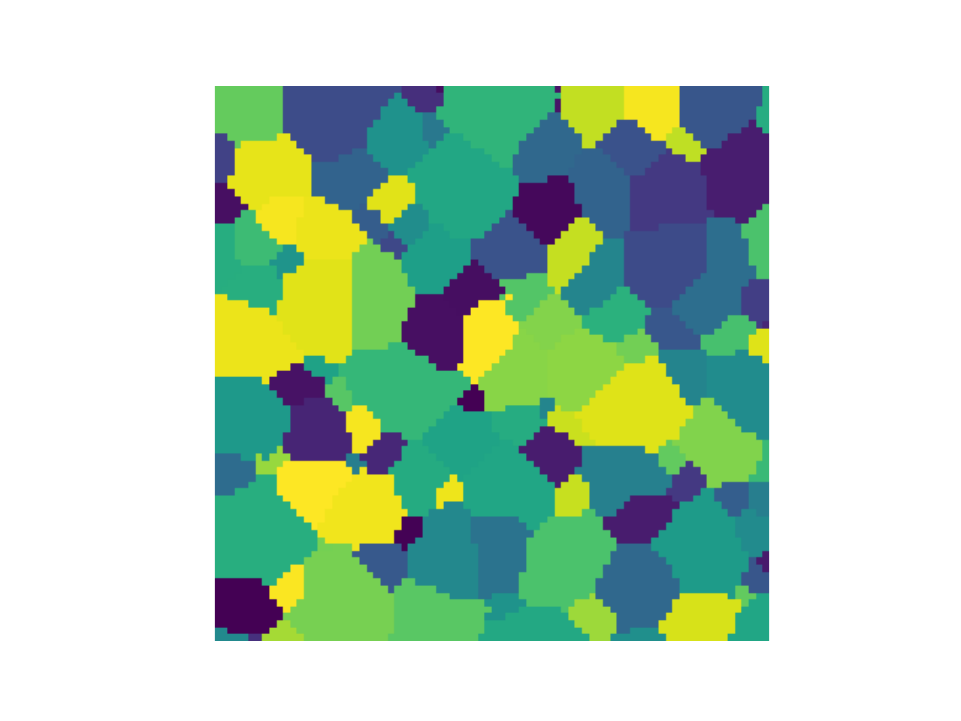}
        \caption{SPPARKS simulated $z$ slice}
    \end{subfigure}
\caption{SPPARKS and GrainPaint isotropic microstructure comparison.}
\label{fig:iso_visual_compare}
\end{figure}


\textcolor{black}{\cref{fig:iso_visual_compare} presents a side-by-side comparison of isotropic microstructure reconstructions generated using GrainPaint (left) and SPPARKS (right), with slices shown along the $x$, $y$, and $z$ directions. Qualitatively, the microstructures in both datasets exhibit isotropy and demonstrate a high degree of similarity.}


\subsection{Anisotropic Microstructure Generation and Evaluation} 
\textcolor{black}{To generate and evaluate the proposed diffusion model on anisotropic microstructures, we again utilize SPPARKS for simulating microstructures in additive manufacturing environment. The likelihood of site $i$ adopting the grain ID of a neighboring site is determined by the Metropolis criterion~\cite{trageser2023bezier}, which defines the probability $P_i$ as:  
\begin{equation}
P_i =  
\begin{cases}  
M(\mathbf{x}) & \text{if } \Delta E_i \leq 0, \\  
M(\mathbf{x}) \exp(-\Delta E_i / k_B T) & \text{if } \Delta E_i > 0,  
\end{cases}  
\label{eq:MetropolisProbability}
\end{equation}
where $\Delta E_i$ denotes the energy of site $i$, $T$ is the numerical temperature, and $M(\mathbf{x}) \in [0,1]$ represents the mobility of the site. 
The mobility $M(\mathbf{x})$ depends on the distance from the melt pool surface and is defined as:  
\begin{equation}
M(\mathbf{x}) =  
\begin{cases}  
1 - \frac{d(\mathbf{x})}{mz} & \text{if } d(\mathbf{x}) \leq mz, \\  
0 & \text{if } d(\mathbf{x}) > mz,  
\end{cases}  
\label{eq:MobilityFunction}
\end{equation}
where $d(\mathbf{x})$ is the distance from site $i$ (located at $\mathbf{x}$) to the nearest point on the melt pool surface, and $mz$ is the threshold distance beyond which mobility is zero. 
This formulation highlights that when a site is farther from the melt pool than the threshold distance $mz$, its mobility becomes zero, effectively halting microstructure evolution in those regions. As a result, changes are confined to areas near the melt pool and the heat-affected zone. 
For more details on the geometric modeling of the melt pool and numerical implementation strategies, readers are referred to~\cite[Section 2.4]{trageser2023bezier}. Similarly, additional insights into the kinetic Monte Carlo model for additive manufacturing simulations, as implemented in SPPARKS, can be found in recent works~\cite{trageser2023bezier,moore2024microstructure,whitney2024solidification}. These studies also delve into temperature modeling using finite-difference methods~\cite{rodgers2021simulation,whitney2024solidification} and provide experimental validations~\cite{adams2015mechanical,soylemez2018modeling}. }

\textcolor{black}{We also evaluate GrainPaint on an anisotropic microstructure dataset generated using this SPPARKS simulation of additive manufacturing. From a dataset of 100 microstructure cubes generated by SPPARKS with side length 100, we sample 40 cubes with side length 32 from uniform random positions within each SPPARKS generated cube. This gives a training set size of 40,000. As the arrangement of the grains in the anisotroipic microstructure is correlated over distances larger than the 32$\times$32$\times$32 window GrainPaint generates in, the generation procedure used for the isotropic microstructures will not work. This is because the isotropic microstructure generation algorithm begins by generating disconnected areas of microstructure, and then connects them. This procedure will not work with the anisotropic microstructure because the rows the grains are arranged in must be aligned. To solve this issue, we use a different generation procedure that first generates the center of the microstructure and then generates new pieces until the edges are reached. The anisotroipc generation algorithm in earlier steps generates a cross-shaped pattern from the center towards the edges of the geometry. This part of the processes uses an overlap of 16 voxels. This overlap is larger than used in the isotropic algorithm, and we believe this larger overlap helps GrainPaint align the orientations of grains across large distances, though we did not test this. After the cross is generated, the rest of the geometry is filled be iteratively generating toward the edges with an 8 voxel overlap. Note that an 8 voxel overlap is the same used for the isotropic generation algorithm.}

\begin{figure}[!htbp]
    \centering
    \includegraphics[width=\linewidth]{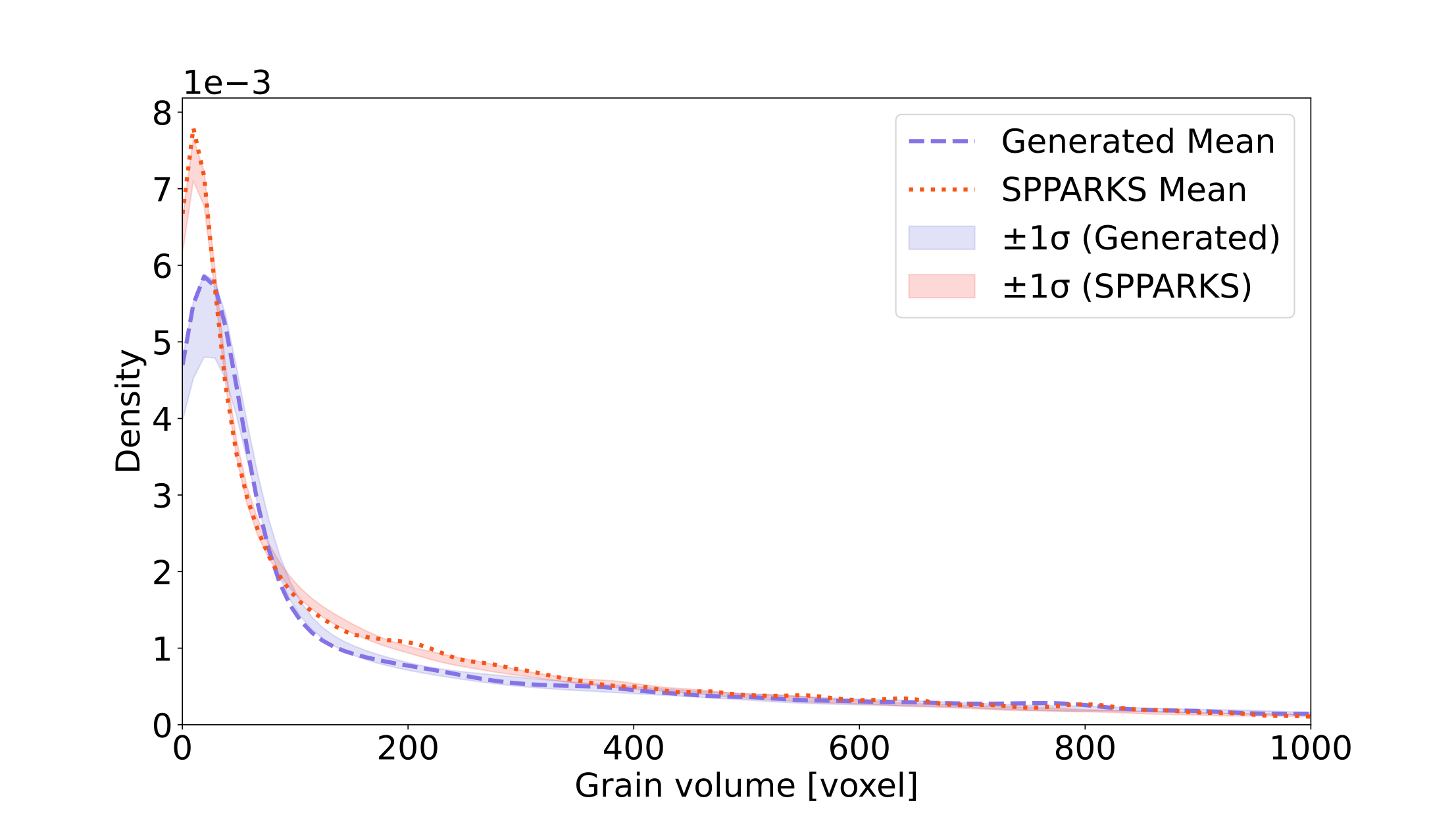}
    \caption{Comparison of grain-size distributions between SPPARKS and GrainPaint for anisotropic microstructures.}
    \label{fig:grain_size_dist_aniso}
\end{figure}

\textcolor{black}{\cref{fig:grain_size_dist_aniso} shows the grain size distribution of SPPARKS and our proposed diffusion model, which shows a reasonable agreement. The tail of both distributions are quantitatively similar, while microstructures generated from SPPARKS has more smaller grains. The mode of these distributions are similar. Therefore, despite a difference in terms of magnitude of small grains, they agree relatively well with each other.}

\begin{figure}[!htbp]
    \centering
    \begin{subfigure}{0.32\textwidth}
        \centering
        \includegraphics[width=\linewidth]{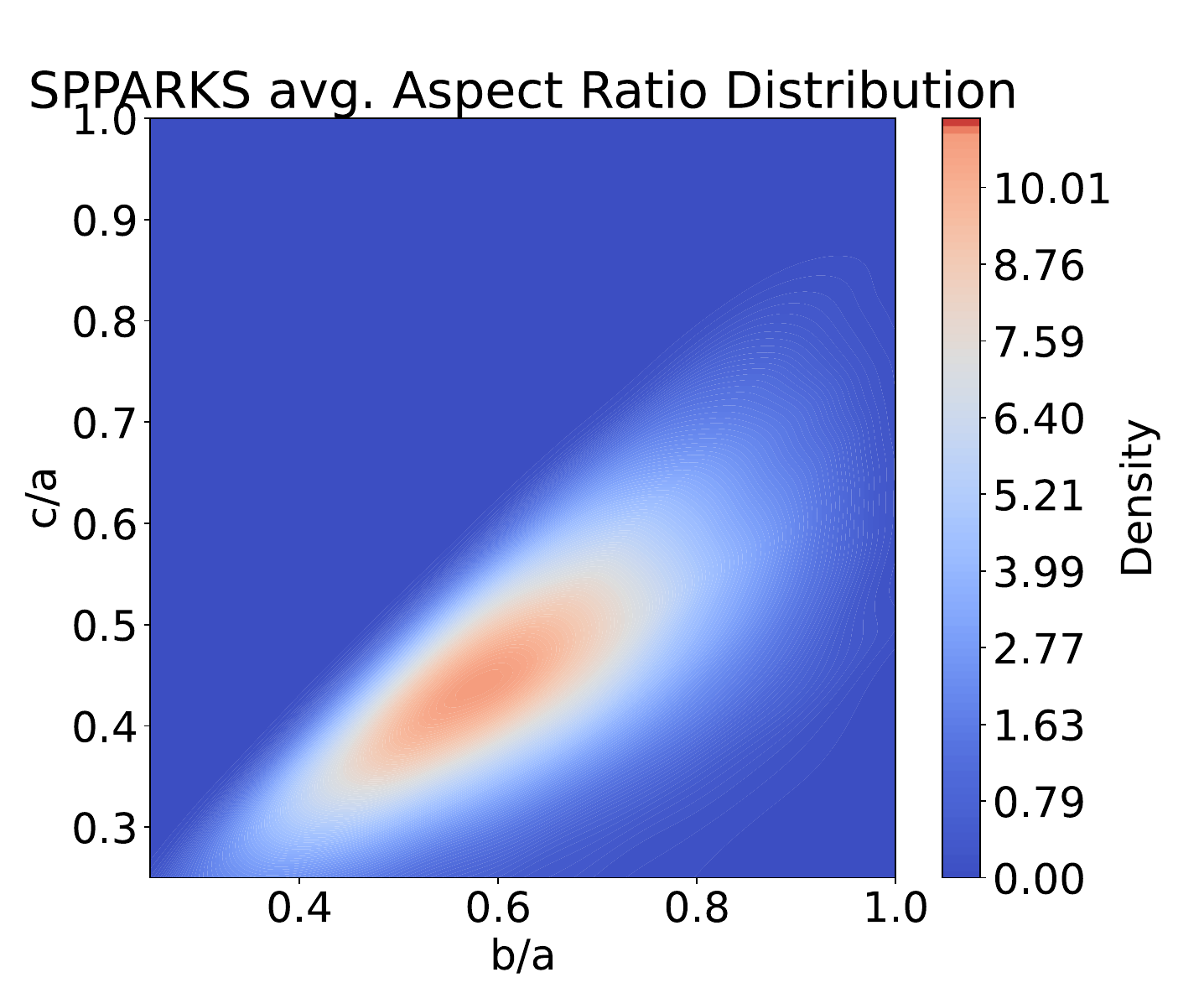}
        \caption{SPPARKS aspect ratio}
    \end{subfigure}
    \hfill
    \begin{subfigure}{0.32\textwidth}
        \centering
        \includegraphics[width=\linewidth]{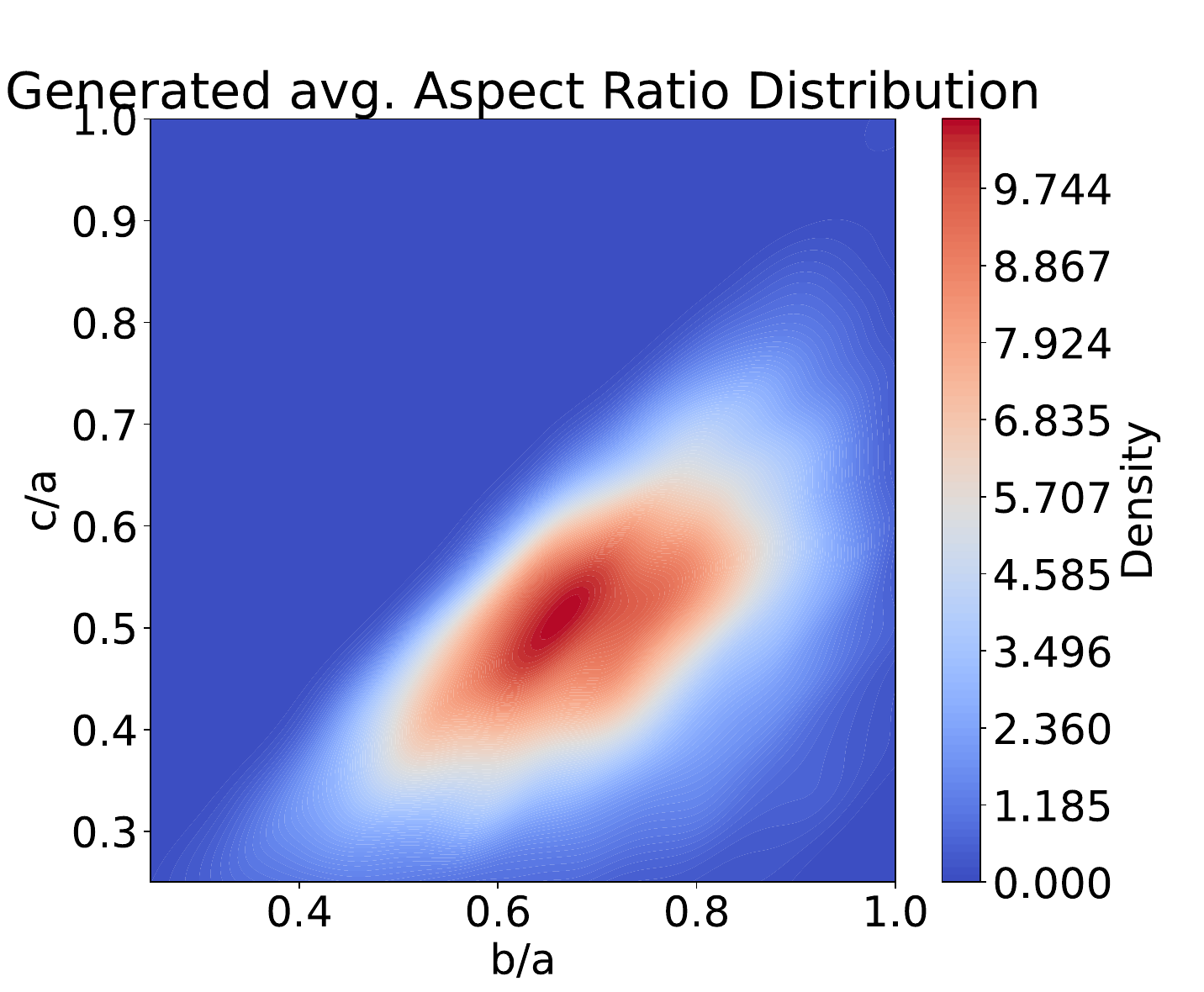}
        \caption{Diffusion aspect ratio}
    \end{subfigure}
    \hfill
    \begin{subfigure}{0.32\textwidth}
        \centering
        \includegraphics[width=\linewidth]{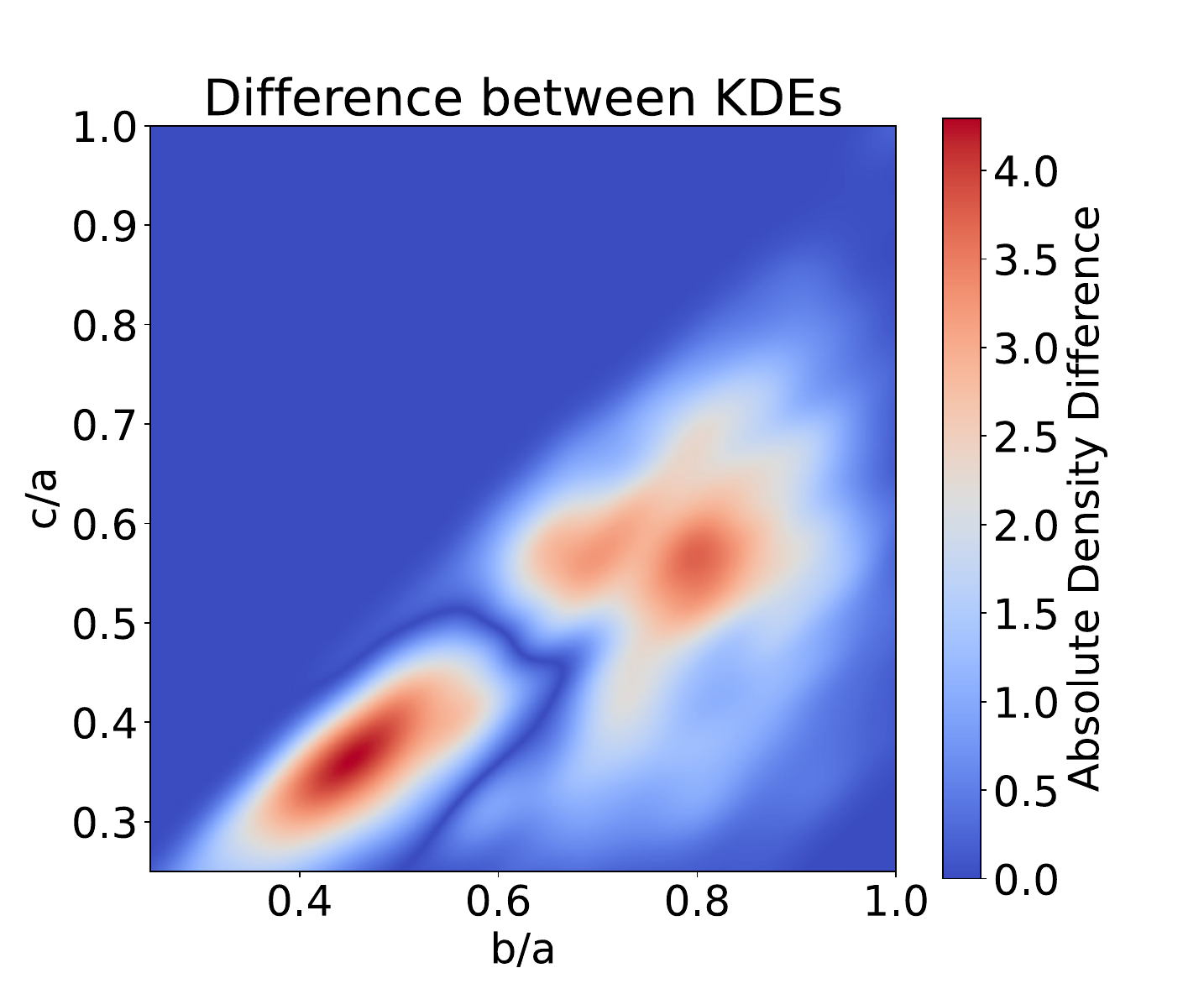}
        \caption{Absolute difference $L_1$.}
    \end{subfigure}
\caption{Grain aspect ratio comparison for anisotropic microstructure dataset.}
\label{fig:aspect_ratio_aniso}
\end{figure}

\textcolor{black}{\cref{fig:aspect_ratio_aniso} shows the grain aspect ratio comparison between SPPARKS and our proposed diffusion model. Both exhibit a single modal distribution with a similar concentrated area. Our diffusion model differs to SPPARKS in the sense that GrainPaint favors less elongated grain with low aspect ratios (rod-like grains), whereas SPPARKS generates more grains with low aspect ratios. Both distributions share a similar support.}

\begin{figure}[!htbp]
    \centering
    \includegraphics[width=\linewidth]{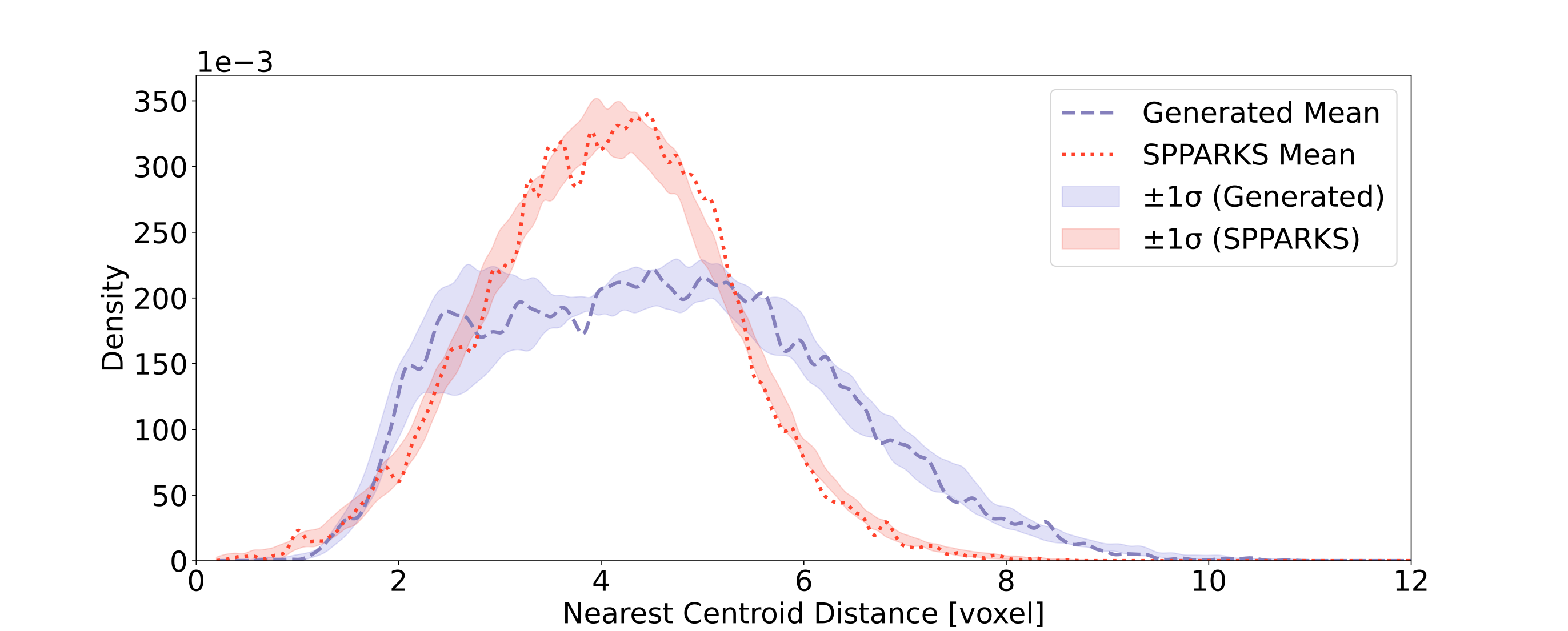}
    \caption{\black{A comparison of the nearest neighboring centroid distance distribution between microstructures simulated with SPPARKS and generated with our diffusion model, GrainPaint, for anisotropic microstructures.}}
    \label{fig:neigh_centroid_dist_aniso}
\end{figure}

\textcolor{black}{\cref{fig:neigh_centroid_dist_aniso} shows the nearest centroid distance distribution between SPPARKS and our proposed diffusion model. While both distributions are somewhat similar (single modal, significant overlap), there are some substantial differences. The distributions from SPPARKS resembles a normal distribution, while the one from GrainPaint is slightly unsymmetrical. Moreover, there is no obvious mode for microstructures generated from GrainPaint, whereas there is an obvious mode for microstructures generated from SPPARKS. This suggests that there is a limitation in our proposed model that does not capture the neighboring relationship well.}

\begin{figure}[!p]
    \centering
    \begin{subfigure}{0.49\textwidth}
        \centering        
        \includegraphics[width=1\linewidth]{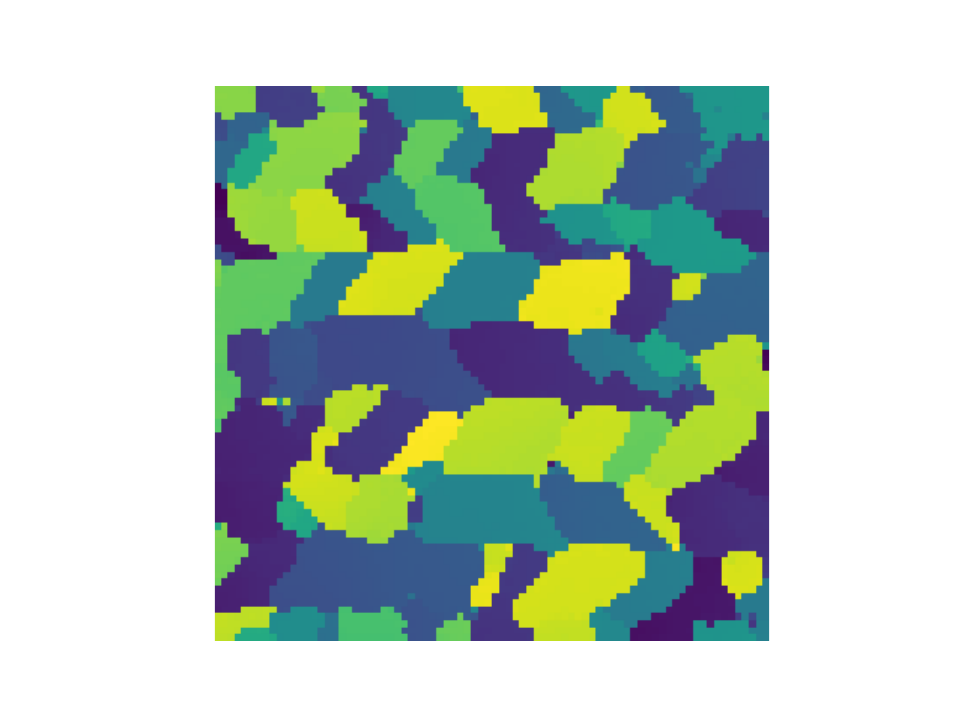}
        \caption{GrainPaint generated $x$ slice}
    \end{subfigure}
    \hfill
    \begin{subfigure}{0.49\textwidth}
        \centering        
        \includegraphics[width=1\linewidth]{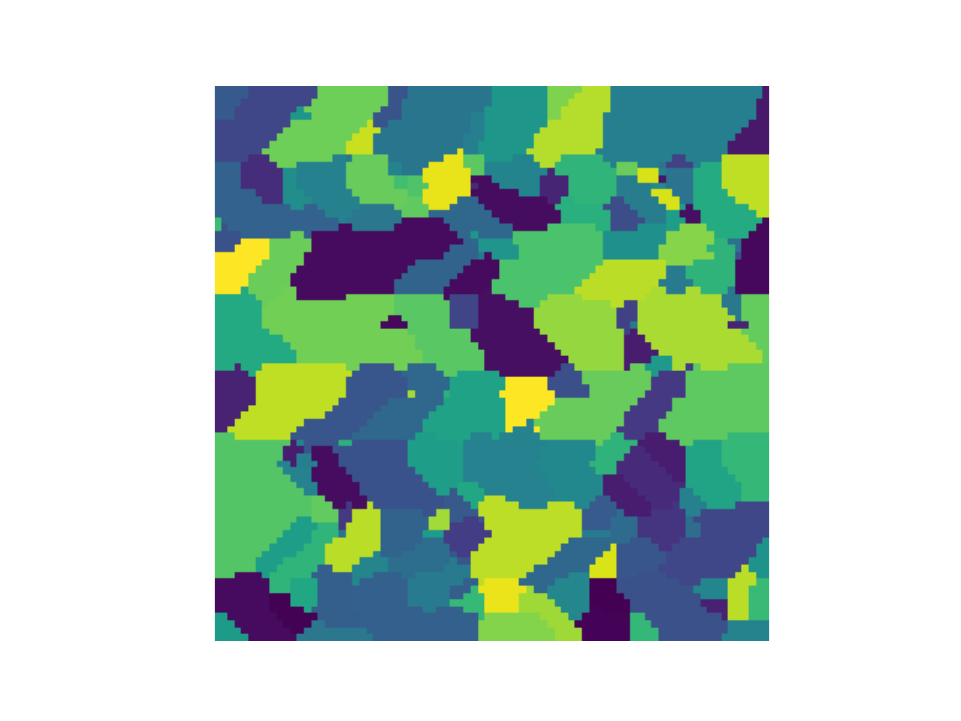}
        \caption{SPPARKS simulated $x$ slice}
    \end{subfigure}
    
    \begin{subfigure}{0.49\textwidth}
        \centering        
        \includegraphics[width=1\linewidth]{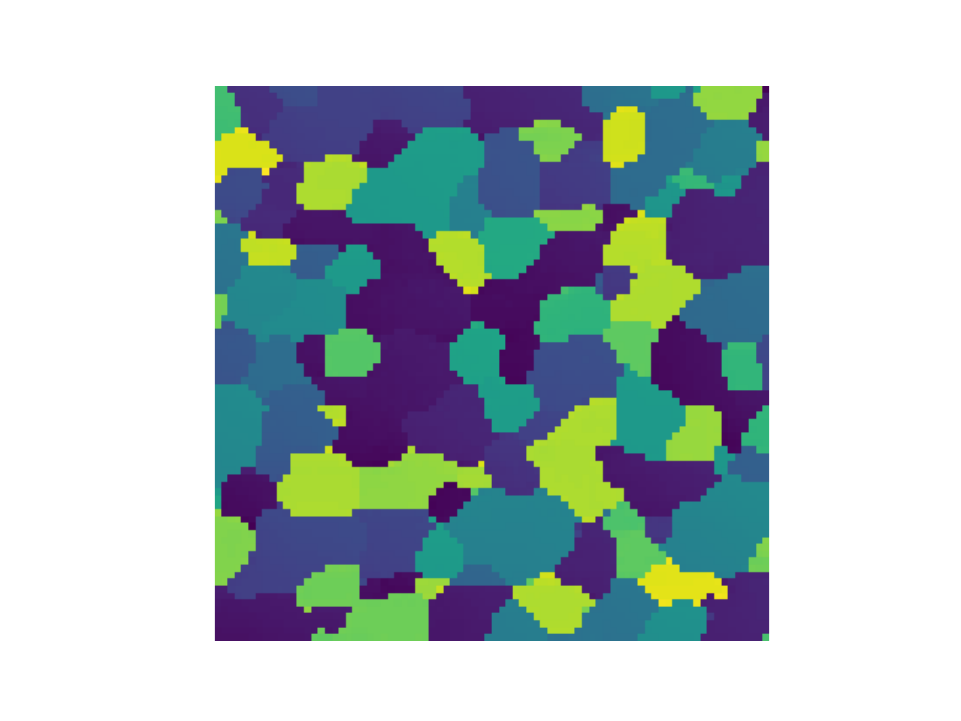}
        \caption{GrainPaint generated $y$ slice}
    \end{subfigure}
    \hfill
    \begin{subfigure}{0.49\textwidth}
        \centering        
        \includegraphics[width=1\linewidth]{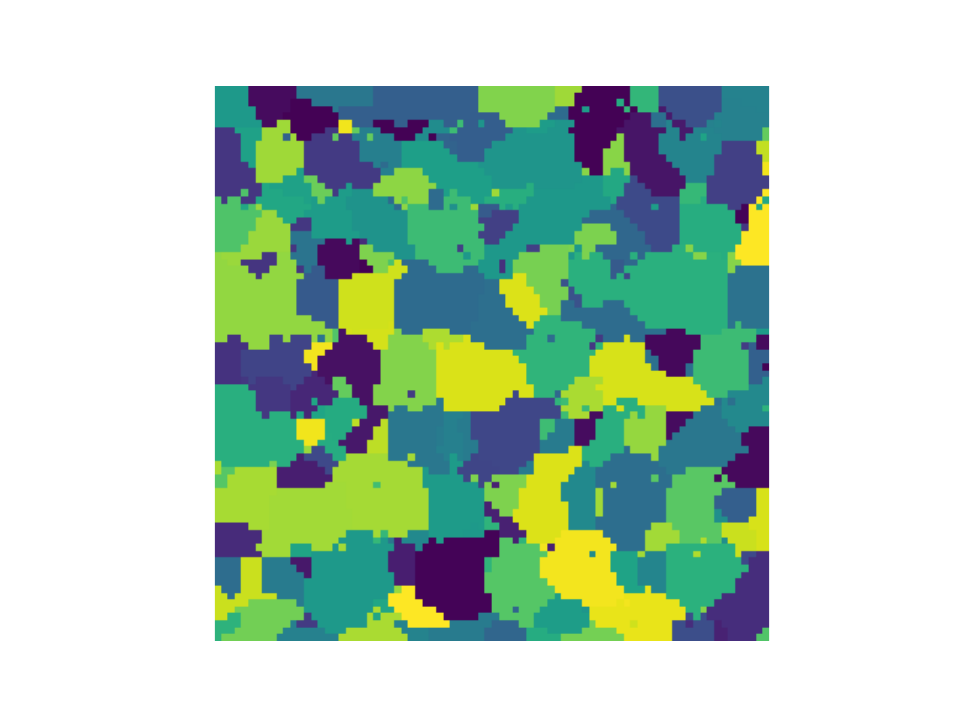}
        \caption{SPPARKS simulated $y$ slice}
    \end{subfigure}

    \begin{subfigure}{0.49\textwidth}
        \centering        
        \includegraphics[width=1\linewidth]{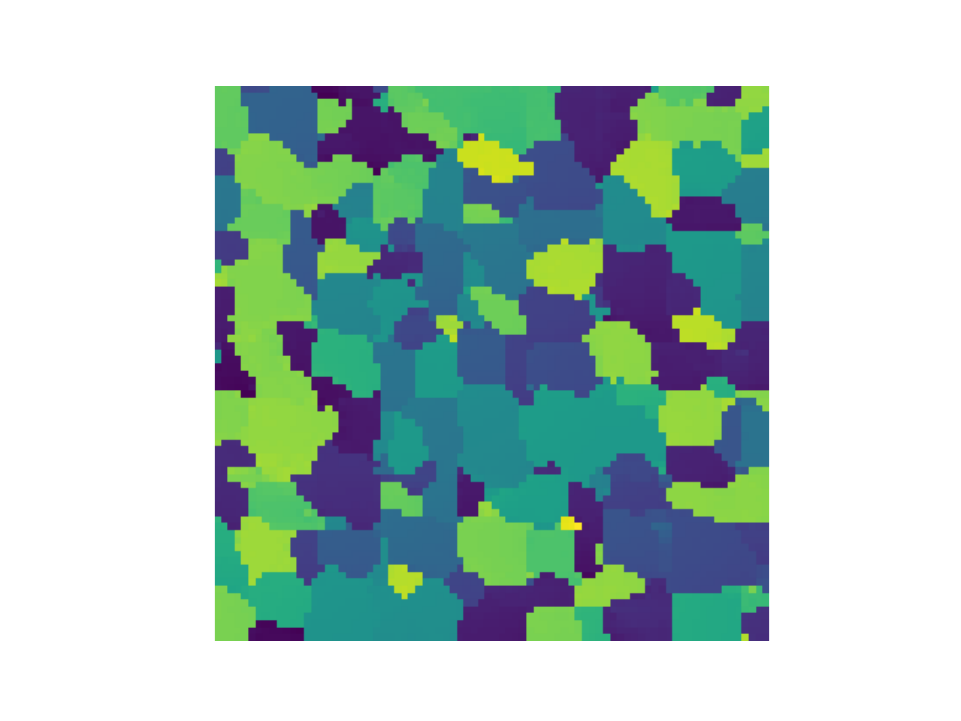}
        \caption{GrainPaint generated $z$ slice}
    \end{subfigure}
    \hfill
    \begin{subfigure}{0.49\textwidth}
        \centering        
        \includegraphics[width=1\linewidth]{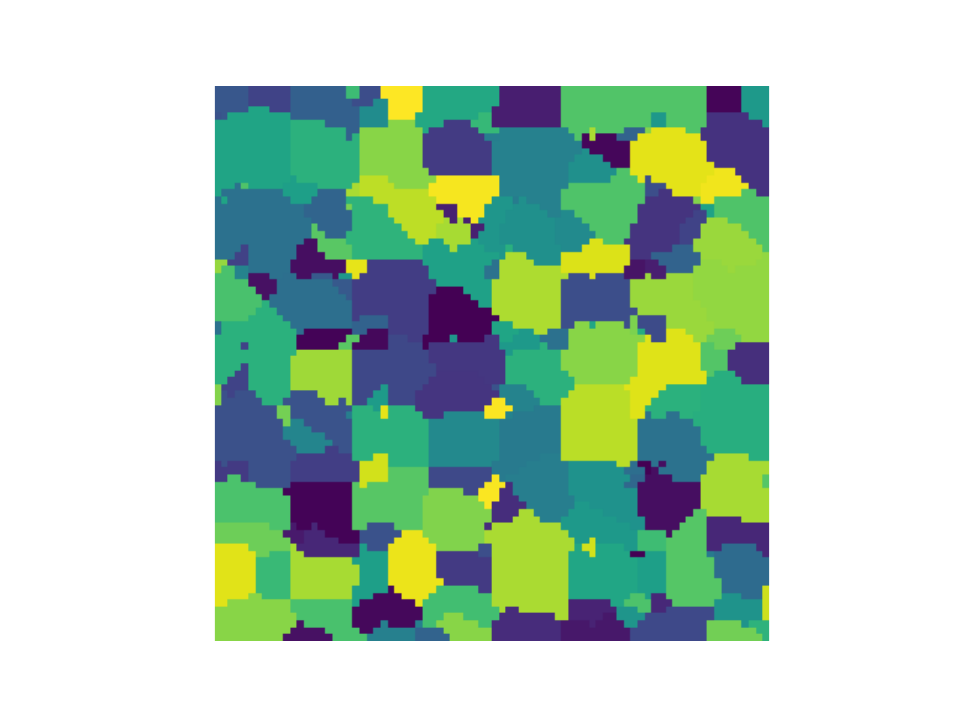}
        \caption{SPPARKS simulated $z$ slice}
    \end{subfigure}
\caption{SPPARKS and GrainPaint anisotropic microstructure comparison.}
\label{fig:aniso_visual_compare}
\end{figure}

\textcolor{black}{A side-by-side comparison of the microstructures generated by SPPARKS and GrainPaint in \cref{fig:aniso_visual_compare} show that GrainPaint can capture some features of the anisotropic microstructure, but not others. GrainPaint appears to be capable of maintaining the alignment of rows of grains across the entire microstructure. However, GrainPaint also appears to favor generating larger grains, and is particularly unlikely to generate the smallest grains. A statistical comparison shows that compared to the SPPARKS training data, GrainPaint generates fewer grains with a volume less than about 50, more between 50 and 100, and fewer between 100 and 400. The centroid distance distribution shows that GrainPaint generates a wider distribution and favors larger centroid distances in comparison to the SPPARKS training set. }

\section{Discussion} \label{sec:discussion}
The diffusion model presented in this work can generate realistic microstructures of arbitrary size. Our results have been validated by microstructure statistics comparison between the diffusion model and SPPARKS, a kinetic Monte Carlo simulation. While we do not see an obvious advantage in computational cost for the diffusion model compared to the kinetic Monte Carlo simulation with SPPARKS, we believe there could be an advantage on other datasets created by a more computationally expensive simulation process such as phase-field modeling. 
We believe there are a variety of potential model inference performance optimizations such as pruning or different model architectures which could significantly reduce computational cost. 
Another area for potential improvement is the segmentation process. We propose two ways to address this: 
\begin{enumerate}
    \item The problem could be converted to binary images. Binarization could be accomplished by representing grain boundaries as voxels.
    \item The training data could be augmented so that there are a low number of grain IDs. This would ensure the values generated by the model are further apart and therefore easier to segment. 
    
\end{enumerate} 
This would reduce the need for a complex segmentation process. The hierarchical segmentation algorithm developed for this work is limited in its practical use by memory usage (approximately 1GB per million voxels). If this algorithm were to be further developed, this issue could likely be resolved with implementation improvements.

The voxelization process that is used to convert the CAD mesh files to voxels has several limitations: first, voxelization can lead to loss of detail, or aliasing artifacts for some features. Second, the simple voxelization algorithm used in this work has performance issues and high memory requirements with large geometries. The first limitation can be ameliorated with the selection of an appropriate resolution for the voxelization. While voxelization will never be a perfect representation of a mesh (due to non-negative numerical approximation errors), it can still provide useful insight. The second limitation was not relevant to the CAD objects used in this work, but we believe it can be addressed with a better voxelization algorithm.

\black{The choice of the number of resampling steps is another area for potential optimization, as the number of resampling steps has a large impact on performance. Identifying an optimal number of resampling steps is challenging as there is a trade-off between computational cost and quality. Furthermore, the Kullback-Leibler Divergence between distributions calculated for generated and simulated microstructures is only a relative measure of quality. Lugmayr et al. evaluate different numbers of resamplings using the Learned Perceptual Image Patch Similarity (LPIPS) metric, which uses a neural network to predict the result of a human similarity rating. LPIPS is well aligned to the objective of image generation: the creation of realistic images. The problem of microstructure generation is more difficult to evaluate as the objective is to able to accurately model properties that arise from the microstructure which is less subjective than LPIPS. In any case, the LPIPS model cannot be used to evaluate GrainPaint as LPIPS is trained on images instead of volumes. Perhaps a LPIPS-like metric could be developed for microstructures which estimates a variety of properties of interest that could be evaluated  depending on the microstructure or the application. In addition, we note that Lugmayr et al. state that the benefits of resampling saturate at about 10 resamplings~\cite{repaint}. As we made a similar observation on a very different dataset, it is possible that 10 resamplings saturates the benefits of resampling on all datasets.}

Diffusion models are known to be capable of a wide variety of tasks, so we expect that the process presented in the work could be used not only with other 3D normal grain growth models, including phase-field and cellular automata, but also with other types of microstructures. One major limitation of the process presented in this work is that any feature the diffusion model will generate must be homogeneous and must fit within the area the model generates in. Many microstructures have multi-scale or non-homogeneous features. We expect that these problems can be addressed by supplying a conditioning vector to the diffusion model and the use of multiple models for different scales. 

\textcolor{black}{From a CAD standpoint for large-scale objects, generating microstructures for an arbitrary CAD object, as demonstrated in \cref{fig:diffusionCAD} and \cref{fig:spparksCAD}, can be done relatively straightforward by masking the object and extracting only the regions of interest. While obviously, this approach does not fully account for boundary conditions, the resulting microstructures are comparable to experimental microstructures in practice. Moreover, one can model \textit{any} object of interest, as long as the CAD object can be voxelized.}

\textcolor{black}{The decision to utilize CPUs for SPPARKS and GPUs for GrainPaint is primarily rooted in historical developments. During the 2000s and early 2010s, as computer hardware continued to advance and scientific computing gained traction, many ICME models~\cite{horstemeyer2012integrated, allison2013implementing} were developed using languages such as Fortran, C, and C++. These models employed OpenMP and MPI parallelism to distribute computational workloads across multiple cores and nodes effectively. In the late 2010s, the rapid rise of ML~\cite{lecun2015deep} brought GPUs to the forefront, thanks to their superior performance in parallel processing, which has significantly advanced scientific computing and scientific ML~\cite{cuomo2022scientific}. Efforts to modernize legacy ICME codes and leverage heterogeneous computing infrastructures, such as Kokkos~\cite{edwards2013kokkos, edwards2014kokkos, trott2021kokkos, trott2021kokkos2}, aim to integrate the strengths of various hardware architectures. However, substantial work remains to achieve a fair comparison between CPUs and GPUs in these contexts.
}

\textcolor{black}{The computational speedup factor for adopting ML to accelerate ICME depends on several factors, based on applications at hand. 
First, it depends on the computational cost of simulating the ICME model, which varies depending on the detail of the physics. 
For example, an additive manufacturing simulation~\cite{rodgers2017simulation} is substantially more expensive, and accounting for thermo-mechanical loading is possible~\cite{moore2024microstructure,whitney2024solidification} through finite difference Monte Carlo, but it would even be more expensive. 
Since DDPM is purely data-driven, its training cost is constant, while the cost to generate the training dataset is different. 
For applications with more physics, such as temperature, phase, composition, would certainly increase the cost efficiency of adopting DDPM for microstructure generation. 
Second, SPPARKS is a highly efficient stochastic ICME model with three solvers, including one dimensionally independent, constant time $\mathcal{O}(1)$ solver~\cite{slepoy2008constant} with rigorous strong and weak scaling on a large CPU clusters~\cite{mitchell2023parallel}, which is hard to compete computationally with the current GrainPaint DDPM model, particularly for simple application such as normal grain growth.}

\textcolor{black}{In materials science, where microstructures solely depends on the chemical compositions and process conditions, varying either could result in completely different microstructure. In this paper, we aim to establish a ML approach that is capable of large-scale microstructure reconstruction with the same process conditions. A conditional model, e.g.~\cite{iyer2019conditional}, may be a potential future work to address various process conditions.}

\textcolor{black}{Experimentally, the microstructures produced through additive manufacturing are path-dependent, meaning that variations in the printing path result in different microstructures and, consequently, distinct material properties. While modeling part-scale systems with mesoscale fidelity to capture microstructural details is feasible~\cite{whitney2024solidification,whitney2024part}, this poses a multi-scale, computationally intensive challenge. Current state-of-the-art methods can effectively handle millimeter-scale components~\cite{bishop2015direct,bishop2016direct,rodgers2018direct,brown2019modeling}, but addressing complex CAD models in practical, real-world scenarios remains out of reach. This limitation represents a significant, unresolved challenge, leaving the field open for future research.}

Voronoi diagrams have long been used as a low computation cost method for generating grain structures~\cite{weyer2002automatic}. While Voronoi diagrams could have been used to generate grain structures similar to the one considered in this work, the Voronoi approach has several limitations. First, they cannot model anisotropic and complicated microstructures such as those found in 3D-printed objects. 
Second, it is difficult to adopt the Voronoi approach to model the process-structure relationship associated with a specific manufacturing process. 

\section{Conclusion} \label{sec:conclusion}

This work demonstrates generating realistic grain structures of arbitrary size using a diffusion model.
Unlike SPPARKS and other simulation software, which are limited by geometry constraints inherent to simulation, \textit{i.e.}, constraints imposed by the necessity of boundary conditions, our method can generate in a wider variety of shapes. SPPARKS also needs to keep the entire microstructure in memory for the type of problem in this work while a diffusion model does not, making it feasible to generate larger microstructures. 
Inference using GrainPaint model has a high computational cost, but the model can scale across a large number of GPUs so that runtime is reduced to a reasonable level. While the process simulated by SPPARKS to produce the dataset had lower computational cost than the GrainPaint model, the GrainPaint model will have the same computational cost for any dataset. Therefore, the GrainPaint model may have a computational cost advantage for processes that are expensive to simulate.


\section{Acknowledgment}


Sandia National Laboratories is a multi-mission laboratory managed and operated by National Technology \& Engineering Solutions of Sandia, LLC (NTESS), a wholly owned subsidiary of Honeywell International Inc., for the U.S. Department of Energy’s National Nuclear Security Administration (DOE/NNSA) under contract DE-NA0003525. This written work is authored by an employee of NTESS. The employee, not NTESS, owns the right, title, and interest in and to the written work and is responsible for its contents. Any subjective views or opinions that might be expressed in the written work do not necessarily represent the views of the U.S. Government. The publisher acknowledges that the U.S. Government retains a non-exclusive, paid-up, irrevocable, world-wide license to publish or reproduce the published form of this written work or allow others to do so, for U.S. Government purposes. The DOE will provide public access to results of federally sponsored research in accordance with the DOE Public Access Plan.

\section{Data availability}

The STL files concerned in this paper are available at \href{https://github.com/anhvt2/spparks-hackathon}{https://github.com/anhvt2/spparks-hackathon}. The dataset used in this work is available at \href{https://zenodo.org/record/8241535}{https://zenodo.org/record/8241535}. 
\section{Code availability}

The code for GrainPaint and our analysis is available at \href{https://github.com/njhoffman11/GrainPaint}{https://github.com/njhoffman11/GrainPaint}.
\section{Author contributions}
\textbf {All}: Review \& Editing \textbf {Nathan Hoffman}: Conceptualization, Data curation, Formal analysis, Visualization, Original Draft, Software, Methodology, Investigation \textbf {Cashen Diniz}:  Conceptualization, Data curation, Formal analysis, Visualization, Original Draft, Software, Methodology, Investigation \textbf {Anh Tran}:  Data curation, Visualization, Original Draft, Software, Methodology \textbf{Dehao Liu}: Original Draft,  Methodology \textbf{Theron Rodgers}: Supervision \textbf{Mark Fuge}: Supervision 


\printbibliography

@article{soylemez2018modeling,
  title={Modeling the melt pool of the laser sintered {Ti6Al4V} layers with {G}oldak's double-ellipsoidal heat source},
  author={Soylemez, E},
  year={2018},
  publisher={University of Texas at Austin}
}

@techreport{adams2015mechanical,
  title={Mechanical Response of Additively Manufactured ({AM}) Stainless Steel {304L} across a Wide Range of Strain Rates.},
  author={Adams, David P and Carpenter, John and Reedlunn, Benjamin and Bishop, Joseph E and Palmer, Todd and Kilgo, Alice and Wise, Jack and Song, Bo and Brown, Don and Clausen, Bjorn and Carroll, Jay and Maguire, Michael Christopher},
  year={2015},
  institution={Sandia National Lab.(SNL-NM), Albuquerque, NM (United States)}
}

@article{rodgers2021simulation,
  title={Simulation of powder bed metal additive manufacturing microstructures with coupled finite difference-{Monte Carlo} method},
  author={Rodgers, Theron M and Moser, Daniel and Abdeljawad, Fadi and Jackson, Olivia D Underwood and Carroll, Jay D and Jared, Bradley H and Bolintineanu, Dan S and Mitchell, John A and Madison, Jonathan D},
  journal={Additive Manufacturing},
  volume={41},
  pages={101953},
  year={2021},
  publisher={Elsevier}
}

@article{trageser2023bezier,
  title={A B{\'e}zier curve fit to melt pool geometry for modeling additive manufacturing microstructures},
  author={Trageser, Jeremy E and Mitchell, John A and Johnson, Kyle L and Rodgers, Theron M},
  journal={Computer Methods in Applied Mechanics and Engineering},
  volume={415},
  pages={116208},
  year={2023},
  publisher={Elsevier}
}

@article{brown2019modeling,
  title={Modeling mechanical behavior of an additively manufactured metal structure with local texture variations: a study on model form error},
  author={Brown, Judith A and Bishop, Joseph E},
  journal={Modelling and Simulation in Materials Science and Engineering},
  volume={27},
  number={2},
  pages={025003},
  year={2019},
  publisher={IOP Publishing}
}

@article{cuomo2022scientific,
  title={Scientific machine learning through physics--informed neural networks: {W}here we are and what's next},
  author={Cuomo, Salvatore and Di Cola, Vincenzo Schiano and Giampaolo, Fabio and Rozza, Gianluigi and Raissi, Maziar and Piccialli, Francesco},
  journal={Journal of Scientific Computing},
  volume={92},
  number={3},
  pages={88},
  year={2022},
  publisher={Springer}
}

@article{trott2021kokkos,
  title={Kokkos 3: {P}rogramming model extensions for the exascale era},
  author={Trott, Christian R and Lebrun-Grandi{\'e}, Damien and Arndt, Daniel and Ciesko, Jan and Dang, Vinh and Ellingwood, Nathan and Gayatri, Rahulkumar and Harvey, Evan and Hollman, Daisy S and Ibanez, Dan and others},
  journal={IEEE Transactions on Parallel and Distributed Systems},
  volume={33},
  number={4},
  pages={805--817},
  year={2021},
  publisher={IEEE}
}

@article{trott2021kokkos2,
  title={The Kokkos ecosystem: {C}omprehensive performance portability for high performance computing},
  author={Trott, Christian and Berger-Vergiat, Luc and Poliakoff, David and Rajamanickam, Sivasankaran and Lebrun-Grandie, Damien and Madsen, Jonathan and Al Awar, Nader and Gligoric, Milos and Shipman, Galen and Womeldorff, Geoff},
  journal={Computing in Science \& Engineering},
  volume={23},
  number={5},
  pages={10--18},
  year={2021},
  publisher={IEEE}
}

@article{edwards2014kokkos,
  title={Kokkos: {E}nabling manycore performance portability through polymorphic memory access patterns},
  author={Edwards, H Carter and Trott, Christian R and Sunderland, Daniel},
  journal={Journal of parallel and distributed computing},
  volume={74},
  number={12},
  pages={3202--3216},
  year={2014},
  publisher={Elsevier}
}

@inproceedings{edwards2013kokkos,
  title={Kokkos: {E}nabling performance portability across manycore architectures},
  author={Edwards, H Carter and Trott, Christian R},
  booktitle={2013 Extreme Scaling Workshop (xsw 2013)},
  pages={18--24},
  year={2013},
  organization={IEEE}
}

@article{moore2024microstructure,
  title={Microstructure-based modeling of laser beam shaping during additive manufacturing},
  author={Moore, Robert and Orlandi, Giovanni and Rodgers, Theron and Moser, Daniel and Murdoch, Heather and Abdeljawad, Fadi},
  journal={JOM},
  volume={76},
  number={3},
  pages={1726--1736},
  year={2024},
  publisher={Springer}
}

@article{whitney2024part,
  title={Part-scale microstructure prediction for laser powder bed fusion Ti-6Al-4V using a hybrid mechanistic and machine learning model},
  author={Whitney, Bonnie C and Spangenberger, Anthony G and Rodgers, Theron M and Lados, Diana A},
  journal={Additive Manufacturing},
  volume={94},
  pages={104500},
  year={2024},
  publisher={Elsevier}
}

@article{whitney2024solidification,
  title={Solidification and crystallographic texture modeling of laser powder bed fusion {Ti-6Al-4V} using finite difference-{Monte Carlo} method},
  author={Whitney, Bonnie C and Rodgers, Theron M and Spangenberger, Anthony G and Rezwan, Aashique and de Oca Zapiain, David Montes and Lados, Diana A},
  journal={Materialia},
  pages={102279},
  year={2024},
  publisher={Elsevier}
}

@article{slepoy2008constant,
  title={A constant-time kinetic {Monte Carlo} algorithm for simulation of large biochemical reaction networks},
  author={Slepoy, Alexander and Thompson, Aidan P and Plimpton, Steven J},
  journal={The journal of chemical physics},
  volume={128},
  number={20},
  year={2008},
  publisher={AIP Publishing}
}

@article{holm1991effects,
  title={Effects of lattice anisotropy and temperature on domain growth in the two-dimensional {P}otts model},
  author={Holm, Elizabeth A and Glazier, James A and Srolovitz, David J and Grest, Gary S},
  journal={Physical Review A},
  volume={43},
  number={6},
  pages={2662},
  year={1991},
  publisher={APS}
}

@article{holm2001computer,
  title={The computer simulation of microstructural evolution},
  author={Holm, Elizabeth A and Battaile, Corbett C},
  journal={JOM},
  volume={53},
  pages={20--23},
  year={2001},
  publisher={Springer}
}

@article{bostanabad2018computational,
title={Computational microstructure characterization and reconstruction: {R}eview of the state-of-the-art techniques},
  author={Bostanabad, Ramin and Zhang, Yichi and Li, Xiaolin and Kearney, Tucker and Brinson, L Catherine and Apley, Daniel W and Liu, Wing Kam and Chen, Wei},
  journal={Progress in Materials Science},
  volume={95},
  pages={1--41},
  year={2018},
  publisher={Elsevier}
}

@article{kingma2013auto,
  title={Auto-encoding variational bayes},
  author={Kingma, Diederik P and Welling, Max},
  journal={arXiv preprint arXiv:1312.6114},
  year={2013}
}

@article{vahdat2020nvae,
  title={NVAE: A deep hierarchical variational autoencoder},
  author={Vahdat, Arash and Kautz, Jan},
  journal={Advances in neural information processing systems},
  volume={33},
  pages={19667--19679},
  year={2020}
}

@article{gulrajani2017improved,
  title={Improved training of wasserstein gans},
  author={Gulrajani, Ishaan and Ahmed, Faruk and Arjovsky, Martin and Dumoulin, Vincent and Courville, Aaron C},
  journal={Advances in neural information processing systems},
  volume={30},
  year={2017}
}

@article{mirza2014conditional,
  title={Conditional generative adversarial nets},
  author={Mirza, Mehdi and Osindero, Simon},
  journal={arXiv preprint arXiv:1411.1784},
  year={2014}
}

@article{chiu2023designing,
  title={Designing Bioinspired Composite Structures via Genetic Algorithm and Conditional Variational Autoencoder},
  author={Chiu, Yi-Hung and Liao, Ya-Hsuan and Juang, Jia-Yang},
  journal={Polymers},
  volume={15},
  number={2},
  pages={281},
  year={2023},
  publisher={MDPI}
}

@article{sun2022variational,
  title={Variational autoencoder-based topological optimization of an anechoic coating: An efficient-and neural network-based design},
  author={Sun, Yiping and Li, Zhaoyu and Chen, Jiadui and Zhao, Xuefeng and Tao, Meng},
  journal={Materials Today Communications},
  volume={32},
  pages={103901},
  year={2022},
  publisher={Elsevier}
}

@article{attari2023towards,
  title={Towards inverse microstructure-centered materials design using generative phase-field modeling and deep variational autoencoders},
  author={Attari, Vahid and Khatamsaz, Danial and Allaire, Douglas and Arroyave, Raymundo},
  journal={Acta Materialia},
  volume={259},
  pages={119204},
  year={2023},
  publisher={Elsevier}
}

@article{kim2021exploration,
  title={Exploration of optimal microstructure and mechanical properties in continuous microstructure space using a variational autoencoder},
  author={Kim, Yongju and Park, Hyung Keun and Jung, Jaimyun and Asghari-Rad, Peyman and Lee, Seungchul and Kim, Jin You and Jung, Hwan Gyo and Kim, Hyoung Seop},
  journal={Materials \& Design},
  volume={202},
  pages={109544},
  year={2021},
  publisher={Elsevier}
}

@article{xue2020machine,
  title={Machine learning generative models for automatic design of multi-material 3D printed composite solids},
  author={Xue, Tianju and Wallin, Thomas J and Menguc, Yigit and Adriaenssens, Sigrid and Chiaramonte, Maurizio},
  journal={Extreme Mechanics Letters},
  volume={41},
  pages={100992},
  year={2020},
  publisher={Elsevier}
}

@article{mao2020designing,
  title={Designing complex architectured materials with generative adversarial networks},
  author={Mao, Yunwei and He, Qi and Zhao, Xuanhe},
  journal={Science advances},
  volume={6},
  number={17},
  pages={eaaz4169},
  year={2020},
  publisher={American Association for the Advancement of Science}
}

@article{qian2022design,
  title={Design of architectured composite materials with an efficient, adaptive artificial neural network-based generative design method},
  author={Qian, Chao and Tan, Ren Kai and Ye, Wenjing},
  journal={Acta Materialia},
  volume={225},
  pages={117548},
  year={2022},
  publisher={Elsevier}
}

@article{liu2023reconstruction,
  title={Reconstruction of the meso-scale concrete model using a deep convolutional generative adversarial network (DCGAN)},
  author={Liu, Yifan and Zhang, Jie and Zhao, Tingting and Wang, Zhiyong and Wang, Zhihua},
  journal={Construction and Building Materials},
  volume={370},
  pages={130704},
  year={2023},
  publisher={Elsevier}
}

@article{lambard2023generation,
  title={Generation of highly realistic microstructural images of alloys from limited data with a style-based generative adversarial network},
  author={Lambard, Guillaume and Yamazaki, Kazuhiko and Demura, Masahiko},
  journal={Scientific Reports},
  volume={13},
  number={1},
  pages={566},
  year={2023},
  publisher={Nature Publishing Group UK London}
}

@article{nguyen2022synthesizing,
  title={Synthesizing controlled microstructures of porous media using generative adversarial networks and reinforcement learning},
  author={Nguyen, Phong CH and Vlassis, Nikolaos N and Bahmani, Bahador and Sun, WaiChing and Udaykumar, HS and Baek, Stephen S},
  journal={Scientific reports},
  volume={12},
  number={1},
  pages={9034},
  year={2022},
  publisher={Nature Publishing Group UK London}
}

@article{woldseth2022use,
  title={On the use of artificial neural networks in topology optimisation},
  author={Woldseth, Rebekka V and Aage, Niels and B{\ae}rentzen, J Andreas and Sigmund, Ole},
  journal={Structural and Multidisciplinary Optimization},
  volume={65},
  number={10},
  pages={294},
  year={2022},
  publisher={Springer}
}

@article{arjovsky2017towards,
  title={Towards principled methods for training generative adversarial networks},
  author={Arjovsky, Martin and Bottou, L{\'e}on},
  journal={arXiv preprint arXiv:1701.04862},
  year={2017}
}

@article{salimans2016improved,
  title={Improved techniques for training gans},
  author={Salimans, Tim and Goodfellow, Ian and Zaremba, Wojciech and Cheung, Vicki and Radford, Alec and Chen, Xi},
  journal={Advances in neural information processing systems},
  volume={29},
  year={2016}
}

@article{dhariwal2021diffusion,
  title={Diffusion models beat gans on image synthesis},
  author={Dhariwal, Prafulla and Nichol, Alexander},
  journal={Advances in neural information processing systems},
  volume={34},
  pages={8780--8794},
  year={2021}
}

@article{ho2020denoising,
  title={Denoising diffusion probabilistic models},
  author={Ho, Jonathan and Jain, Ajay and Abbeel, Pieter},
  journal={Advances in neural information processing systems},
  volume={33},
  pages={6840--6851},
  year={2020}
}

@inproceedings{rombach2022high,
  title={High-resolution image synthesis with latent diffusion models},
  author={Rombach, Robin and Blattmann, Andreas and Lorenz, Dominik and Esser, Patrick and Ommer, Bj{\"o}rn},
  booktitle={Proceedings of the IEEE/CVF conference on computer vision and pattern recognition},
  pages={10684--10695},
  year={2022}
}

@article{ramesh2022hierarchical,
  title={Hierarchical text-conditional image generation with clip latents},
  author={Ramesh, Aditya and Dhariwal, Prafulla and Nichol, Alex and Chu, Casey and Chen, Mark},
  journal={arXiv preprint arXiv:2204.06125},
  volume={1},
  number={2},
  pages={3},
  year={2022}
}

@article{vlassis2023denoising,
  title={Denoising diffusion algorithm for inverse design of microstructures with fine-tuned nonlinear material properties},
  author={Vlassis, Nikolaos N and Sun, WaiChing},
  journal={Computer Methods in Applied Mechanics and Engineering},
  volume={413},
  pages={116126},
  year={2023},
  publisher={Elsevier}
}

@article{buehler2023computational,
  title={A computational building block approach towards multiscale architected materials analysis and design with application to hierarchical metal metamaterials},
  author={Buehler, Markus J},
  journal={Modelling and Simulation in Materials Science and Engineering},
  volume={31},
  number={5},
  pages={054001},
  year={2023},
  publisher={IOP Publishing}
}

@inproceedings{rastegarzadeh2022multi,
  title={Multi-scale topology optimization with neural network-assisted optimizer},
  author={Rastegarzadeh, Sina and Wang, Jun and Huang, Jida},
  booktitle={International Design Engineering Technical Conferences and Computers and Information in Engineering Conference},
  volume={86212},
  pages={V002T02A041},
  year={2022},
  organization={American Society of Mechanical Engineers}
}

@article{wang2022ih,
  title={IH-GAN: A conditional generative model for implicit surface-based inverse design of cellular structures},
  author={Wang, Jun and Chen, Wei Wayne and Da, Daicong and Fuge, Mark and Rai, Rahul},
  journal={Computer Methods in Applied Mechanics and Engineering},
  volume={396},
  pages={115060},
  year={2022},
  publisher={Elsevier}
}

@article{chang2022machine,
  title={Machine learning-based inverse design of auxetic metamaterial with zero Poisson's ratio},
  author={Chang, Yafeng and Wang, Hui and Dong, Qinxi},
  journal={Materials Today Communications},
  volume={30},
  pages={103186},
  year={2022},
  publisher={Elsevier}
}

@article{plimpton2009crossing,
  title={Crossing the mesoscale no-man’s land via parallel kinetic {Monte Carlo}},
  author={Plimpton, Steve and Battaile, Corbett and Chandross, Mike and Holm, Liz and Thompson, Aidan and Tikare, Veena and Wagner, Greg and Webb, E and Zhou, X and Cardona, C Garcia and others},
  journal={Sandia Report SAND2009-6226},
  year={2009}
}

@article{lecun2015deep,
  title={Deep learning},
  author={LeCun, Yann and Bengio, Yoshua and Hinton, Geoffrey},
  journal={nature},
  volume={521},
  number={7553},
  pages={436},
  year={2015},
  publisher={Nature Publishing Group}
}

@article{anderson1989computer,
  title={Computer simulation of normal grain growth in three dimensions},
  author={Anderson, MP and Grest, GS and Srolovitz, DJ},
  journal={Philosophical Magazine B},
  volume={59},
  number={3},
  pages={293--329},
  year={1989},
  publisher={Taylor \& Francis}
}

@article{bishop2015direct,
  title={Direct numerical simulations in solid mechanics for understanding the macroscale effects of microscale material variability},
  author={Bishop, Joseph E and Emery, John M and Field, Richard V and Weinberger, Christopher R and Littlewood, David J},
  journal={Computer Methods in Applied Mechanics and Engineering},
  volume={287},
  pages={262--289},
  year={2015},
  publisher={Elsevier}
}

@article{bishop2016direct,
  title={Direct numerical simulations in solid mechanics for quantifying the macroscale effects of microstructure and material model-form error},
  author={Bishop, Joseph E and Emery, John M and Battaile, Corbett C and Littlewood, David J and Baines, Andrew J},
  journal={JOM},
  volume={68},
  number={5},
  pages={1427--1445},
  year={2016},
  publisher={Springer}
}

@article{garcia2008three,
  title={Three-dimensional simulation of grain growth in a thermal gradient with non-uniform grain boundary mobility},
  author={Garcia, Anthony L and Tikare, Veena and Holm, Elizabeth A},
  journal={Scripta Materialia},
  volume={59},
  number={6},
  pages={661--664},
  year={2008},
  publisher={Elsevier}
}

@article{rodgers2018direct,
  title={Direct numerical simulation of mechanical response in synthetic additively manufactured microstructures},
  author={Rodgers, Theron M and Bishop, Joseph E and Madison, Jonathan D},
  journal={Modelling and Simulation in Materials Science and Engineering},
  volume={26},
  number={5},
  pages={055010},
  year={2018},
  publisher={IOP Publishing}
}

@article{rodgers2017simulation,
  title={Simulation of metal additive manufacturing microstructures using {kinetic Monte Carlo}},
  author={Rodgers, Theron M and Madison, Jonathan D and Tikare, Veena},
  journal={Computational Materials Science},
  volume={135},
  pages={78--89},
  year={2017},
  publisher={Elsevier}
}

@article{rodgers2017monte,
  title={{A Monte Carlo model for 3D grain evolution during welding}},
  author={Rodgers, Theron M and Mitchell, John A and Tikare, Veena},
  journal={Modelling and Simulation in Materials Science and Engineering},
  volume={25},
  number={6},
  pages={064006},
  year={2017},
  publisher={IOP Publishing}
}

@article{rodgers2016predicting,
  title={Predicting mesoscale microstructural evolution in electron beam welding},
  author={Rodgers, Theron M and Madison, Jonathan D and Tikare, Veena and Maguire, Michael C},
  journal={JOM},
  volume={68},
  number={5},
  pages={1419--1426},
  year={2016},
  publisher={Springer}
}

@book{national2011materials,
  title={{Materials Genome Initiative} for global competitiveness},
  author={{US NSTC}},
  year={2011},
  publisher={Executive Office of the President, National Science and Technology Council}
}

@article{holdren2014materials,
  title={Materials genome initiative strategic plan (2014)},
  author={Holdren, John P and Kalil, Thomas and Wadia, Cyrus and Locascio, Laurie and Kung, Harriet and Horton, Linda and Warren, James},
  journal={National Science And Technology Council},
  year={2014}
}

@article{lander2021materials,
  title={Materials genome initiative strategic plan (2021)},
  author={Lander, Eric and Koizumi, Kei and Christodoulou, Julie and Sapochak, Linda and Friedersdorf, Lisa E. and Warren, James},
  journal={National Science And Technology Council},
  year={2021}
}

@article{iyer2019conditional,
  title={A conditional generative model for predicting material microstructures from processing methods},
  author={Iyer, Akshay and Dey, Biswadip and Dasgupta, Arindam and Chen, Wei and Chakraborty, Amit},
  journal={arXiv preprint arXiv:1910.02133},
  year={2019}
}

@article{rodgers2021fast,
  title={Fast three-dimensional rules-based simulation of thermal-sprayed microstructures},
  author={Rodgers, Theron M and Mitchell, John A and Olson, Aaron and Bolintineanu, Dan S and Vackel, Andrew and Moore, Nathan W},
  journal={Computational Materials Science},
  volume={194},
  pages={110437},
  year={2021},
  publisher={Elsevier}
}

@article{de2019new,
  title={New frontiers for the materials genome initiative},
  author={de Pablo, Juan J. and Jackson, Nicholas E. and Webb, Michael A. and Chen, Long-Qing and Moore, Joel E. and Morgan, Dane and Jacobs, Ryan and Pollock, Tresa and Schlom, Darrell G. and Toberer, Eric S. and Analytis, James and Dabo, Ismaila and DeLongchamp, Dean M. and Fiete, Gregory A. and Grason, Gregory M. and Hautier, Geoffroy and Mo, Yifei and Rajan, Krishna and Reed, Evan J. and Rodriguez, Efrain and Stevanovic, Vladan and Suntivich, Jin and Thornton, Katsuyo and Zhao, Ji-Cheng},
  journal={npj Computational Materials},
  volume={5},
  number={1},
  pages={1--23},
  year={2019},
  publisher={Nature Publishing Group}
}

@techreport{allison2013implementing,
  title={{Implementing ICME in the Aerospace, Automotive, and Maritime Industries}},
  author={Allison, John and Cowles, Brad and DeLoach, John and Pollock, Tresa and Spanos, George and others},
  year={2013},
  institution={The Minerals Metals and Materials Society}
}

@article{mitchell2023parallel,
  title={Parallel simulation via {SPPARKS} of on-lattice kinetic and {Metropolis Monte Carlo} models for materials processing},
  author={Mitchell, John A and Abdeljawad, Fadi and Battaile, Corbett and Garcia-Cardona, Cristina and Holm, Elizabeth A and Homer, Eric R and Madison, Jon and Rodgers, Theron M and Thompson, Aidan P and Tikare, Veena and Webb, Ed and Plimpton, Steven J},
  journal={Modelling and Simulation in Materials Science and Engineering},
  volume={31},
  number={5},
  pages={055001},
  year={2023},
  publisher={IOP Publishing}
}

@article{luo2022understanding,
  title={Understanding diffusion models: A unified perspective},
  author={Luo, Calvin},
  journal={arXiv preprint arXiv:2208.11970},
  year={2022}
}

@inproceedings{diniz2024optimizing,
  title={Optimizing Diffusion to Diffuse Optimal Designs},
  author={Diniz, Cashen and Fuge, Mark},
  booktitle={AIAA SCITECH 2024 Forum},
  pages={2013},
  year={2024}
}

@inproceedings{ester1996density,
  title={A density-based algorithm for discovering clusters in large spatial databases with noise},
  author={Ester, Martin and Kriegel, Hans-Peter and Sander, J{\"o}rg and Xu, Xiaowei and others},
  booktitle={kdd},
  volume={96},
  number={34},
  pages={226--231},
  year={1996}
}

@misc{ronneberger_u-net_2015,
    title = {U-Net: Convolutional Networks for Biomedical Image Segmentation},
    url = {http://arxiv.org/abs/1505.04597},
    shorttitle = {U-Net},
    number = {{arXiv}:1505.04597},
    publisher = {{arXiv}},
    author = {Ronneberger, Olaf and Fischer, Philipp and Brox, Thomas},
    urldate = {2024-01-23},
    date = {2015-05-18},
    langid = {english},
    eprinttype = {arxiv},
    eprint = {1505.04597 [cs]},
}

@INPROCEEDINGS{repaint,
  author={Lugmayr, Andreas and Danelljan, Martin and Romero, Andres and Yu, Fisher and Timofte, Radu and Van Gool, Luc},
  booktitle={2022 IEEE/CVF Conference on Computer Vision and Pattern Recognition (CVPR)}, 
  title={RePaint: Inpainting using Denoising Diffusion Probabilistic Models}, 
  year={2022},
  volume={},
  number={},
  pages={11451-11461},
  doi={10.1109/CVPR52688.2022.01117}}

@inproceedings{zhang2023towards,
  title={Towards Coherent Image Inpainting Using Denoising Diffusion Implicit Models},
  author={Zhang, Guanhua and Ji, Jiabao and Zhang, Yang and Yu, Mo and Jaakkola, Tommi S and Chang, Shiyu},
  booktitle={ICML},
  volume={202},
  pages={1--30},
  year={2023}
}

@article{kanwar_novel_2022,
	title = {A novel method to design biomimetic, 3D printable stochastic scaffolds with controlled porosity for bone tissue engineering},
	volume = {220},
	issn = {02641275},
	url = {https://linkinghub.elsevier.com/retrieve/pii/S0264127522004798},
	doi = {10.1016/j.matdes.2022.110857},
	pages = {110857},
	journaltitle = {Materials \& Design},
	shortjournal = {Materials \& Design},
	author = {Kanwar, Susheem and Al-Ketan, Oraib and Vijayavenkataraman, Sanjairaj},
	urldate = {2024-03-22},
	date = {2022-08},
	langid = {english},
}

@article{wang_data-driven_2019,
	title = {A Data-Driven Approach for Process Optimization of Metallic Additive Manufacturing Under Uncertainty},
	volume = {141},
	issn = {1087-1357, 1528-8935},
	url = {https://asmedigitalcollection.asme.org/manufacturingscience/article/doi/10.1115/1.4043798/726777/A-DataDriven-Approach-for-Process-Optimization-of},
	doi = {10.1115/1.4043798},
	pages = {081004},
	number = {8},
	journaltitle = {Journal of Manufacturing Science and Engineering},
	author = {Wang, Zhuo and Liu, Pengwei and Xiao, Yaohong and Cui, Xiangyang and Hu, Zhen and Chen, Lei},
	urldate = {2024-03-22},
	date = {2019-08-01},
	langid = {english},
}

@article{karaki_optimizing_2023,
	title = {Optimizing the Microstructure and Processing Parameters for Lithium‐Ion Battery Cathodes: A Use Case Scenario with a Digital Manufacturing Platform},
	volume = {11},
	issn = {2194-4288, 2194-4296},
	url = {https://onlinelibrary.wiley.com/doi/10.1002/ente.202201032},
	doi = {10.1002/ente.202201032},
	shorttitle = {Optimizing the Microstructure and Processing Parameters for Lithium‐Ion Battery Cathodes},
	pages = {2201032},
	number = {5},
	journaltitle = {Energy Technology},
	shortjournal = {Energy Tech},
	author = {Karaki, Hassan and Thomitzek, Matthias and Obermann, Tim and Herrmann, Christoph and Schröder, Daniel},
	urldate = {2024-03-22},
	date = {2023-05},
	langid = {english},
}

@article{mitchell_parallel_2023,
	title = {Parallel simulation via {SPPARKS} of on-lattice kinetic and Metropolis Monte Carlo models for materials processing},
	volume = {31},
	issn = {0965-0393, 1361-651X},
	url = {https://iopscience.iop.org/article/10.1088/1361-651X/accc4b},
	doi = {10.1088/1361-651X/accc4b},
	pages = {055001},
	number = {5},
	journaltitle = {Modelling and Simulation in Materials Science and Engineering},
	shortjournal = {Modelling Simul. Mater. Sci. Eng.},
	author = {Mitchell, John A and Abdeljawad, Fadi and Battaile, Corbett and Garcia-Cardona, Cristina and Holm, Elizabeth A and Homer, Eric R and Madison, Jon and Rodgers, Theron M and Thompson, Aidan P and Tikare, Veena and Webb, Ed and Plimpton, Steven J},
	urldate = {2024-05-29},
	date = {2023-07-01},
	langid = {english},
}

@article{fernandez4698278denoising,
  title={Denoising Diffusion Probabilistic Models for Generative Alloy Design},
  author={Fernandez Zelaia, Patxi and Thapliyal, Saket and Kannan, Rangasayee and Nandwana, Peeyush and Yamamoto, Yukinori and Paquit, Vincent and Nycz, Andrzej and Kirka, Michael M},
  journal={Available at SSRN 4698278}
}

@article{fernandez-zelaia_digital_2024,
	title = {Digital polycrystalline microstructure generation using diffusion probabilistic models},
	volume = {33},
	issn = {25891529},
	url = {https://linkinghub.elsevier.com/retrieve/pii/S2589152923003034},
	doi = {10.1016/j.mtla.2023.101976},
	pages = {101976},
	journaltitle = {Materialia},
	shortjournal = {Materialia},
	author = {Fernandez-Zelaia, Patxi and Cheng, Jiahao and Mayeur, Jason and Ziabari, Amir Koushyar and Kirka, Michael M.},
	urldate = {2024-06-26},
	date = {2024-03},
	langid = {english},
}

@article{dureth_conditional_2023,
	title = {Conditional diffusion-based microstructure reconstruction},
	volume = {35},
	issn = {23524928},
	url = {https://linkinghub.elsevier.com/retrieve/pii/S2352492823002982},
	doi = {10.1016/j.mtcomm.2023.105608},
	pages = {105608},
	journaltitle = {Materials Today Communications},
	shortjournal = {Materials Today Communications},
	author = {Düreth, Christian and Seibert, Paul and Rücker, Dennis and Handford, Stephanie and Kästner, Markus and Gude, Maik},
	urldate = {2024-06-26},
	date = {2023-06},
	langid = {english},
}

@article{vlassis_denoising_2023,
	title = {Denoising diffusion algorithm for inverse design of microstructures with fine-tuned nonlinear material properties},
	volume = {413},
	issn = {00457825},
	url = {https://linkinghub.elsevier.com/retrieve/pii/S0045782523002505},
	doi = {10.1016/j.cma.2023.116126},
	pages = {116126},
	journaltitle = {Computer Methods in Applied Mechanics and Engineering},
	shortjournal = {Computer Methods in Applied Mechanics and Engineering},
	author = {Vlassis, Nikolaos N. and Sun, {WaiChing}},
	urldate = {2024-06-26},
	date = {2023-08},
	langid = {english},
}

@article{lee_data-driven_2024,
	title = {A data-driven framework for designing microstructure of multifunctional composites with deep-learned diffusion-based generative models},
	volume = {129},
	issn = {09521976},
	url = {https://linkinghub.elsevier.com/retrieve/pii/S0952197623017748},
	doi = {10.1016/j.engappai.2023.107590},
	pages = {107590},
	journaltitle = {Engineering Applications of Artificial Intelligence},
	shortjournal = {Engineering Applications of Artificial Intelligence},
	author = {Lee, Kang-Hyun and Lim, Hyoung Jun and Yun, Gun Jin},
	urldate = {2024-06-26},
	date = {2024-03},
	langid = {english},
}

@article{azqadan_predictive_2023,
	title = {Predictive microstructure image generation using denoising diffusion probabilistic models},
	volume = {261},
	issn = {13596454},
	url = {https://linkinghub.elsevier.com/retrieve/pii/S135964542300736X},
	doi = {10.1016/j.actamat.2023.119406},
	pages = {119406},
	journaltitle = {Acta Materialia},
	shortjournal = {Acta Materialia},
	author = {Azqadan, Erfan and Jahed, Hamid and Arami, Arash},
	urldate = {2024-06-26},
	date = {2023-12},
	langid = {english},
}

@article{buzzy_statistically_2024,
	title = {Statistically conditioned polycrystal generation using denoising diffusion models},
	volume = {267},
	issn = {13596454},
	url = {https://linkinghub.elsevier.com/retrieve/pii/S1359645424000995},
	doi = {10.1016/j.actamat.2024.119746},
	pages = {119746},
	journaltitle = {Acta Materialia},
	shortjournal = {Acta Materialia},
	author = {Buzzy, Michael O. and Robertson, Andreas E. and Kalidindi, Surya R.},
	urldate = {2024-06-26},
	date = {2024-04},
	langid = {english},
}

@book{horstemeyer2012integrated,
  title={{Integrated Computational Materials Engineering (ICME)} for metals: using multiscale modeling to invigorate engineering design with science},
  author={Horstemeyer, Mark F},
  year={2012},
  publisher={John Wiley \& Sons}
}

@article{weyer2002automatic,
  title={Automatic finite element meshing of planar Voronoi tessellations},
  author={Weyer, Stefan and Fr{\"o}hlich, Andreas and Riesch-Oppermann, Heinz and Cizelj, Leon and Kovac, Marko},
  journal={Engineering fracture mechanics},
  volume={69},
  number={8},
  pages={945--958},
  year={2002},
  publisher={Elsevier}
}

@article{phan2024generating,
  title={Generating 3D images of material microstructures from a single 2D image: a denoising diffusion approach},
  author={Phan, Johan and Sarmad, Muhammad and Ruspini, Leonardo and Kiss, Gabriel and Lindseth, Frank},
  journal={Scientific Reports},
  volume={14},
  number={1},
  pages={6498},
  year={2024},
  publisher={Nature Publishing Group UK London}
}

@article{lee2024multi,
  title={Multi-plane denoising diffusion-based dimensionality expansion for 2D-to-3D reconstruction of microstructures with harmonized sampling},
  author={Lee, Kang-Hyun and Yun, Gun Jin},
  journal={npj Computational Materials},
  volume={10},
  number={1},
  pages={99},
  year={2024},
  publisher={Nature Publishing Group UK London}
}

@article{bentamou20253d,
  title={3D Denoising Diffusion Probabilistic Models for 3D microstructure image generation of fuel cell electrodes},
  author={Bentamou, Abdelouahid and Chr{\'e}tien, St{\'e}phane and Gavet, Yann},
  journal={Computational Materials Science},
  volume={248},
  pages={113596},
  year={2025},
  publisher={Elsevier}
}

\end{document}